\documentclass[aps,prx,floatfix,twocolumn,superscriptaddress,longbibliography,reprint]{revtex4-1}

\usepackage[%
	colorlinks=true,
	urlcolor=blue,
	linkcolor=blue,
	citecolor=blue,
	breaklinks=true
]{hyperref}
\usepackage[english]{babel} 
\usepackage{bm}
\usepackage{amsmath}
\usepackage{graphicx} 
\graphicspath{ {./Figures/} } 
\usepackage{color}
\usepackage{siunitx} 
\usepackage[all]{hypcap} 
\usepackage[normalem]{ulem} 

\newcommand{\pdag}{{\phantom\dagger}}

\begin{document}

\title{Determining the Electron-Phonon Coupling in Superconducting Cuprates by Resonant Inelastic X-ray Scattering: Methods and Results on Nd$_{1+x}$Ba$_{2-x}$Cu$_3$O$_{7-\delta}$}

\author{Lucio~Braicovich}
\email{lucio.braicovich@polimi.it}
\affiliation{Dipartimento di Fisica, Politecnico di Milano, Piazza Leonardo da Vinci 32, I-20133 Milano, Italy}
\affiliation{ESRF -- The European Synchrotron, 71 Avenue des Martyrs, CS 40220, F-38043 Grenoble, France}

\author{Matteo~Rossi}
\email{rossim@stanford.edu}
\altaffiliation[Present address: ]{Stanford Institute for Materials and Energy Sciences, SLAC National Accelerator Laboratory, 2575 Sand Hill Road, Menlo Park, CA 94025, USA}
\affiliation{Dipartimento di Fisica, Politecnico di Milano, Piazza Leonardo da Vinci 32, I-20133 Milano, Italy}

\author{Roberto~Fumagalli}
\affiliation{Dipartimento di Fisica, Politecnico di Milano, Piazza Leonardo da Vinci 32, I-20133 Milano, Italy}

\author{Yingying~Peng}
\altaffiliation[Present address: ]{International Center for Quantum Materials, School of Physics, Peking University, Beijing 100871, China}
\affiliation{Dipartimento di Fisica, Politecnico di Milano, Piazza Leonardo da Vinci 32, I-20133 Milano, Italy}

\author{Yan~Wang}
\affiliation{Department of Physics and Astronomy, The University of Tennessee, Knoxville, TN 37996, USA}

\author{Riccardo~Arpaia}
\affiliation{Dipartimento di Fisica, Politecnico di Milano, Piazza Leonardo da Vinci 32, I-20133 Milano, Italy}
\affiliation{Department of Microtechnology and Nanoscience, Chalmers University of Technology, SE-41296 G\"oteborg, Sweden}

\author{Davide~Betto}
\altaffiliation[Present address: ]{Max Planck Institut f\"{u}r Festk\"{o}rperforschung, Heisenbergstrasse 1, D-70569 Stuttgart, Germany}
\affiliation{ESRF -- The European Synchrotron, 71 Avenue des Martyrs, CS 40220, F-38043 Grenoble, France}

\author{Gabriella~M.~{De~Luca}}
\affiliation{Dipartimento di Fisica ``E. Pancini'', Universit\`{a} degli Studi di Napoli ``Federico II'', Complesso Monte Sant'Angelo - Via Cinthia, I-80126 Napoli, Italy}
\affiliation{CNR-SPIN, Complesso Monte Sant'Angelo - Via Cinthia, I-80126 Napoli, Italy}

\author{Daniele~{Di~Castro}}
\affiliation{CNR-SPIN and Dipartimento di Ingegneria Civile e Ingegneria Informatica, Universit\`{a} di Roma Tor Vergata, Via del Politecnico 1, I-00133 Roma, Italy}

\author{Kurt~Kummer}
\affiliation{ESRF -- The European Synchrotron, 71 Avenue des Martyrs, CS 40220, F-38043 Grenoble, France}

\author{Marco~{Moretti~Sala}}
\affiliation{Dipartimento di Fisica, Politecnico di Milano, Piazza Leonardo da Vinci 32, I-20133 Milano, Italy}

\author{Mattia~Pagetti}
\affiliation{Dipartimento di Fisica, Politecnico di Milano, Piazza Leonardo da Vinci 32, I-20133 Milano, Italy}

\author{Giuseppe~Balestrino}
\affiliation{CNR-SPIN and Dipartimento di Ingegneria Civile e Ingegneria Informatica, Universit\`{a} di Roma Tor Vergata, Via del Politecnico 1, I-00133 Roma, Italy}

\author{Nicholas~B.~Brookes}
\affiliation{ESRF -- The European Synchrotron, 71 Avenue des Martyrs, CS 40220, F-38043 Grenoble, France}

\author{Marco~Salluzzo}
\affiliation{CNR-SPIN, Complesso Monte Sant'Angelo - Via Cinthia, I-80126 Napoli, Italy}

\author{Steven~Johnston}
\affiliation{Department of Physics and Astronomy, The University of Tennessee, Knoxville, TN 37996, USA}
\affiliation{Joint Institute for Advanced Materials, The University of Tennessee, Knoxville, TN 37996, USA}
\affiliation{Institute for Theoretical Solid State Physics, IFW Dresden, Helmholtzstrasse 20, D-01069 Dresden, Germany}

\author{Jeroen~{van~den~Brink}}
\affiliation{Institute for Theoretical Solid State Physics, IFW Dresden, Helmholtzstrasse 20, D-01069 Dresden, Germany}
\affiliation{Department of Physics, Technical University Dresden, D-01062 Dresden, Germany}
\affiliation{Department of Physics, Washington University, St.~Louis, MO 63160, USA}

\author{Giacomo~Ghiringhelli}
\email{giacomo.ghiringhelli@polimi.it}
\affiliation{Dipartimento di Fisica, Politecnico di Milano, Piazza Leonardo da Vinci 32, I-20133 Milano, Italy}
\affiliation{CNR-SPIN, Dipartimento di Fisica, Politecnico di Milano, Piazza Leonardo da Vinci 32, I-20133 Milano, Italy}

\date{\today}

\begin{abstract}
The coupling between lattice vibration quanta and valence electrons can induce charge density modulations and decisively influence the transport properties of materials, e.g., leading to conventional superconductivity. In high critical temperature superconductors, where electronic correlation is the main actor, the actual role of electron-phonon coupling (EPC) is being intensely debated theoretically and investigated experimentally. We present an in-depth study of how the EPC strength can be obtained directly from resonant inelastic x-ray scattering (RIXS) data through the theoretical approach derived by Ament \emph{et al}. [\href{http://stacks.iop.org/0295-5075/95/i=2/a=27008}{EPL {\bf 95}, 27008 (2011)}]. The role of the model parameters (e.g., phonon energy $\omega_0$, intermediate state lifetime $1/\Gamma$, EPC matrix element $M$, and detuning energy $\Omega$) is thoroughly analyzed, providing general relations among them that can be used to make quantitative estimates of the dimensionless EPC $g = (M/\omega_0)^2$ without detailed microscopic modeling. We then apply these methods to very high resolution Cu $L_3$ edge RIXS spectra of three Nd$_{1+x}$Ba$_{2-x}$Cu$_3$O$_{7-\delta}$ films. For the insulating antiferromagnetic parent compound, the value of $M$ as a function of the in-plane momentum transfer is obtained for Cu-O bond-stretching (breathing) and bond-bending (buckling) phonon branches. For the underdoped and the nearly optimally doped samples, the effects of Coulomb screening and of charge-density-wave correlations on $M$ are assessed. In  light of the anticipated further improvements of the RIXS experimental resolution, this work provides a solid framework for an exhaustive investigation of the EPC in cuprates and other quantum materials. 
\end{abstract}

\maketitle

\section{Introduction}
\label{sec:introduction}

The role of the electron-phonon coupling (EPC) in the high critical temperature (high-$T_\mathrm{c}$) superconducting cuprates is still an open problem deserving further research. Indeed, even if pairing is not of the phonon-mediated BCS type~\cite{Scalapino2012}, the question of a potential role for the EPC remains extremely interesting. For instance, it has been suggested theoretically \cite{Savrasov1996,Andersen1996,Sakai1997,Jepsen1998,Ishihara2004,Johnston2010} that a synergy between a suitable phonon and other pair-driving excitations can greatly enhance the critical temperature $T_\mathrm{c}$. In particular, in the case of magnetic excitations, even a small amount of EPC (\emph{per se} irrelevant) should be sufficient to considerably increase $T_\mathrm{c}$ \cite{Johnston2010}. Another reason for studying the EPC is the recent observation of transient superconductivity induced in cuprates by strong illumination with mid-infrared pulses \cite{Fausti2011,Kaiser2014,Hu2014,Liu2019}, i.e., more precisely, the increase of $T_\mathrm{c}$ under optical pumping of apical oxygen phonons in the non-linear regime. Moreover, the interplay of phonons with the electronic states is very important in the presence of charge-density-wave (CDW) correlations. Indeed, the low-energy acoustic \cite{LeTacon2013,Miao2018} and high-energy Cu-O bond-stretching \cite{McQueeney1999,Pintschovius2004,Uchiyama2004,Reznik2006,Graf2008,Chaix2017} phonon branches are strongly modulated in intensity and energy around the CDW wave vector. These examples, including results from the recent literature, provide new clues towards understanding the role of the EPC in the cuprates, on top of the long debated polaronic behavior of carriers in this class of materials~\cite{Kastner1998,Alexandrov2000,Rosch2004,Shen2007}.

Clarifying the role of the EPC in the cuprates and other strongly correlated systems is a formidable task because the EPC acts on top of strong electron-electron interactions, giving rise to a correlated electron liquid where phononic and electronic degrees of freedom are highly entangled. This difficulty is reflected in broadly discrepant estimates of the EPC, not only between theory \cite{Bohnen2003,Giustino2008,Reznik2008,Heid2008,Veenstra2010} and experiment \cite{Lanzara2001,Pintschovius2004,Zhou2005,Cuk2005,Reznik2006,Johnston2012} but also among different theoretical approaches~\cite{Savrasov1996,Bohnen2003,Giustino2008,Reznik2008}. For example, the Cu-O bond-stretching modes have been proposed to mediate both an effective attractive \cite{Shen2002,Ishihara2004} and, more recently, repulsive \cite{Bulut1996,Sandvik2004,Johnston2010} interaction. 

The experimental verification of the phenomenology of the EPC, possibly with momentum resolution, is crucial in this context. This goal calls for the improvement of the existing techniques and for the introduction of new methods. Among traditional measurements, bulk-sensitive neutron scattering \cite{Pintschovius2005,Reznik2010} suffers from sensitivity limitations as it requires massive homogeneous samples, whereas surface-sensitive angle-resolved photoemission spectroscopy \cite{Damascelli2003,Cuk2005} and scanning tunneling microscopy \cite{Lee2006} need high quality surfaces that are not always available. Moreover, the comparison with bulk sensitive results is a nontrivial task. In this field, resonant inelastic soft x-ray scattering (RIXS) is likely to offer an important contribution, made possible by the recent progress in instrumentation~\cite{Brookes2018}. For instance, state-of-the-art RIXS can be performed with a combined energy resolution of $\sim 30$ -- 40~meV at the Cu $L_3$ edge, which facilitates studies of high-energy phonons ($\gtrsim 30$~meV). For detailed investigations, a further increase of the resolution is needed, which is a realistic prospect in the next three to five years. What is needed in this context is a well defined procedure for extracting quantitative information about the EPC from the available RIXS data, preferably in a model independent way.

Determining the EPC from RIXS data was first proposed by Ament \emph{et al.}~\cite{Ament2011} (for a detailed account see Ref.~\onlinecite{Ament2010}). This idea has also been reinforced by the theoretical treatment by Devereaux \emph{et al.}~\cite{Devereaux2016}, based on the lowest order Feynman diagrams. Devereaux \emph{et al.}'s calculations on single layer Bi-based cuprates show very distinct RIXS signals for the main phonons excited at the Cu $L_3$ edge. The relationship between the EPC and the RIXS signal has been expanded recently by Geondzhian and Gilmore \cite{Geondzhian2018} with a model conceptually similar to Ref.~\onlinecite{Ament2011}.

Due to the limited energy resolution, RIXS experiments on lattice excitations were mainly carried out at low incident energies, i.e., the O $K$ edge in cuprates and other functional oxides and $L$ edges of systems containing elements at the beginning of the $3d$ series. A convenient option are the quasi-one-dimensional compounds~\cite{Lee2013,Johnston2016}, having in general more pronounced phonon peaks and van Hove singularities. A recent remarkable work has also reported the first application to $n$SrIrO$_3$/$m$SrTiO$_3$ multilayers~\cite{Meyers2018}. For the low-Z $3d$ oxides, data on Ti $L$ edge are also available~\cite{Moser2015,Fatale2016}.

From a conceptual point of view, state-of-the-art RIXS research on the EPC has proceeded along two main lines: \emph{i)} the search for a simplified universal scheme allowing experimentalists to address a variety of cases without detailed calculations; \emph{ii)} the development of more specific treatments that exploit numerical computation. While both approaches are useful and cross fertilize each other, in the following we focus on the first one. Specifically, we introduce innovative ways of using the theory by Ament \emph{et al.} \cite{Ament2011} and identify some scaling laws with a wide range of application. The proposed methods are then employed on the cuprate Nd$_{1+x}$Ba$_{2-x}$Cu$_3$O$_{7-\delta}$ (NBCO). We examine the EPC in the antiferromagnetic (AF) compound and the effect of charge order \cite{Ghiringhelli2012,daSilvaNeto2014,daSilvaNeto2015,Comin2016,Arpaia2019} on the EPC at different doping levels. From the RIXS data, we quantitatively determine the EPC strength in NBCO-AF and qualitatively discuss its doping dependence in relation to the presence of CDW correlations in the system. We selected a compound of the so-called ``123'' family of cuprates since the CDW signal is perhaps the strongest in this case~\cite{Comin2016}. Our work also paves the way for more advanced experiments in the future.

The article is organized as follows. Section~\ref{sec:exp_details} describes the samples and the experimental setup. Section~\ref{sec:EPC_extraction} provides an in-depth discussion of our theoretical framework. We first review the main results of the theory as developed previously. We then thoroughly explore the origin of lattice excitations during the RIXS process and discuss the role of the various parameters that govern the phonon intensity. Importantly, we unveil universal scaling laws that involve the main parameters of the theory. We also examine three methods for determining the EPC from RIXS data. Although some of them have been already introduced elsewhere \cite{Ament2011,Rossi2019}, we provide novel and efficient ways to exploit them based on some general scaling laws. We also compare the methods and point out their advantages and limitations. Moreover, we show how to extract the momentum dependence of the EPC by combining experimental and theoretical information. Finally, we conclude Section~\ref{sec:EPC_extraction} by briefly addressing the case of absorption edges other than the Cu $L_3$ edge. Section~\ref{sec:results} presents the experimental RIXS results on the bilayer cuprate NBCO, with emphasis on the identification of the phonon modes, determination of the EPC on a quantitative basis, and an inspection of doping effects. Finally, Section~\ref{sec:discussion} provides some additional discussion of our results.

\section{Experimental details}
\label{sec:exp_details}

\begin{figure}[t]
	\centering
	\includegraphics[width=\columnwidth]{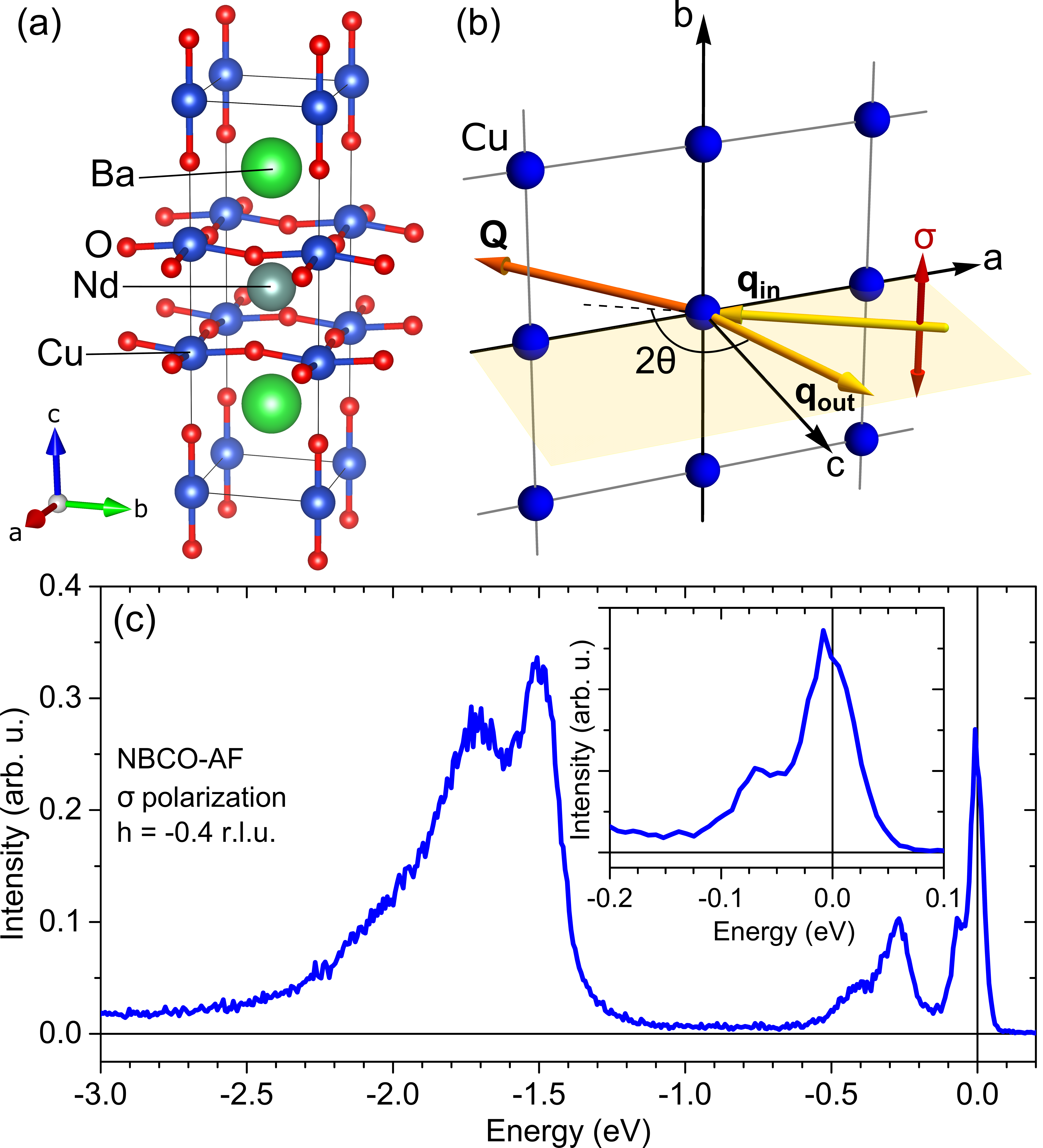}
	\caption{\label{fig:fig1} (a) The crystal structure of undoped antiferromagnetic NBCO. (b) The layout of the experiment. (c) An example RIXS spectrum of NBCO collected at $\mathbf{Q}_\parallel = (\frac{2\pi}{a}h,0)$ with $h = -0.4$ and $T = 20$ K. A wide energy range covering the $dd$ excitations is displayed. The inset shows a close-up of the region of interest to the present work, which clearly reveals} a phonon feature at approximately 70~meV.
\end{figure}

\subsection{The NBCO samples}
\label{sec:samples}

The crystal structure of  Nd$_{1+x}$Ba$_{2-x}$Cu$_3$O$_{7-\delta}$ (NBCO) is displayed in Fig.~\ref{fig:fig1}(a) for the undoped antiferromagnetic parent compound ($x = 0$ and $\delta = 1$). NBCO belongs to the so-called ``123'' family of high $T_\mathrm{c}$ superconductors and is isostructural to YBa$_2$Cu$_3$O$_{7-\delta}$ (YBCO). Of primary importance in the centrosymmetric unit cell is the presence of the CuO$_2$ bilayer where superconductivity occurs. The hole density per CuO$_2$ plane can be changed by modifying the oxygen content $\delta$ (similarly to YBCO \cite{Arpaia2018}) or by introducing excess Nd $x$ at the Ba sites. In the latter case, 0.5 holes are removed from the CuO$_2$ planes for each excess Nd~\cite{Salluzzo2002,Salluzzo2005}. At optimal doping, $T_\mathrm{c}$ reaches 93 K. NdBa$_2$Cu$_3$O$_6$ has a tetragonal crystal structure;  however, the system becomes orthorhombic upon doping. For the purpose of the present study, we neglect orthorhombicity and adopt a tetragonal description. The CuO$_2$ planes are characterized by a buckling of the Cu-O bonds, which depends on the overall stoichiometry. As a general rule, the buckling angle increases from underdoped to overdoped NBCO~\cite{Chmaissem1999}.

Epitaxial NBCO films were deposited on a $(001)$-oriented SrTiO$_3$ substrate with an almost perfect in-plane matching of the $a$ and $b$ lattice parameters and the buckling practically unchanged with respect to the corresponding bulk structure~\cite{Salluzzo2002,Salluzzo2005}. The thickness of the NBCO films is around 150 nm, hence they can be considered infinitely thick for the scattering angles used in our experiment (where the incidence angle with respect to the surface ranged from \SI{25}{\degree} to \SI{65}{\degree}). 

We investigated three NBCO samples:
\begin{itemize}
	\item NBCO-AF: the sample is antiferromagnetic with $x \approx 0$, $\delta \approx 0.9$, and with a Cu-O-Cu buckling angle of \SI{6.35}{\degree} \cite{Takita1988,Kramer1994};
	\item NBCO-UD: the sample is underdoped with $T_\mathrm{c} = 63$ K,  hole doping of 0.11, $x \approx 0.2$, $\delta \approx 0$, and a buckling angle of \SI{6.74}{\degree} \cite{Izumi1987}; 
	\item NBCO-OP: the sample is very close to optimal doping with $T_\mathrm{c} = 90$ K, $x \approx 0$, $\delta \approx 0$, and a buckling angle of \SI{7.75}{\degree} \cite{Takita1988}. The slightly lower $T_\mathrm{c}$ compared to optimal doping is due to a slight overdoping, which has been identified by measuring the $c$ lattice parameter.
\end{itemize}

While the superconducting NBCO films are stable in time, the AF samples are more likely to grow in doping by capturing oxygen from the atmosphere when stored for weeks or months at room temperature. As a result, the actual doping of the sample here named AF was not perfectly zero, and the upper limit for its hole doping level is $p\simeq 0.04$. In RIXS spectra, a fingerprint of the doping level appears in the formation of a continuum in the otherwise empty region around 0.8~eV energy loss. We were thus able to monitor the sample during the experiment to ensure that it stayed in the AF insulating region of the phase diagram.  In any case, the exact doping level of this sample is not relevant for the present work. The UD and OP samples are at risk of loosing oxygen if kept in vacuum at room temperature. This risk is why we have performed all the measurements at 20 K, where radiation induced and spontaneous loss of oxygen is minimized.

\subsection{Experimental layout and energy resolution}

The measurements were performed at beam line ID32 of the European Synchrotron (ESRF, France), which is equipped with the new spectrometer ERIXS (European RIXS)~\cite{Brookes2018}. The layout of the experiment is sketched in Fig.~\ref{fig:fig1}(b). The sample is mounted on the in-vacuum diffractometer with the $ab$ planes perpendicular to the horizontal scattering plane, defined by the incoming and outgoing photon wave vectors ($\mathbf{q}_\mathrm{in}$ and $\mathbf{q}_\mathrm{out}$, respectively). The scattering angle $2\theta$ was fixed to \SI{149.5}{\degree}. Most of the measurements were performed with a combined (beam line $+$ spectrometer) energy resolution of $\sim 40$~meV at the Cu $L_3$ edge ($\sim 930$~eV). The contribution to the resolution from the incident beam is $\sim 25$~meV. One spectrum was collected with high statistics and high energy resolution (32~meV) to be used as benchmark.  The sample was kept at a temperature of 20K during the measurements.

The momentum transferred to the sample $\mathbf{Q} = \mathbf{q}_\mathrm{in} - \mathbf{q}_\mathrm{out}$ and projected on the basal $ab$ plane is the relevant observable in dispersion studies of two-dimensional cuprates. The projected in-plane momentum $\mathbf{Q}_\parallel$ is scanned in the $\Gamma$-$X$ direction of the Brillouin zone (BZ) by rotating the sample around the $b$ axis, i.e., $\mathbf{Q}_\parallel = \left(\frac{2\pi}{a} h, 0\right)$. The rotation direction was chosen in order to minimize the incidence angle to the surface, which reduces self-absorption effects, one of the experimental difficulties of RIXS. The incident beam polarization was perpendicular to the scattering plane [$\sigma$ incident polarization, red arrow in Fig.~\ref{fig:fig1}(b)]. Thus, the electric field vector lay in the basal plane, simplifying the problem (see Sec.~\ref{sec:theory}). We did not investigate small momentum transfers $|\mathbf{Q}_\parallel| < 0.1$~r.l.u. because the phonon and magnon signals begin to overlap in this region, making it difficult to identify the phonon contribution. In principle the lattice and magnetic excitations can be separated by measuring the scattered beam polarization. For example, such an analysis has been performed in Ref.~\onlinecite{Hepting2018}, where the use of the polarimeter allowed the authors to isolate the plasmon signal from the magnetic one. The present case is technically more complex, however, and so we have postponed this part of the work until after the commissioning of the new storage ring at the ESRF is complete. Note that the analysis of the scattered beam polarization in the soft x-ray range is possible at present only at beamline ID32 of the ESRF.

Figure~\ref{fig:fig1}(c) displays an example of RIXS spectrum of NBCO-AF over an extended energy range. The inset provides a close-up of the region of interest to the present work. Note that a phonon peak around 70~meV is clearly visible in the data.

\section{Tools to determine the electron-phonon coupling from RIXS spectra}
\label{sec:EPC_extraction}

\subsection{Theoretical background}
\label{sec:theory}

The present work is based on the theory by Ament \emph{et al.}~\cite{Ament2010,Ament2011}, which was laid down for localized electrons coupled to dispersive phonons via a momentum-dependent EPC $M({\bf k},{\bf q})$ and then simplified to the case of coupling to local phonons. The approximation of Einstein phonons is appropriate for most of the medium- and high-energy optical branches in cuprates, which are generally found to be weakly dispersing~\cite{Pintschovius2005,Reznik2010}. Moreover, the assumption of a 
localized electrons can be justified by appealing to the strong core-hole potential 
that will have a tendency to localize the excited core electron. We will come back to this issue in section \ref{sec:discussion}.
Here, we do not introduce any changes in the theory but, instead, propose an innovative way of using it. For completeness, we now summarize the main theoretical results of this approach. The following equations are rather intuitive in spite of their complexity at first sight. For a rigorous derivation, the reader is referred to Ref.~\onlinecite{Ament2010}. 

In the case of dispersionless phonons -- which is the situation we are interested in -- the starting Hamiltonian is
\begin{equation}
\mathcal{H} = \sum_{i} \omega^\pdag_0 b_i^\dagger b^\pdag_i + M \sum_i d_i^\dagger d^\pdag_i (b_i^\dagger + b^\pdag_i),
\label{eq:Hamiltonian}
\end{equation}
where $\omega_0$ is the phonon energy, $M$ is the EPC matrix element, and $b_i^\dagger$ ($b^\pdag_i$) and $d_i^\dagger$ ($d^\pdag_i$) are the creation (annihilation) operators for phonons and electrons, respectively, at site $i$. We have neglected the spin index for brevity. Eq.~\eqref{eq:Hamiltonian} can be diagonalized exactly using a canonical Lang-Firsov transformation $\mathcal{H^\prime} = e^S\mathcal{H}e^{-S}$, where $S = \sum_i S_i$ and $S_i = \frac{M}{\omega_0} d^\dagger_id^\pdag_i\left(b^\dagger_i - b^\pdag_i\right)$. This same transformation can be used to derive an exact expression for the scattering amplitude:
\begin{equation}
\mathcal{F} = T_\mathrm{el}(\bm{\epsilon}^\prime, \bm{\epsilon}) \sum_{i} e^{i\mathbf{Q}\cdot\mathbf{R}_i}\sum_{n_i=0}^{\infty}\frac{\langle n_i^\prime | e^{-S_i} | n_i\rangle \langle n_i | e^{S_i} | n_i^0 \rangle}{z + M^2/\omega_0 - n_i \omega_0}.
\label{eq:scattering_amplitude}
\end{equation}
Here, $T_\mathrm{el}$ is the polarization-dependent atomic elastic scattering factor; $z = \Omega + i\Gamma$ is a complex number whose real part corresponds to the detuning energy $\Omega$ (i.e., the difference between the incoming photon energy and the resonance energy) and whose imaginary part $\Gamma$ is related to inverse lifetime of the core hole. Note that $\Gamma$ is  the Half Width at Half Maximum (HWHM) of the core line. For simplicity we will refer to $1/\Gamma$ as the inverse lifetime.  $|n_i^0\rangle$ denotes the initial ground state of the Hamiltonian, which corresponds to the ground state of a shifted Harmonic oscillator. 

By introducing the Franck-Condon (FC) coefficients
\begin{equation}
B_{ab} = \sqrt{e^{-g}a!b!} \sum_{l=0}^{b}\frac{(-1)^a(-g)^l g^{(a-b)/2}}{(b-l)!l!(a-b+l)!},
\end{equation}
the cross section can be written as:
\begin{widetext}
	\begin{equation}
		\frac{\mathrm{d}^2 \sigma}{\mathrm{d}\Omega\mathrm{d}\omega} \propto \sum_{f}|\mathcal{F}|^2\delta(\omega-n^\prime\omega_0) = N|T_\mathrm{el}(\bm{\epsilon}^\prime, \bm{\epsilon})|^2 \sum_{n^\prime=0}^{\infty}\left|\sum_{n=0}^{n^\prime}\frac{B_{n^\prime n}(g)B_{n0}(g)}{z + (g-n)\omega_0} + \sum_{n=n^\prime+1}^{\infty}\frac{B_{nn^\prime}(g)B_{n0}(g)}{z + (g-n)\omega_0}\right|^2 \delta(\omega - n^\prime\omega_0),
		\label{eq:cross_section}
	\end{equation}
\end{widetext}
where $g = (M/\omega_0)^2$ is a dimensionless coupling constant and the index $n^\prime$ identifies the number of phonons in each final state $f$ indexing  the first summation. We note that a sum with index $n$, the number of phonons excited in the intermediate state, has to be extended to a potentially very large number (in principle to infinity).

For convenience, we report the exact expressions for the one-phonon intensity $I_1$ and two-phonon intensity $I_2$, corresponding to the $n^\prime=1,2$ final states,  respectively:
\begin{eqnarray} \label{Eq5}
    I_1 & \propto & \frac{e^{-2g}}{g}\left| \sum_{n=0}^{\infty} \frac{g^n (n - g)}{n! \left(\Omega + i\Gamma + (g - n)\omega_0 \right)} \right|^2, \\
    \label{Eq6}
    I_2 & \propto & \frac{e^{-2g}}{2g^2}\left| \sum_{n=0}^{\infty} \frac{g^n \left((n - g)^2 - n \right)}{n! \left(\Omega + i\Gamma + (g - n)\omega_0 \right)} \right|^2.
\end{eqnarray}
Note that a common prefactor has been dropped from the two equations.

\subsection{The genesis of the phonon excitations during the RIXS process}
\label{sec:phonon_genesis}

\begin{figure}
	\centering
	\includegraphics[width=1.0\columnwidth]{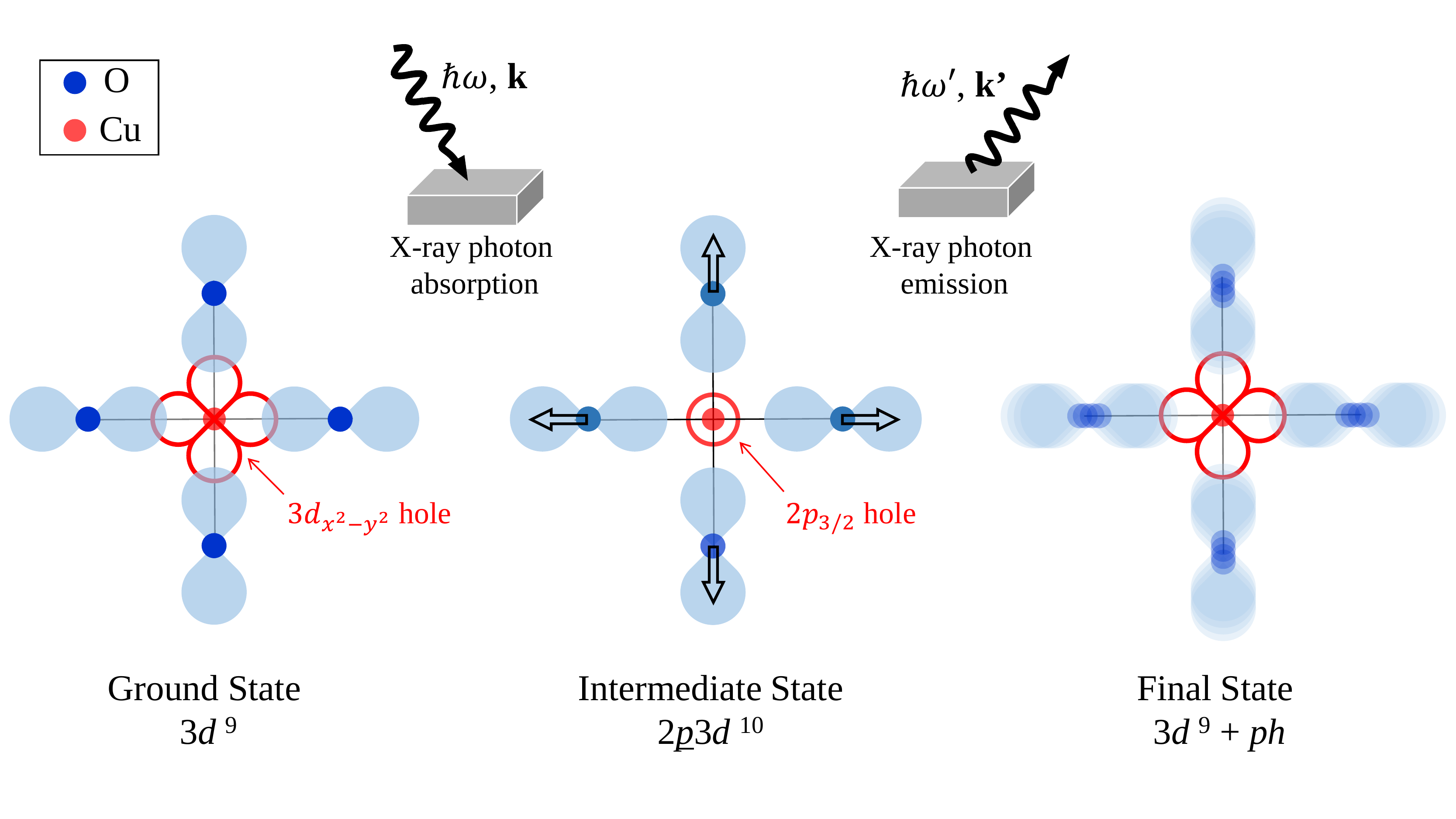}
	\caption{\label{fig:fig2} A sketch of the RIXS process leading to low-energy phonon excitations at the Cu $L_3$ edge. Upon absorption of a photon, a $2p_{3/2}$ electron is promoted into an empty $3d$ state and a core hole is left in the Cu ion. Due to the attraction between the core hole and the excited electron, the intermediate charge distribution has an excitonic character. When viewed from the oxygen ion, the core hole is well screened while the weakened Cu-O bond pushes the oxygen ions towards an equilibrium position at farther distance. In the intermediate state, several phonons are thus excited. When the core hole is filled by the electron that was initially promoted to the valence state and a photon is emitted (radiative decay), the system reaches its original electronic ground state, but one or more phonons are left behind in the sample.}
\end{figure}

The excitation of a phonon during the Cu $L$-edge RIXS process is sketched in Fig.~\ref{fig:fig2}. Upon photon absorption at the Cu $L_3$ resonance, an excited state is created having a $2p_{3/2}$ core hole and an extra electron in a $3d$ orbital. In layered cuprates, the ground state has a $3d_{x^2-y^2}$ symmetry of a single local hole, corresponding to an empty antibonding molecular orbital, which is  temporarily filled in the intermediate state. The intermediate state is, therefore, not an eigenstate of the electron-lattice coupled system, and the lattice will deform towards the new equilibrium structure where the oxygen atoms are at a different distance from the central Cu ion. This intermediate state can decay back to the original $3d^9$ configuration but with one or several excited phonons. The detailed nature of the lattice excitations is specific to the intermediate state (and thus depends on the core and valence states involved in the RIXS process) but the general scheme is always the one we have described. Note that the intermediate state has excitonic character, i.e., it is charge neutral. In a sense, we can consider RIXS as a way of introducing a probing charge to measure the EPC (i.e., the value of $g$) while maintaining charge neutrality. In principle, the value of $g$ measured with RIXS is different from the value $g_\mathrm{t}$ involved in the transport measurements, as stressed in Ref.~\onlinecite{Geondzhian2018}, because of the core hole effect and the symmetry of the intermediate state. However, the strong screening of the core hole by the excited electron reduces the difference between $g$ and $g_\mathrm{t}$. The core hole's role in localizing the excited electron is crucial to this process; it is the excitonic nature of the intermediate state that makes $g$ and $g_\mathrm{t}$ close. 

Let us first consider the effect of the core hole lifetime, which introduces a time scale into the problem. It is intuitive that the phonon signal increases with the EPC matrix element $M$ and that it decreases with the core hole lifetime ($1/\Gamma$), as shorter lived core hole states provide less time for the lattice to evolve towards the new equilibrium configuration. The logical consequence is that the phonon excitation efficiency by RIXS depends on the ratio $M/\Gamma$, irrespective of the value of $g$, over a broad range of $M/\Gamma$.

\begin{figure}
	\centering
	\includegraphics[width=0.95\columnwidth]{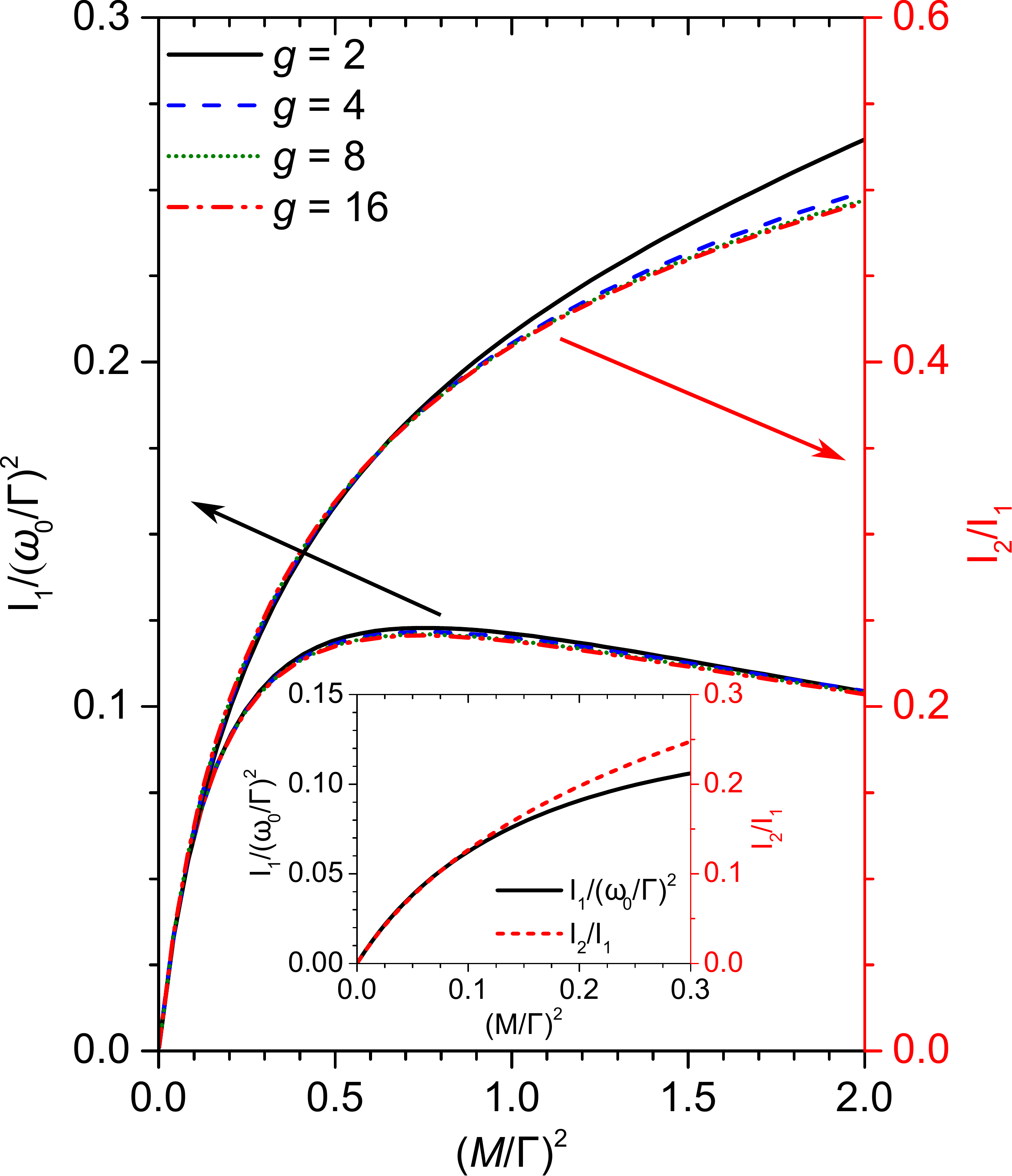}
	\caption{\label{fig:fig3} The universal plots of the phonon excitation intensities based on rescaled dimensionless variables. $M$ is the absolute value of the EPC matrix element, $\Gamma$ is the intrinsic half width at half maximum of the Cu $L_3$ resonance so that $1/\Gamma$ is proportional to the core hole lifetime, and $g = (M/\omega_0)^2$ is a dimensionless coupling constant. Left axis: The intensity of the one-phonon excitation $I_1$ rescaled by $(\omega_0/\Gamma)^2$ and plotted as a function of $(M/\Gamma)^2$. The different lines corresponding to different values of $g$ superimpose almost perfectly, so that the curve is universal. Right axis: The ratio between the two- and one-phonon excitation intensities as a function of $(M/\Gamma)^2$. The behavior is also universal for $(M/\Gamma)^2 \lesssim 1.5$. The inset shows that in the limit of small coupling, the universal curves of $I_1$ (solid black line) and $I_2/I_1$ (dashed red line) share the same behavior as a function of $(M/\Gamma)^2$ apart from an overall factor.}
\end{figure}

The universal dependence of the scattering cross section on $M/\Gamma$ is confirmed by numerical calculations made with Eqs.~\eqref{Eq5} and \eqref{Eq6} and is summarized in Fig.~\ref{fig:fig3}. On the left axis, we plot the intensity of the one-phonon excitation $I_1$, normalized to $(\omega_0/\Gamma)^2$, for several values of the dimensionless coupling constant $g = (M/\omega_0)^2$. It is evident that all curves corresponding to different $g$ values collapse on the same ``universal'' curve when $(M/\Gamma)^2 \lesssim 2$. Above this value, the curves begin to diverge and the behavior is no longer universal. Similar considerations apply to the ratio of the intensities of the one-phonon and two-phonon excitations $I_2/I_1$  (right axis). We note that for an Einstein phonon, the two-phonon excitation has exactly twice the energy of a single phonon, which would not be the case for dispersing phonon branches. In this simplified picture, it is interesting to note that $I_2/I_1$ vs. $(M/\Gamma)^2$ is independent of $g$ for $(M/\Gamma)^2 \lesssim 1$. At the Cu $L_3$ resonance, $\Gamma \approx 250$ -- $280$~meV \cite{KeskiRahkonen1974,Krause1979} so the universality range extends up to about $M \approx 350$~meV. This value is much larger than most estimates for the EPC constants appearing in the literature. Thus, we can safely say that the intensity of the high-energy phonon excitations in cuprates universally scales with $(M/\Gamma)^2$.

For weak interactions, the intensity $I_2$ is directly proportional to $(I_1)^2$ by definition \cite{Ament2011}, so that $I_2/I_1$ coincides with $I_1$ apart from an overall factor. As a matter of fact, the inset of Fig.~\ref{fig:fig3} shows that the two curves are superimposed at low interactions after rescaling. The region of small interaction extends up to $(M/\Gamma)^2 \approx 0.12$. Note that the condition of linearity is even more restrictive, as it is evident from the figure. 

The universal plots can be exploited in several ways to extract the EPC from experimental data. A critical comparison between the different methods is presently lacking in the literature and we summarize a number of key points in the following subsections.

\subsection{The use of $I_2/I_1$}

\begin{figure}
	\centering
	\includegraphics[width=\columnwidth]{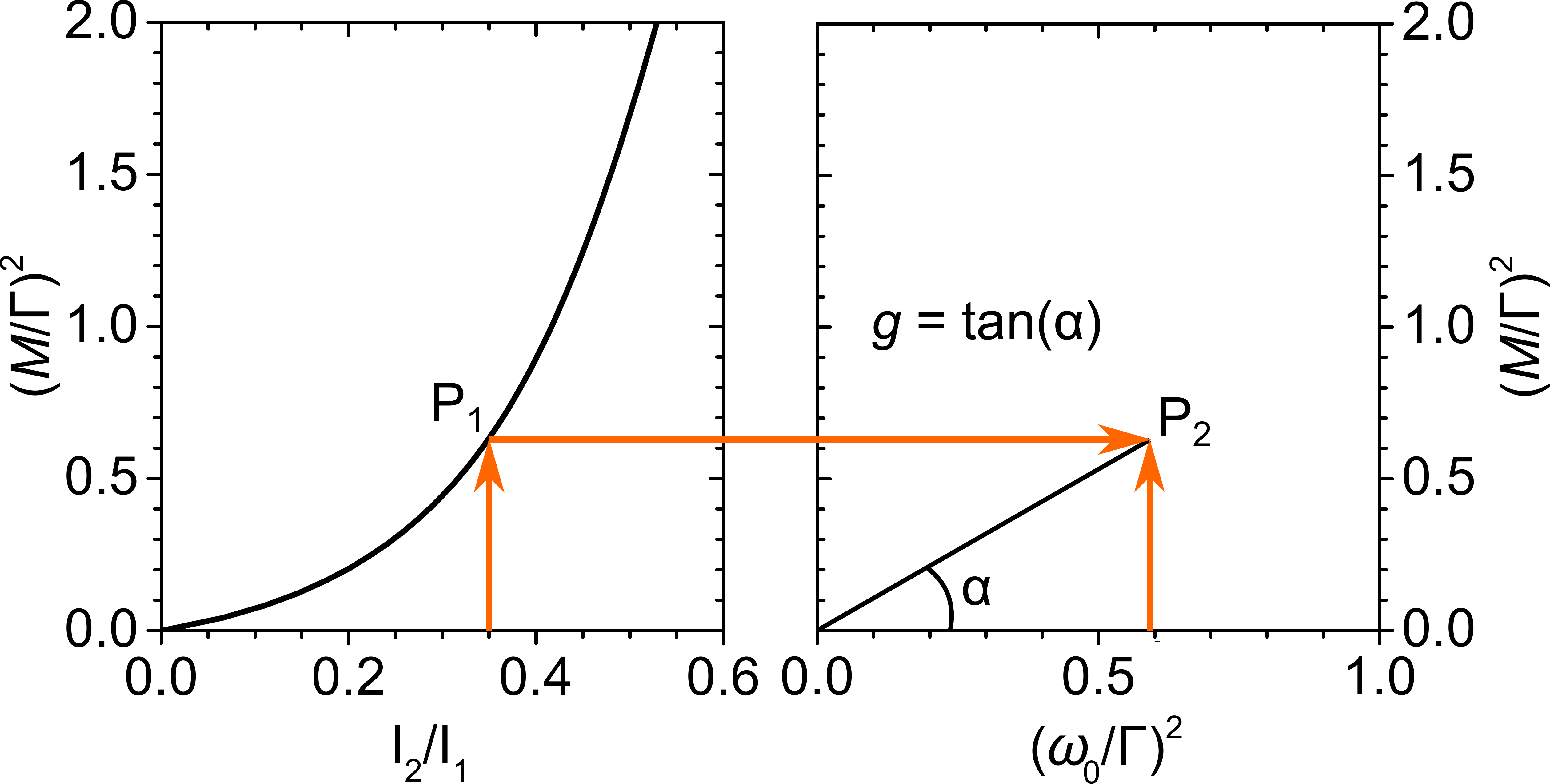}
	\caption{\label{fig:fig4} A geometrical construction showing the relationship between the parameters. The left panel plots the universal curve obtained by inverting the function of Fig.~\ref{fig:fig3}, while the right panel has ordinates $(M/\Gamma)^2$ (in common with the left panel) and $(\omega_0/\Gamma)^2$ as abscissa. The way of using the diagram depends on  the choice of the input data. For instance, suppose we know the intensity ratio $I_2/I_1$, the phonon frequency $\omega_0$, and the core hole lifetime $1/\Gamma$. We enter the left panel with a vertical line at $I_2/I_1$ that intercepts the universal curve at point P$_1$; this gives the value of $(M/\Gamma)^2$. The horizontal line through P$_1$ intersects the vertical line through $(\omega_0/\Gamma)^2$ in the right panel at point P$_2$. The straight line from the origin to P$_2$ has the slope $\tan\alpha = M^2/\omega_0^2 = g$.}
\end{figure}

When the phonon peak and its overtone can be detected in the experimental RIXS spectra, one can directly determine $I_2/I_1$ and use the universal plot of Fig.~\ref{fig:fig3} to obtain $(M/\Gamma)^2$. This approach has been used in early literature at the O $K$ edge \cite{Lee2013,Johnston2016,Meyers2018} and at the Ti $L_3$ edge \cite{Moser2015,Fatale2016} with some uncertainty due to the modest resolving power. In order to make the application more transparent, we present a simple geometrical construction allowing at a glance to capture the interplay between the different parameters (see Fig.~\ref{fig:fig4}). Let us suppose we can determine experimentally the intensity ratio $I_2/I_1$, the phonon energy $\omega_0$, and the core hole lifetime $1/\Gamma$. The value of $I_2/I_1$ corresponds to the point P$_1$ on the universal curve (left panel) and thus to a specific value of $M/\Gamma$. The horizontal line through P$_1$ intersects the value of $(\omega_0/\Gamma)^2$, which is the entrance value of the right panel. This defines the point P$_2$. The straight line from the origin through P$_2$ defines the angle $\alpha$ whose tangent is $(M/\Gamma)^2/(\omega_0/\Gamma)^2 = g$. The $I_2/I_1$ approach does not require the knowledge of the absolute efficiency of the instrument since it is based on a ratio of intensities.

The $I_2/I_1$ method may seem ideal, but this approach has a very serious drawback: it is very difficult and often impossible to experimentally identify the two-phonon spectral feature and to determine its intensity. As a matter of fact, at the Cu $L_3$ edge, this is only possible with great difficulty in parent compounds, as demonstrated below. In the doped cuprates, the broadening of the features and the presence of a continuum prevent the detection of the overtones in the majority of cases. This method does, however, work particularly well in the case of quasi-1D systems characterized by a single non-dispersing lattice mode, which can give rise to sharp and intense multi-phonon peaks \cite{Lee2013,Johnston2016,Vale2019}.  

\subsection{The use of $I_1$ on resonance}

The universal curve of Fig.~\ref{fig:fig3} can also be used to recover $(M/\Gamma)^2$ from the intensity of the single phonon excitation $I_1$. The geometrical construction shown in Fig.~\ref{fig:fig4} is formally the same, but the ratio $I_2/I_1$ is substituted by $I_1$ in the left panel. However, the use of this method requires the measurement of the absolute value of $I_1$ and RIXS cross sections are not typically measured in an absolute way. Hence, this approach can be exploited only if the relation between $I_1$ and the EPC is determined at least at one point by another method,  and if a relative measurement of $I_1$ can be made as a function of the relevant parameter (momentum transfer, temperature, doping, etc.).  Fig.~\ref{fig:fig3} also highlights the non-proportionality and non-monotonicity of $I_1$ as a function of the EPC. In particular, the broad maximum of $I_1$ is due to the increase of $I_2$ spoiling the intensity $I_1$ at large values of the electron-lattice interaction.  The existence of a maximum for $I_1$ means that the plot equivalent to that of Fig.~\ref{fig:fig4} based on $I_1$ would no longer be single valued, creating some ambiguities for strong EPC. One might be able to resolve this issue by considering the nature of the physical problem.

\subsection{The use of $I_1$ upon detuning below threshold}

This method has been recently introduced by some of us in a concise way~\cite{Rossi2019}. We add here some relevant information, including a systematic overview of the method, a straightforward way to retrieve the EPC from energy-detuned RIXS spectra, and the identification of an approximate scaling law. On this basis we also introduce a remarkably simple rule allowing, in many cases, an immediate approximate estimate of the EPC. The detuning approach is based on two observations: \emph{i)} upon detuning below threshold, i.e., using an incident photon energy slightly lower than the absorption resonance, the phonon signal evolves differently with respect to the other features in the same RIXS spectrum; \emph{ii)} the difference in the phonon behavior depends on the strength of the EPC. Therefore, it is possible to recover the EPC from a suitable set of spectra measured as a function of the detuning. An inspection of the scattering amplitude [Eq.~\eqref{eq:scattering_amplitude}] helps one understand why this is the case. The denominator is the sum of the detuning energy $\Omega = \omega - \omega_\mathrm{res}$, of $i\Gamma$ and of $M^2/\omega_0$. Thus, detuning has a larger effect if $M$ is small. Indeed, there is a kind of trade-off between detuning and $M^2/\omega_0$.

\begin{figure}
	\centering
	\includegraphics[width=0.85\columnwidth]{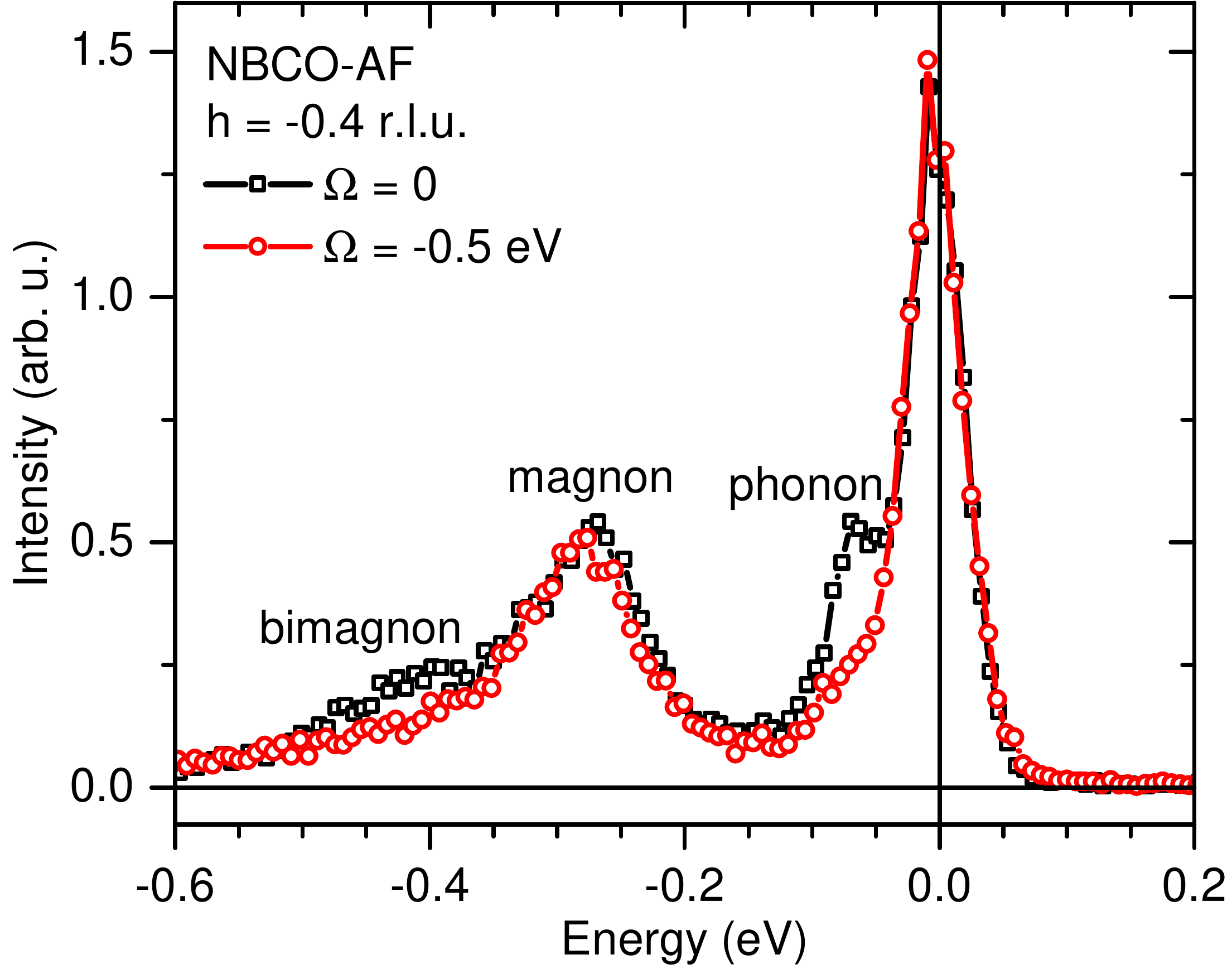}
	\caption{\label{fig:fig5} Comparison between the RIXS spectrum of NBCO-AF measured at in-plane momentum transfer $h = -0.4$~r.l.u. at resonance (black squares) and energy-detuned by $\Omega = -0.5$~eV (red circles). The spectra are normalized to the $dd$ excitations. The decrease in intensity of the phonon and of the bimagnon excitations is evident.}
\end{figure}

\begin{figure}
	\centering
	\includegraphics[width=\columnwidth]{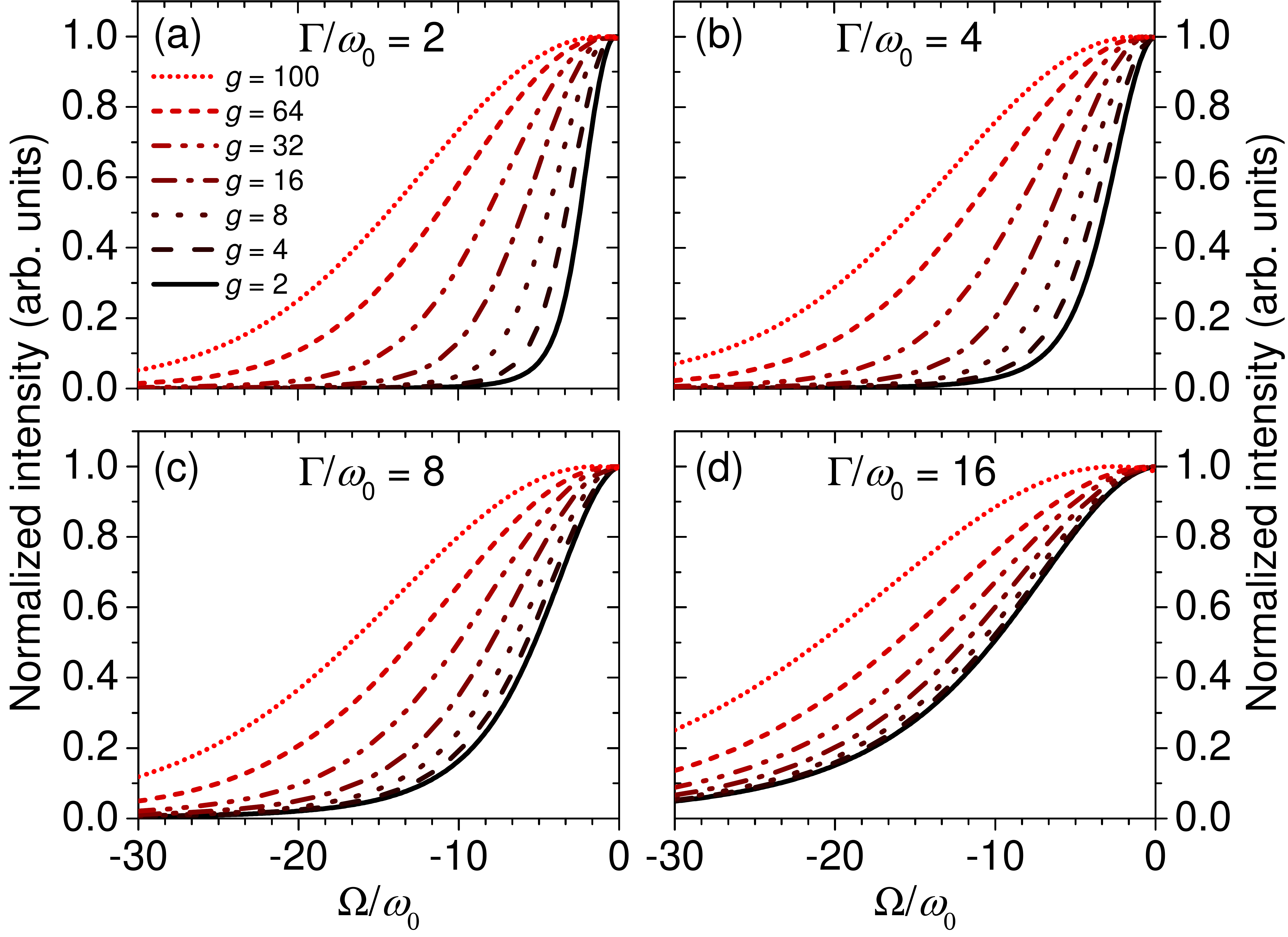}
	\caption{\label{fig:fig6} Theoretical intensity of local phonon excitations upon detuning. The different panels correspond to specific values of $\Gamma/\omega_0$. Within each panel, the curves are labeled with the dimensionless coupling constant $g$ (from left to right: $g = 100$, 64, 32, 16, 8, 4, 2). The curves are normalized to the value at resonance.}
\end{figure}

\begin{figure}
	\centering
	\includegraphics[width=0.85\columnwidth]{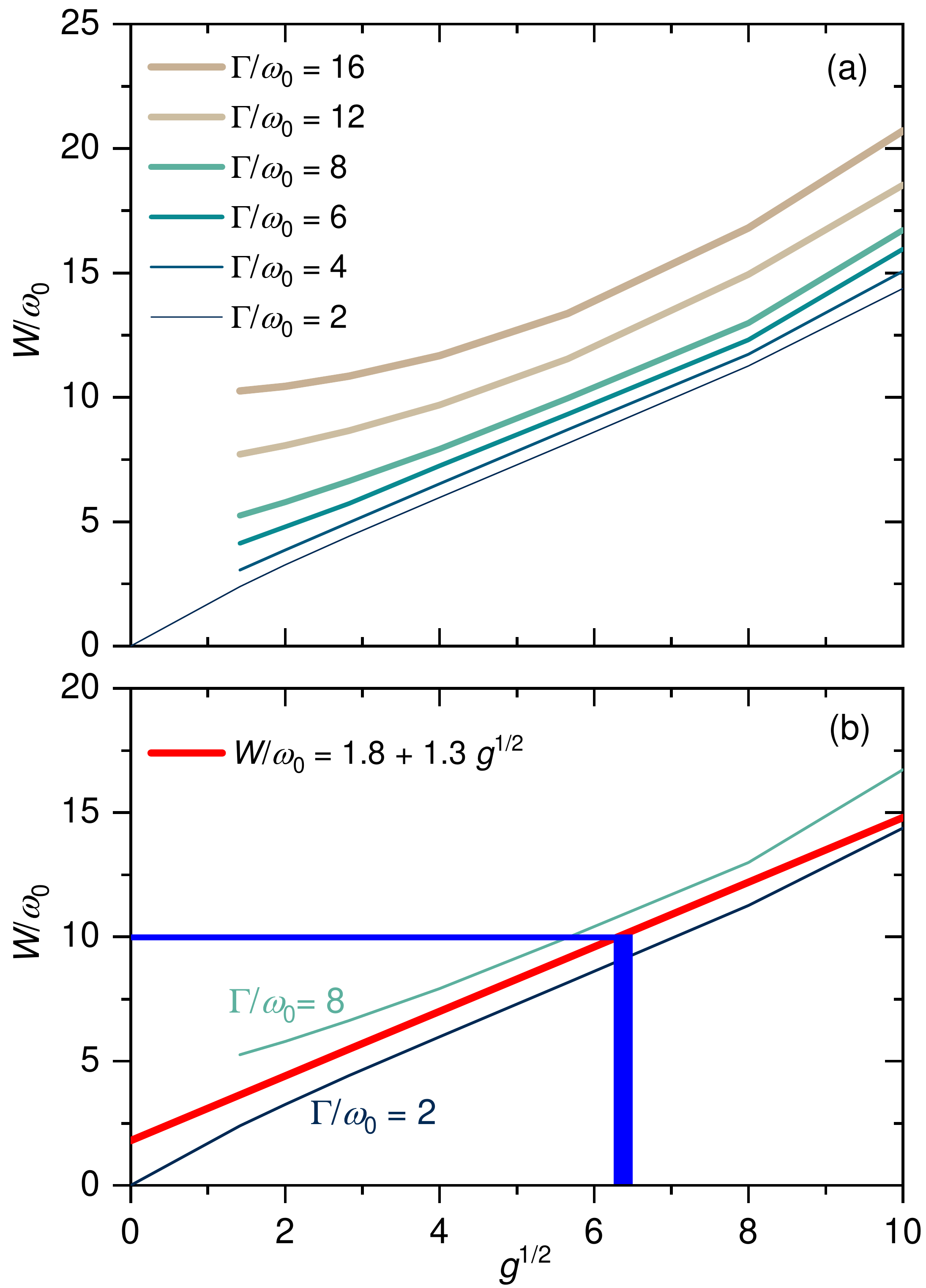}
	\caption{\label{fig:fig7} (a) Width at half maximum of the theoretical detuning curves as a function of $\sqrt{g}$ with $\Gamma/\omega_0$ as a parameter. The width $W$ is plotted in units of $\omega_0$. Note the linear dependence of the width over a wide range of $\sqrt{g}$. The lines at small values of the parameter $\Gamma/\omega_0$ are very close one to the other suggesting an approximate scaling law. (b) An example of the use of the simplified detuning rule. Suppose that the red straight line approximates the system in the parameter range $2 \leq  \Gamma / \omega_0 \leq 8$. If the width $W$ of the detuning curve is measured, the intersection of the horizontal line with height $W$ with the red line approximates the value of $g$ without numerical calculations.}
\end{figure}

An example of the effect of detuning on the RIXS spectrum of NBCO-AF is shown in Fig.~\ref{fig:fig5}, where we compare the spectrum measured at $h = -0.4$~r.l.u. and with an incident photon energy tuned at the Cu $L_3$ resonance (black squares) and detuned to $-0.5$~eV (red circles). It is evident that the phonon and the bimagnon become weaker upon detuning; their processes of excitation are ``slow'' and the intensity depends on the time duration set by the core hole decay. It is sufficient to fit the measured detuning effect to the cross section to obtain $g$. More precisely, we define the detuning curve as the phonon intensity as a function of the excitation energy, normalized to the value at resonance. An overview of the theoretical detuning curves is given in Fig.~\ref{fig:fig6}, where each panel is identified by the value of $\Gamma/\omega_0$ and contains the detuning curves for different values of $g$. Since $\Gamma/\omega_0$ is known for a given absorption edge and phonon mode, by comparing the experimental data to this set of theoretical curves, one can identify the appropriate detuning curve, and thus the corresponding value of $g$. The sensitivity of the method is clearly decreasing at higher values of $\Gamma/\omega_0$, where the approach is useful only in the cases of large $g$. A very convenient way of handling the data is presented in Fig.~\ref{fig:fig7}(a), which reports the width $W$ (in units of $\omega_0$) of the detuning curve, defined as the detuning value of the point at half height. The width is displayed as a function of $\sqrt{g}$, with $\Gamma/\omega_0$ as a parameter. The lines are remarkably linear in a wide range and at lower values of $\Gamma/\omega_0$ they are very close to each other, suggesting an approximate scaling law.

The procedure based on Fig. \ref{fig:fig7}(a) is exact within the limits of Ament {\it et al.'s} model, since no further approximation has been introduced. However, if one accepts a modest inaccuracy, it is possible to further simplify the procedure in many cases.  Here, we take advantage of the fact that the lines in Fig \ref{fig:fig7}(a) become close to each other when $\Gamma / \omega_0$ decreases. In this parameter region,  we approximate the state of the system with a single line. This is done in Fig. \ref{fig:fig7}(b) by introducing the red straight line to represent the system in the interval $2 \leq  \Gamma / \omega_0 \leq 8$. By using the red line the scaling becomes universal, i.e., independent of $\Gamma$. The red line has the equation:
\begin{equation}
  \frac{W}{\omega_0}  =  1.8 + 1.3 ~ g^{\frac{1}{2}} = 1.8 + 1.3 ~ \frac{M}{\omega_0}.  
\end{equation}
If one enters the diagram with a value of $W$, the working point is the intersection of the horizontal line representing $W$ with the red line in place of the exact point, as illustrated in Fig. \ref{fig:fig7}(b) for the case $W=10$. The inaccuracy is of the order of 10\%. Obviously the error is maximum at $\Gamma / \omega_0$ equal to 2 and 8.
This astonishingly simple rule can be extremely useful during experiments since it offers a shortcut to estimate the EPC from the raw data. We will refer to this rule  as the Simplified Detuning Rule (SDR).

The method based on detuning is advantageous because it uses the value of $I_1$ normalized to its value at resonance, which is easily measured. In essence, the detuning method retains the advantages of the two approaches discussed above without their limitations. The primary drawback of this method is that the detuning curves require long acquisition times due to the loss of intensity below threshold.

\subsection{Access to momentum-dependent EPC}

As mentioned above, the direct estimation of the EPC from our spectra is based on the important assumptions made by Ament \emph{et al.}; in particular, the electronic state involved in the RIXS process is localized (i.e. an excitonic intermediate state) and the phonon is assumed to have a local character. We will address later the effect of the localized vs itinerant nature of the electron states in connection with EPC. Here, we limit ourselves to note that the two assumptions are less stringent than it might seem. Very often the RIXS spectra of correlated electron systems involve an excitonic intermediate state, even when the ground state electronic structure implies delocalized states, such as Hubbard bands in cuprates. Therefore, the RIXS process is non-selective on the wave vector $\mathbf{k}$ of the electron for which one wants to determine the EPC and the result is related to weighted integrals over the Brillouin zone. On the contrary, the RIXS experiment is intrinsically selective on the transferred momentum, which is supplied to the phonons excited in the final state. In particular, in experimental RIXS spectra, the one-phonon peak disperses and changes its intensity as the momentum transfer changes, following the phonon branch dispersion and the evolution of the EPC in the reciprocal space. Therefore, the use of the methods presented here at each momentum point is particularly well suited for optical phonon branches with little energy dispersion which correspond to good approximation to Einstein phonons. The method is as follows: for each transferred momentum, dictated by the energy of the photons and the experimental geometry, one can use one of the methods presented above or a combination of them to estimate $g$ or $M$. The same theoretical treatment is used to go from the measured quantities ($I_1$, $I_2$) to the EPC parameters. Eventually the momentum dependence of the EPC is thus obtained. We note that the procedure is safe because $\Gamma$ does not depend on the momentum and, $\omega_0$ is practically constant for optical oxygen phonons in the cuprates.

In the experimental results, it is important to separate a true variation of the EPC with momentum transfer from spurious effects like the dependence of atomic RIXS cross sections on the scattering geometry and photon polarization \cite{Devereaux2016}. In our experiment, we used $\sigma$ incident polarization to accomplish this. As shown by measurements with polarization analysis of the scattered beam~\cite{Fumagalli2019}, the $\sigma\sigma^\prime$ channel (with incident and scattered $\sigma$ polarization) has an overwhelming intensity with respect to the $\sigma\pi^\prime$ channel. The ionic cross section for Cu $L_3$ resonant elastic scattering in the $\sigma\sigma^\prime$ channel is isotropic; moreover the $\sigma\pi^\prime$ cross section has negligible variation in our range of experimental momenta~\cite{MorettiSala2011,Fumagalli2019}. For these reasons, we do not expect any sizable angular variation of the phonon intensity due to the resonant form factor. Therefore, the measured change in the phonon intensity as a function of the incidence angle is due to a genuine variation of the EPC influencing $I_1$, $I_2$, and their ratio.

\subsection{Application to other absorption edges}

This work work is focused on the Cu $L_3$ edge due to our interest in understanding the role of the EPC in the high-$T_\mathrm{c}$ cuprates; however, this goal should not obscure the generality of our results and the great amount of work that remains to be done. The case of other absorption edges is briefly discussed here.

A situation very similar to the Cu $L_3$ edge is RIXS at the $L$ edges of other $3d$ transition metals in oxides, a vast and important class of materials. The levels involved and the nature of the cation-ligand bond are the same as in cuprates, so the mechanism governing the phonon excitation by RIXS is also the same. However, by going from Cu to Ti the available momentum decreases, the effects of electronic correlation weaken, and the lifetime of the intermediate state can change case by case. For example, the lifetime of the $L_3$ core hole in Ti$^{4+}$ is longer than Cu by about a factor of 2.5 (tabulated $\Gamma=0.11$~eV for Ti, $\Gamma=0.28$~eV for Cu \cite{KeskiRahkonen1974,Krause1979}), resulting in much stronger peaks for single- and multi-phonon excitations.

The $L$ edges of heavier elements, namely $4d$ and $5d$ transition metals and lanthanides, are characterized by much larger values of $\Gamma$, i.e., shorter living intermediate states, whereas the $K$ edges of C, N, and O are narrower. By looking at Table \ref{table:1}, we reckon that phonon peaks can be very strong at C and O $K$ edges \cite{Lee2013,Johnston2016,Meyers2018}, sizable at $3d$ transition metals $L_3$ edges, and hardly detectable at $L_3$ edges of heavier elements or at $K$ edges of any element heavier than chlorine ($Z=17$). The phonon signal in RIXS spectra at the $M_{4,5}$ edges of lanthanides is also expected to be very weak despite the relatively narrow line width because the EPC of the very localized $4f$ states is notoriously small.

\begin{table}
\caption{\label{table:1}
The theoretical HWHM $\Gamma$ for selected elements and absorption edges in units of eV. Adapted from Refs.~\onlinecite{KeskiRahkonen1974,Krause1979}.}
\begin{ruledtabular}
\begin{tabular}{| c | c | c | c |c | }
  $Z$ & Elements  & $K$ edge & $L_3$ edge & $M_5$ edge \\
\hline
 6 - 9 & C - F & 0.05 - 0.10  &   &   \\ 
 11 - 17 & Na - Cl & 0.15 - 0.34  &   &   \\ 
 21 - 30 & Sc - Zn & 0.43 - 0.85  & 0.09 - 0.33  &   \\ 
 39 - 48 & Y - Cd & 1.7 - 3.6  & 0.07 - 1.2  &   \\ 
 57 - 70 & La - Yb & 7.0 - 16 & 1.7 - 2.3  &  0.15 - 0.30 \\ 
 72 - 80 & Hf - Hg & 18 - 27 & 2.4 - 2.7  &  1.0 - 1.5 \\ 
 
\end{tabular}
\end{ruledtabular}
\end{table}

\section{Experimental results}
\label{sec:results}

\subsection{Overview and assignment of phonon modes}

\begin{figure}[ht!]
	\centering
	\includegraphics[width=0.75\columnwidth]{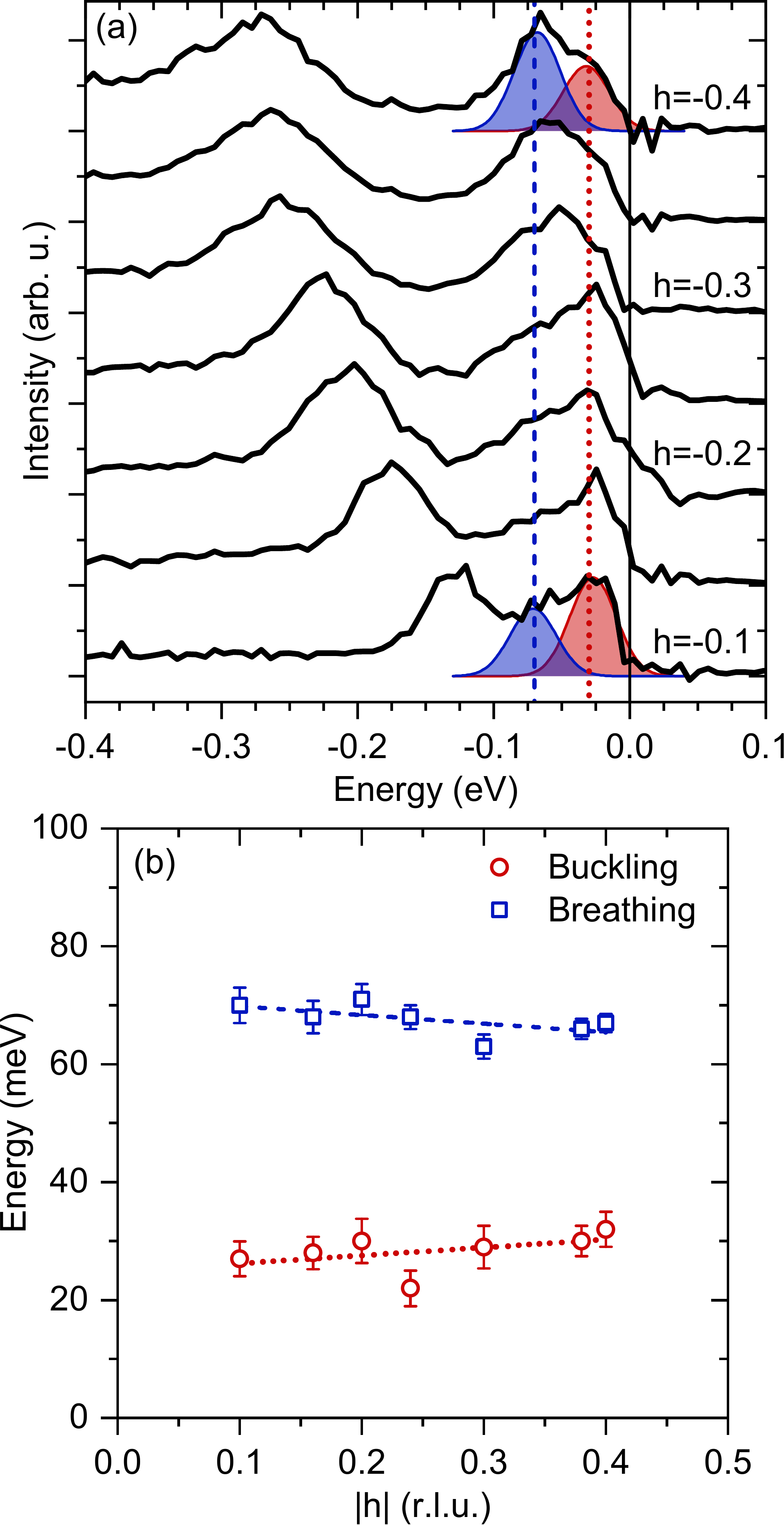}
	\caption{\label{fig:fig9} Experimental results on NBCO-AF and assignments of the phonon modes. (a) Stack of raw RIXS spectra with the elastic line subtracted (black solid lines). The spectra show the presence of two main features in the phonon energy range (shaded areas) whose intensities have opposite momentum dependence. (b) Experimental dispersion relation of the two phonon branches: the Cu-O bond-buckling branch (red circles) at an energy of $\approx 30$~meV and the Cu-O bond-stretching or breathing branch (blue squares) at an energy of $\approx 70$~meV. The lines are linear fits to the dispersion relations.}
\end{figure}

Figure~\ref{fig:fig9}(a) reports an overview of the momentum dependence of the low-energy RIXS features of NBCO-AF. Raw data (black solid lines) are plotted after subtraction of the elastic line. As anticipated, we only consider the momentum range of $|h| \ge 0.1$~r.l.u.~and energy loss range $\lesssim 120$ -- 140~meV to avoid the superposition of the phonon excitations with the rapidly dispersing magnetic excitations. The data are very expressive at a glance: even without optimizing the fitting, two main features are clearly detected with opposite momentum dependence of their intensity. These features are qualitatively plotted as red and blue Gaussian peaks, while the vertical dotted and dashed lines guide the eye across different spectra. While the present resolution allows us to separate these two features, it is insufficient to further decompose each into potential contributions from individual phonon branches that may be overlapping. Nevertheless, the structure of the phonon excitations observed here allows us to divide the excitations into two distinct energy regions, each containing modes that behave qualitatively different as a function of momentum. This observation enables us to assign the two features while keeping in mind that they may represent ``effective branches'' (i.e., the superposition of  different phonons). 

The behavior in high-energy region [50 -- 100~meV, blue shaded area in Fig.~\ref{fig:fig9}(a)] is typical of the breathing (Cu-O in-plane bond-stretching modes) and apical oxygen modes. The EPC of the breathing mode increases on going from the $\Gamma$ to the $X$ point of the BZ, as theoretically shown in Ref.~\onlinecite{Johnston2010}. Also, its intensity increases with momentum according to Ref.~\onlinecite{Devereaux2016} so that the intensity and the EPC go together. The increase of the intensity with momentum is very clear from Fig.~\ref{fig:fig9}(a), thus we assign the higher energy feature to the breathing modes. We note that the apical modes are not expected to influence appreciably the situation within our region of interest: the RIXS signal from the apical branch is expected to be about $1/3$ of that from the breathing branch and, even more importantly, its intensity is concentrated at small momenta outside of our working window \cite{Devereaux2016}. In passing, we note that the separation of the apical and breathing modes in reciprocal space can be very useful in future RIXS experiments in nonlinear conditions at Free Electron Lasers, since the apical mode is often used to pump the sample in pump-probe experiments \cite{Hu2014,Kaiser2014}.

The feature at lower energy [0 -- 50~meV, red shaded area in Fig.~\ref{fig:fig9}(a)] is in the typical energy range of the out-of-plane polarized modes (e.g. the Cu-O bond buckling modes). The EPC of these phonons depends on both phonon \emph{and} electron momentum so that a variety of situations can occur \cite{Johnston2010} (see Appendix~\ref{appendix1}), including a decrease of the intensity with increasing momentum transfer which is, in a sense, a fingerprint of buckling branches. Thus, it is quite natural to assign the lower energy feature to buckling modes. This assignment is consistent with the strong static buckling in NBCO, which breaks inversion symmetry across the CuO$_2$ plane and results in a larger EPC. We note that in the absence of such static buckling, one should expect a much weaker intensity for the buckling phonon modes; the case of Bi2212 is important with this respect, because the buckling mode intensity in experimental RIXS spectra is very small (below the present detection limit) and the static buckling is much smaller and with different local symmetry than in YBCO and NBCO. This family of phonon modes includes both in-phase (A$_1$) and out-of-phase (B$_1$) vibrations of the planar oxygen ions. In our case, the out-of-phase one is silent due to symmetry of the selection rules, while the in-phase one contributes to the RIXS signal (see the calculations in Appendix~\ref{appendix1}). In the model by Devereaux {\it et al.}~\cite{Devereaux2016}, both branches are visible but the B$_1$ is weaker by a factor $\sim 20$ with respect to A$_1$, in agreement with our present assignment.

Further support for our assignments comes from the consistency between our observed phonon dispersion and the measured phonon dispersions in isostructural YBCO-AF \cite{Pintschovius1998}. This analogy is valid for our purposes, as one expects the phonon dispersions of the two compounds differ at mainly at very low energy, where the phonons involving the motion of Y and Nd are found~\cite{Liu1988,Cardona1988,Yoshida1990,Limonov1998,Bohnen2003}. 
The dispersion relations of the buckling and breathing branches of NBCO-AF are plotted in Fig.~\ref{fig:fig9}(b) as red circles and blue squares, respectively. The weak dispersion extracted from the fit of the data also provides {\it post hoc} justification for our use of a local phonon model. 

\subsection{Determining the EPC in NBCO-AF}

\begin{figure}[ht]
	\centering
	\includegraphics[width=0.85\columnwidth]{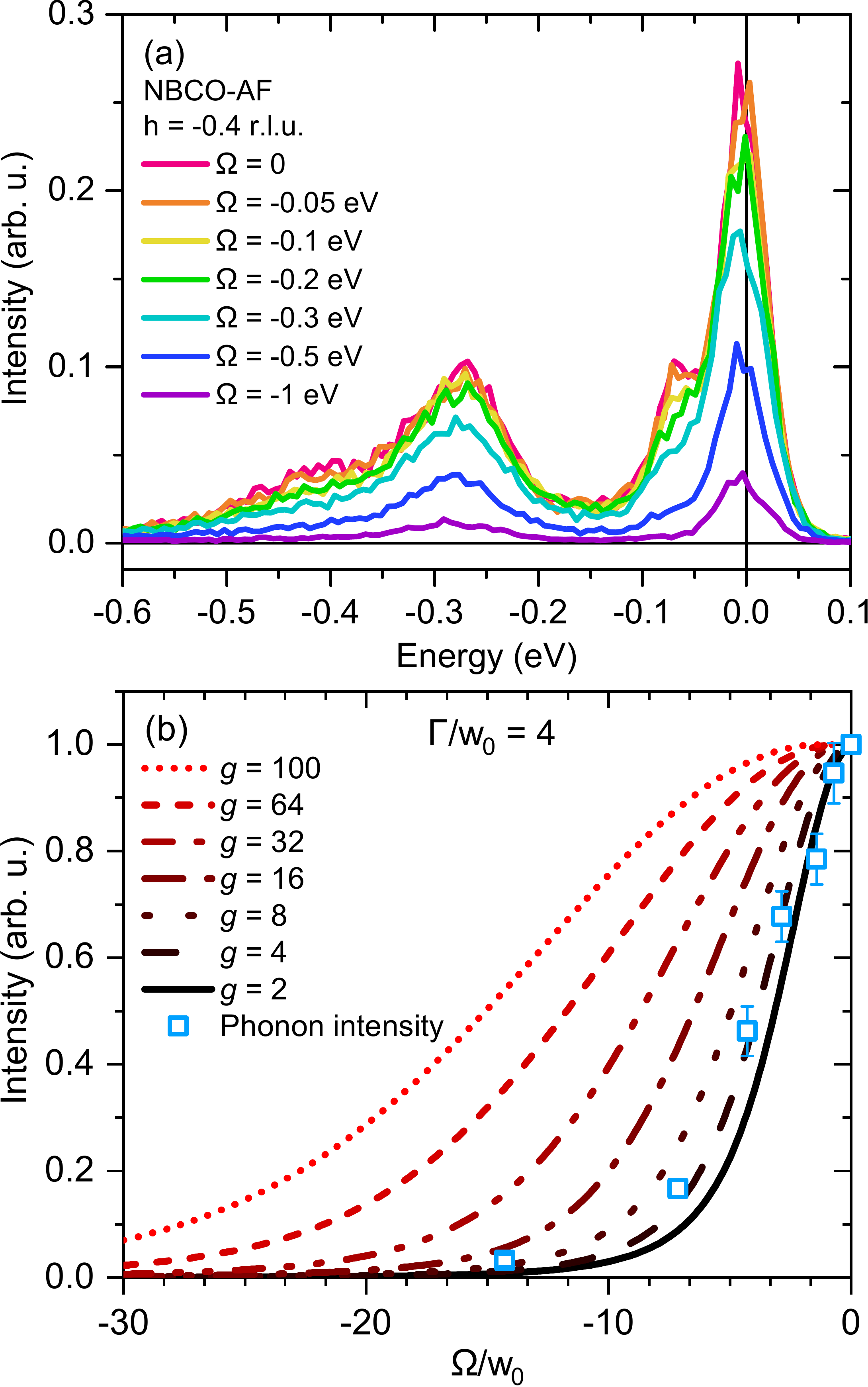}
	\caption{\label{fig:fig10} The application of the detuning approach to the $h = -0.4$~r.l.u. mode of the breathing branch in NBCO-AF. (a) Cu $L_3$ edge RIXS spectra of NBCO-AF as a function of detuning energy $\Omega$. (b) Detuning dependence of the phonon intensity (squares) compared against the theoretical detuning curves. Note that the experimental values discriminate very well among the theoretical curves having different dimensionless coupling constant $g$.}
\end{figure}

We now proceed with a detailed analysis of the breathing branch, which is clearly resolved in the spectra [Fig.~\ref{fig:fig10}(a)]. From energy-detuned RIXS spectra, one can extract the detuning dependence of the phonon intensity. Figure~\ref{fig:fig10}(b) reports a set of theoretical detuning curves for the appropriate value of $\Gamma/\omega_0 = 4$ corresponding to the Cu $L_3$ resonance. The comparison of the detuning curves with the phonon intensity (light blue squares) shows that we can achieve a good agreement with the data with a value of $g \approx 4$, which corresponds to $M = \omega_0\sqrt{g} \approx 0.13$~eV. The value of $\Gamma / \omega_0$ is in the range of validity of the SDR (Simplified Detuning Rule).  It is remarkable  that this rule gives $M = 120$~meV, comparable with the above fitting.
We note that from the fitted momentum dependence of the EPC of NBCO-AF reported by Rossi \emph{et al.}, one estimates $M \approx 0.16$~eV at $h = -0.4$~r.l.u. \cite{Rossi2019}. By taking into account an uncertainty on the fitting results of about 10 -- 15\% (see Appendix~\ref{appendix2}), the two values agree within error bars. We also note that the two different samples employed in the measurements may differ slightly in oxygen content, for the reasons stated in Sec.~\ref{sec:samples}. A slightly larger oxygen content in one of the samples may enhance the screening of the photoexcited electron, whose coupling to phonons may be consequently reduced. 

\begin{figure}
	\centering
	\includegraphics[width=0.8\columnwidth]{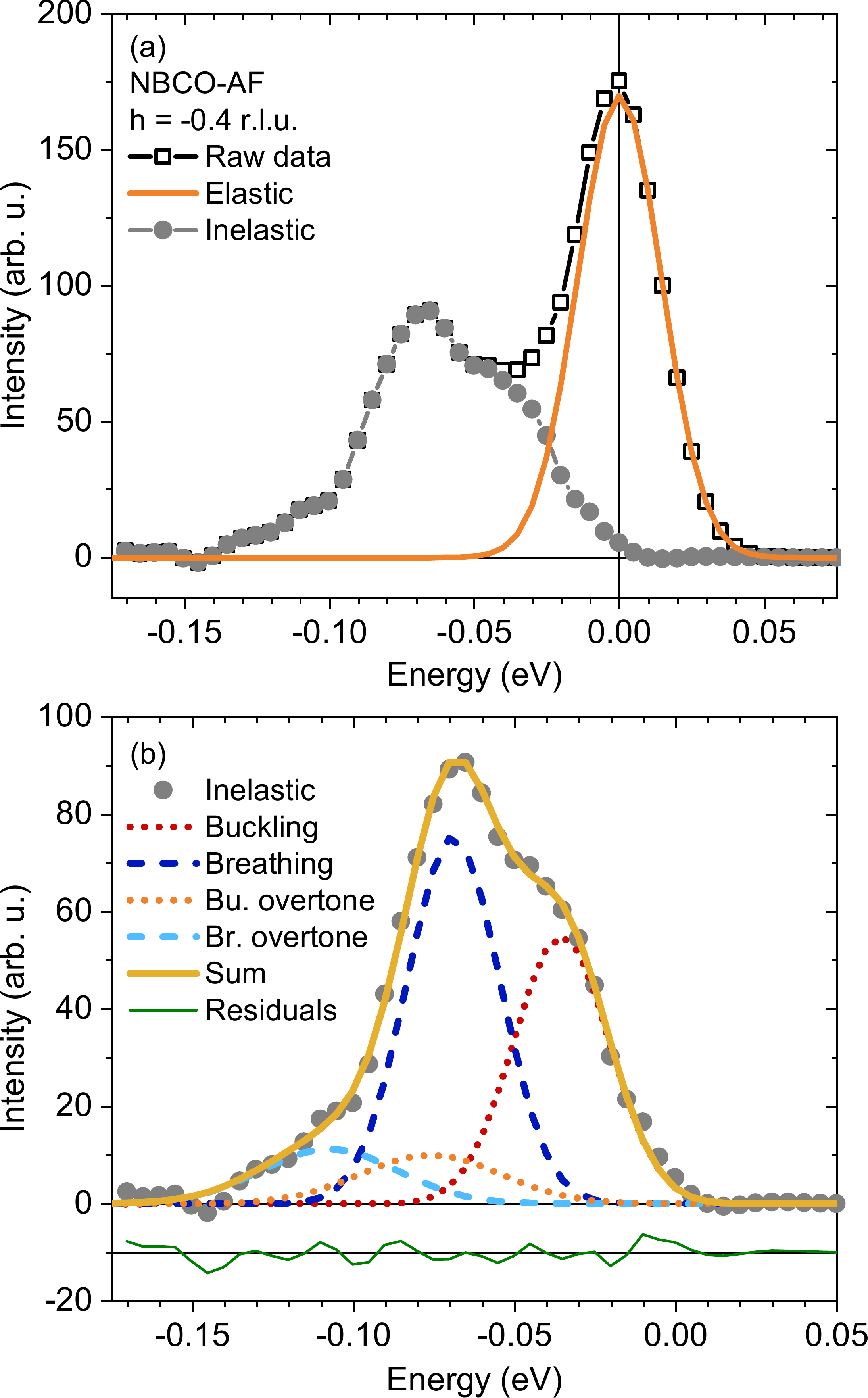}
	\caption{\label{fig:fig11} Fitting and decomposition of the RIXS spectra. The complete presentation is given in APPENDIX \ref{appendix2}, and here we summarize the main points. The decomposition of the high-statistics RIXS spectrum of NBCO-AF at $h = -0.4$~r.l.u., which is used to recover $I_1$ and $I_2$ of the buckling and breathing modes. (a) The RIXS spectrum (black empty squares) is decomposed into an elastic contribution (orange solid line) and an inelastic signal (gray filled circles). A tiny tail arising from magnetic the excitations has been subtracted. (b) Decomposition of the inelastic signal into buckling (red dotted line) and breathing (blue dashed line) modes and their overtones (orange dotted and light blue dashed lines, respectively). The dark yellow line is the sum of the components. The residuals are plotted as a green solid line, vertically offset for clarity. The full decomposition procedure is detailed in Appendix~\ref{appendix2}.} 
\end{figure}

Extracting the EPC strength for the buckling branch is more delicate and gives the us the opportunity to employ the other methods presented in Sec.~\ref{sec:EPC_extraction}. Figure~\ref{fig:fig11}(a) represents the benchmark RIXS spectrum of NBCO-AF measured at $h = -0.4$~r.l.u. (black squares), which is decomposed into an elastic line (orange solid line) and an inelastic signal (gray circles). The latter is further decomposed in Fig.~\ref{fig:fig11}(b) into the buckling (red dotted line) and breathing (blue dashed line) contributions each with their first overtone (orange dotted and light blue dashed lines, respectively).

\begin{figure}
	\centering
	\includegraphics[width=0.8\columnwidth]{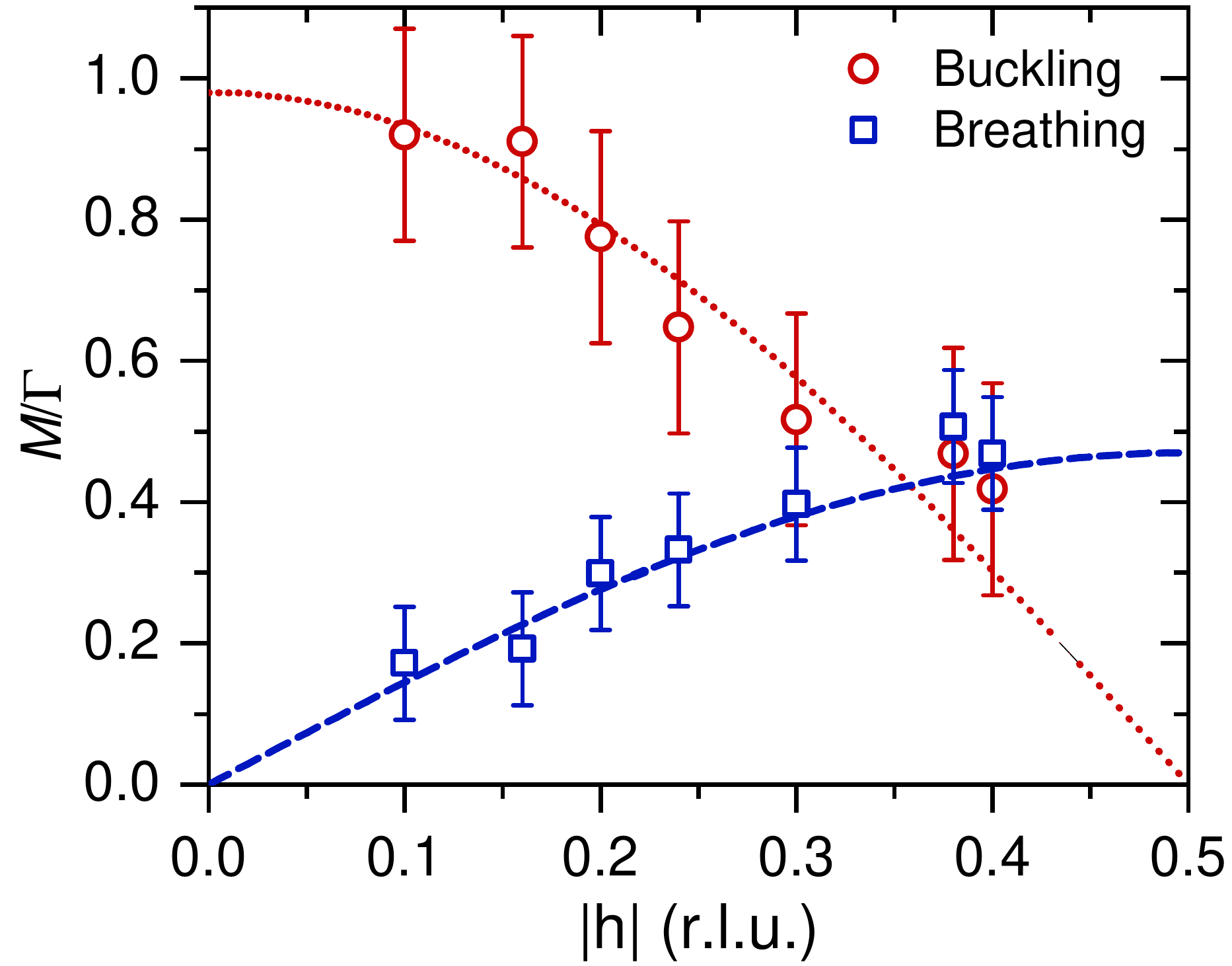}
	\caption{\label{fig:fig12} The final results of the  data analysis of NBCO-AF. The absolute value of the matrix element $M$ of the EPC in units of $\Gamma$ is plotted as a function of the momentum transfer for the buckling (red circles) and breathing (blue squares) branches. Dashed and dotted lines are a fitting with sine and cosine functions, respectively.}
\end{figure}

The step-by-step decomposition of the inelastic spectrum is reported in Appendix~\ref{appendix2}. We note that the robustness of the fit can be improved if we constrain the ratio $I_2/I_1$ for the breathing mode to the value corresponding to $g$ found using the detuning method. Once this is done, the ratio $I_2/I_1$ of the buckling mode is well determined. Moreover, knowledge of $I_2/I_1$ at $h = -0.4$~r.l.u. allows us to calibrate the intensity of the buckling mode excitations $I_1$ such that its momentum dependence can be measured directly with $I_1$ on resonance. We use this approach since we could not measure the detuning curves at all momenta because of the limited availability of beam time. At small momentum transfer the intensity $I_1$ tends to saturate because of the saturation of the universal curve, while  the results of detuning are free of this problem at the Cu $L_3$ edge. Thus we used the detuning result at small $\mathbf{Q}_\parallel$. This is a typical case in which the direct connection between intensity and EPC does not work or works only in a subset of the parameter space. The final results on the EPC (in units of $\Gamma$) are summarized in Fig.~\ref{fig:fig12}. While the EPC of the buckling phonon branch (red circles) decreases going from the BZ center towards $X$, the EPC of the breathing branch (blue squares) shows the opposite trend.

\subsection{Doping effects}

\begin{figure}
	\centering
	\includegraphics[width=0.8\columnwidth]{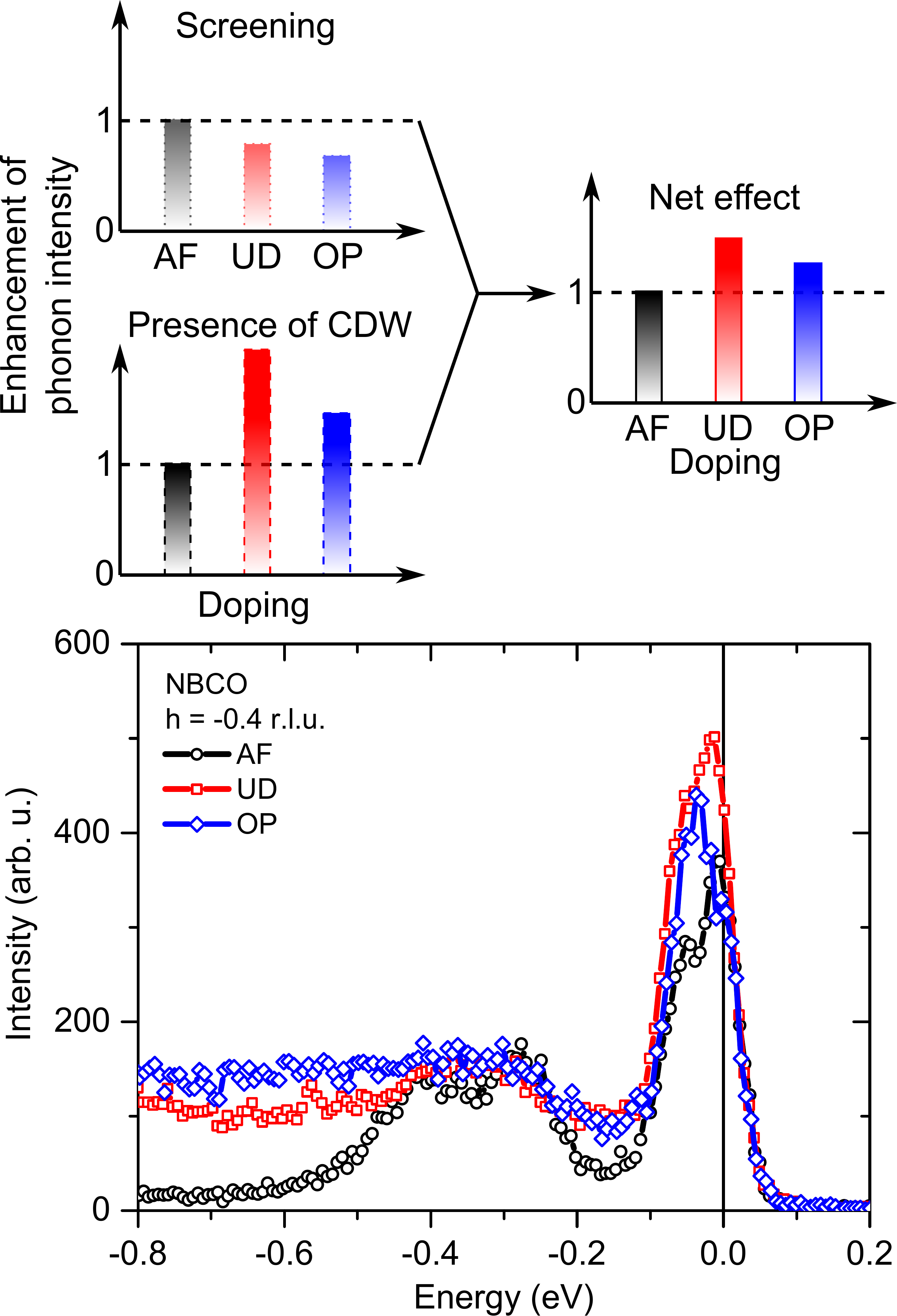}
	\caption{\label{fig:fig13} The upper panel qualitatively shows two situations that modify the RIXS phonon signal: in the absence of charge order a smooth decrease of the EPC with doping is expected due to the increased screening from the free carriers. In the presence of CDWs, instead, CDW-induced effects may prevail so that a non-monotonic trend is found as a function of doping. Indeed, the signal from the CDWs is absent in the AF sample, is the strongest in the UD sample, and is weaker in the OP sample. The non-monotonic trend is not a necessary condition but it is sufficient to demonstrate the prevalence of the CDWs. This is indeed the case, as shown by the experimental results of the bottom panel, where spectra of NBCO-AF (black circles), NBCO-UD (red squares) and NBCO-OP (blue diamonds) are compared. The spectra were collected at $h = -0.4$~r.l.u. and are measured with an energy resolution of 60~meV. Note that in the 0.6 -- 0.8~eV energy range the spectral intensity scales with doping.}
\end{figure}

In the doped samples, one expects that the photoexcited electron and the EPC are  screened by the free carriers. Thus, the phonon intensity is expected to monotonically decrease, as qualitatively shown in the cartoon of Fig.~\ref{fig:fig13}. However, a different scenario arises from the comparison between the measured RIXS spectra for the antiferromagnetic NBCO-AF (black circles), the underdoped NBCO-UD (red squares), and the (nearly) optimally doped NBCO-OP (blue diamonds) samples at $h = -0.4$~r.l.u.  (Fig.~\ref{fig:fig13}). Note that this momentum transfer is well above the CDW wave vector ($\approx 0.31$~r.l.u. for NBCO \cite{Ghiringhelli2012}). The comparison between the intensities (normalized to the photon flux) among the three samples is reliable as can be seen in the spectral region between $0.6$ -- $0.8$~eV, where only the electron-hole continuum contributes to the spectra and scales with the doping. The behavior of the phonon intensity qualitatively mimics the doping dependence of the CDW signal, which is stronger in the UD sample and weaker in the OP sample~\cite{BlancoCanosa2014,Arpaia2019}, as sketched in Fig.~\ref{fig:fig13}. The trend suggests that the CDW effect dominates over the screening even at $\mathbf{Q}_\parallel$ considerably higher than the CDW wave vector.

\begin{figure}
	\centering
	\includegraphics[width=\columnwidth]{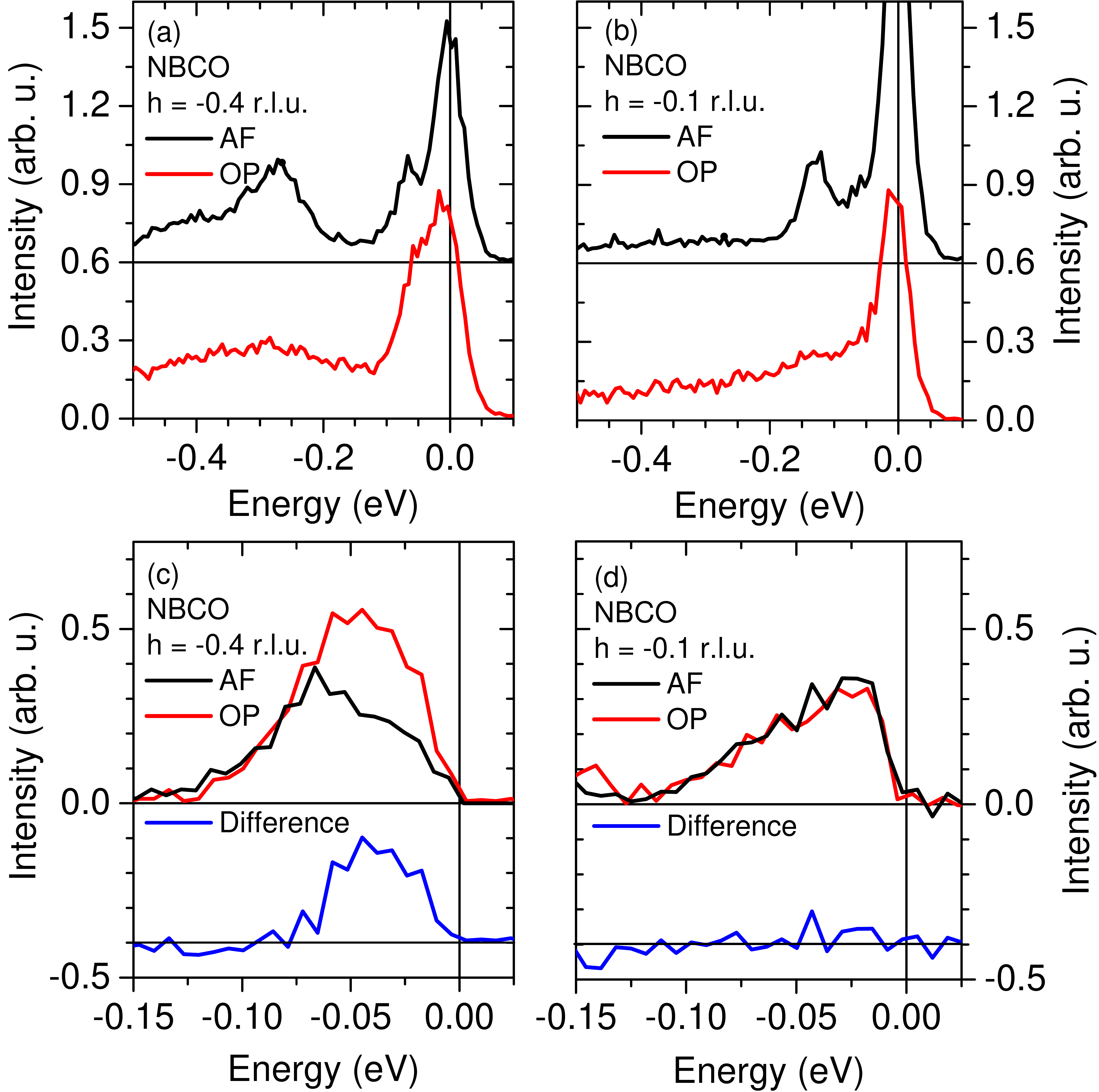}
	\caption{\label{fig:fig14} Evidence for the increase of the phonon signal from the NBCO-AF to the NBCO-OP sample at large momentum transfer $h = -0.4$~r.l.u. (a, c). No difference is detected in the phonon energy range at small momentum transfer $h = -0.1$~r.l.u. (b, d). The bottom panels show a close-up of the inelastic region of the spectra. The RIXS spectrum of NBCO-AF is shown in black, the one of NBCO-OP in red. In the latter, the continuum has been subtracted. Despite the difficulty of the subtraction, the difference spectra (OP minus AF, blue solid lines) are very clear. The difference spectra are vertically offset by -0.4 for clarity.}
\end{figure}

\begin{figure}
	\centering
	\includegraphics[width=0.8\columnwidth]{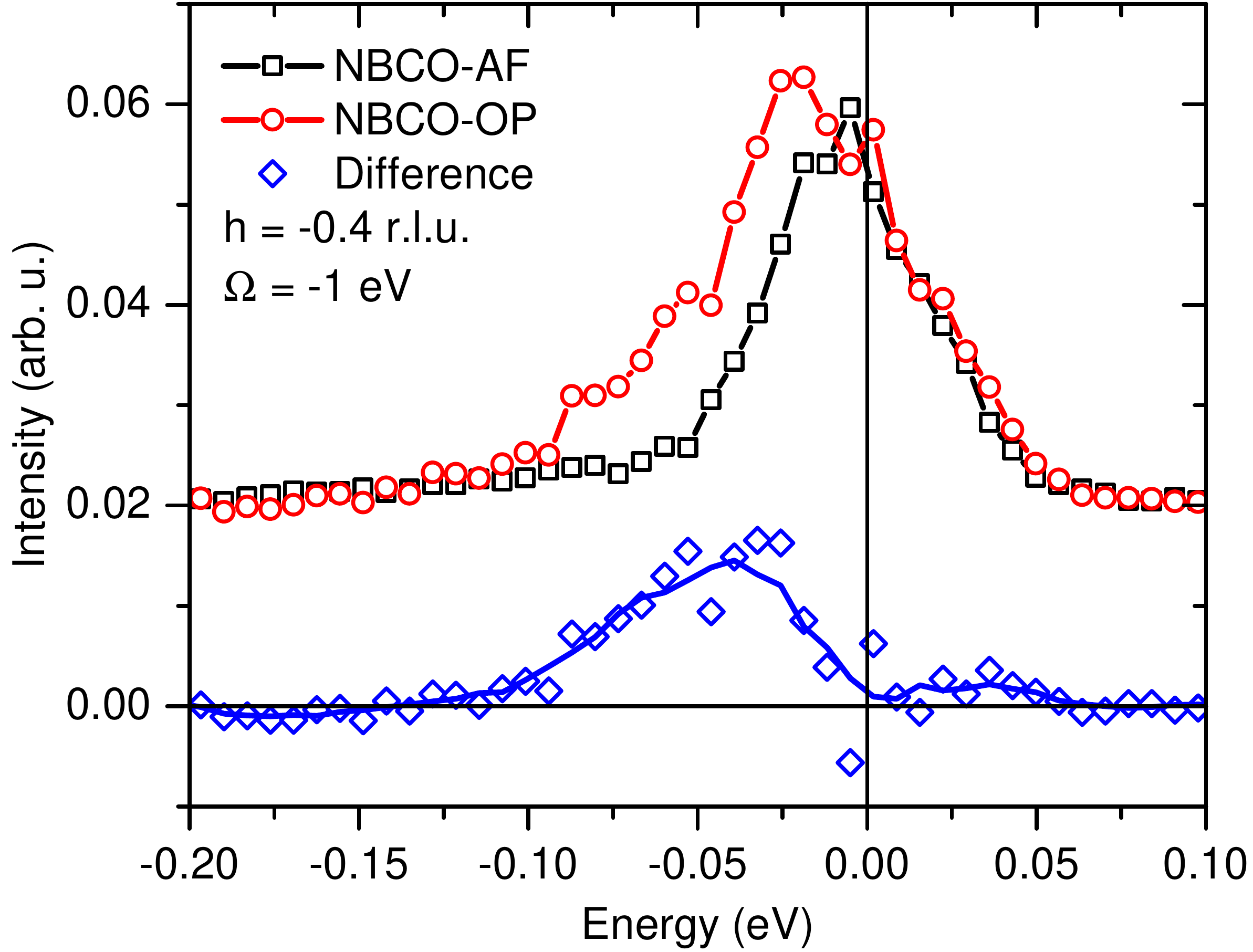}
	\caption{\label{fig:fig15} The different behavior of the NBCO-AF and NBCO-OP samples upon detuning. The doped sample (red circles) is definitely more robust than the undoped one (black squares). The difference spectrum (blue diamonds) shows that both phonon branches detected by RIXS are more robust in the OP sample.}
\end{figure}

As a matter of fact, there is an interplay between CDWs and phonons that strongly depends on the momentum transfer, which is evident from Fig.~\ref{fig:fig14}, where we compare the behavior at $h = -0.4$~r.l.u. (a, c) and $h = -0.1$~r.l.u. (b, d) of the NBCO-AF (black line) and NBCO-OP (red line) samples. After subtraction of the elastic line and the continuum, we find  that the phonon intensity is stronger in the doped system at large momentum transfer [Fig.~\ref{fig:fig14}(c)]. The difference spectrum (blue line) does not vanish in the energy region of the phonon excitations. In contrast, at small momentum transfer, the two dopings are equivalent within our sensitivity [Fig.~\ref{fig:fig14}(d)]. Moreover, the response to detuning is very different between NBCO-AF and NBCO-OP, as shown in Fig.~\ref{fig:fig15}. At $h = -0.4$~r.l.u., the phonon intensity is still sizable in the NBCO-OP compound upon detuning by $\Omega = -1$~eV, whereas it has almost disappeared in the NBCO-AF sample (black squares). This fact is evident from Fig.~\ref{fig:fig15}, in which the spectra have been normalized to the photon flux. The difference spectrum (blue diamonds) shows that both phonon  modes (breathing and buckling) are more robust upon detuning in the doped system.

\section{Discussion}
\label{sec:discussion}

Here, we focus on the experimental results for NBCO since the methods used to recover the EPC from RIXS spectra have been already discussed. Nevertheless, these two aspects are connected so that the discussion of the results on NBCO will also clarify the limits of the methods.

The momentum dependence of the EPC in the parent compound, plotted in Fig.~\ref{fig:fig12}, is fitted with simple trigonometric functions. The coupling to the buckling branch scales as $\cos(\pi h)$ (red dotted line) and the  coupling to the breathing branch as $\sin(\pi h)$ (blue dashed line) (see Appendix~\ref{appendix1}). Once again, we stress that these data may contain averages of unresolved branches. It is remarkable that such simple rules fit the data well, especially for the breathing mode whose error bars are smaller. This result is important because the simple dependence of $M/\Gamma$ on the momentum is theoretically obtained from models that consider only nearest neighbor interactions. The amplitude of the trigonometric functions has been adjusted by tuning the scaling factor so that:
\begin{align}
	& \left(\frac{M}{\Gamma}\right)_\mathrm{breath} = 0.45 \sin(\pi h),\\
	& \left(\frac{M}{\Gamma}\right)_\mathrm{buckle} = 0.98 \cos(\pi h).
\end{align}

These results demonstrate that RIXS provides easy and direct access to the momentum-dependent EPC. The analysis of the RIXS data of NBCO-AF and the internal consistency of the results obtained with different methods demonstrate that our model involving the starting Hamiltonian [Eq.~\eqref{eq:Hamiltonian}] and Einstein phonons is useful to study high energy phonons in undoped cuprates. It is remarkable that the intensity of the in-phase buckling branch is high and has a strong EPC; this is consistent with the large static buckling in the system, which is known to enhance the coupling to the bond-buckling phonons~\cite{Devereaux1995,Andersen1996}.

Until now, we have been using a strictly local perspective, where both the phonon and electron degrees of freedom are localized. While this approach produces good agreement with experimental data, as shown in Fig. \ref{fig:fig12}, it is important to also 
consider what happens when itinerant description of the electron states is used instead. 
Interestingly, in the RIXS studies of EPC, the theoretical results turn out to be quite similar in the two approaches. The comparison is shown in Fig.~\ref{fig:fig16}, where the experimental data are also reproduced (the calculations are presented and compared in Appendix A). At least in this case the sensitivity to the electron states description is rather modest. This observation has profound implications on our work since it legitimates the use of the same theoretical tools in treating AF compounds and doped superconducting cuprates. The limits of applicability to other systems will have to be assessed with further work, since the results depend on a delicate combination of many factors, such as electronic screening, excitonic effects and time scales.  

\begin{figure}
	\centering
	\includegraphics[width=0.8\columnwidth]{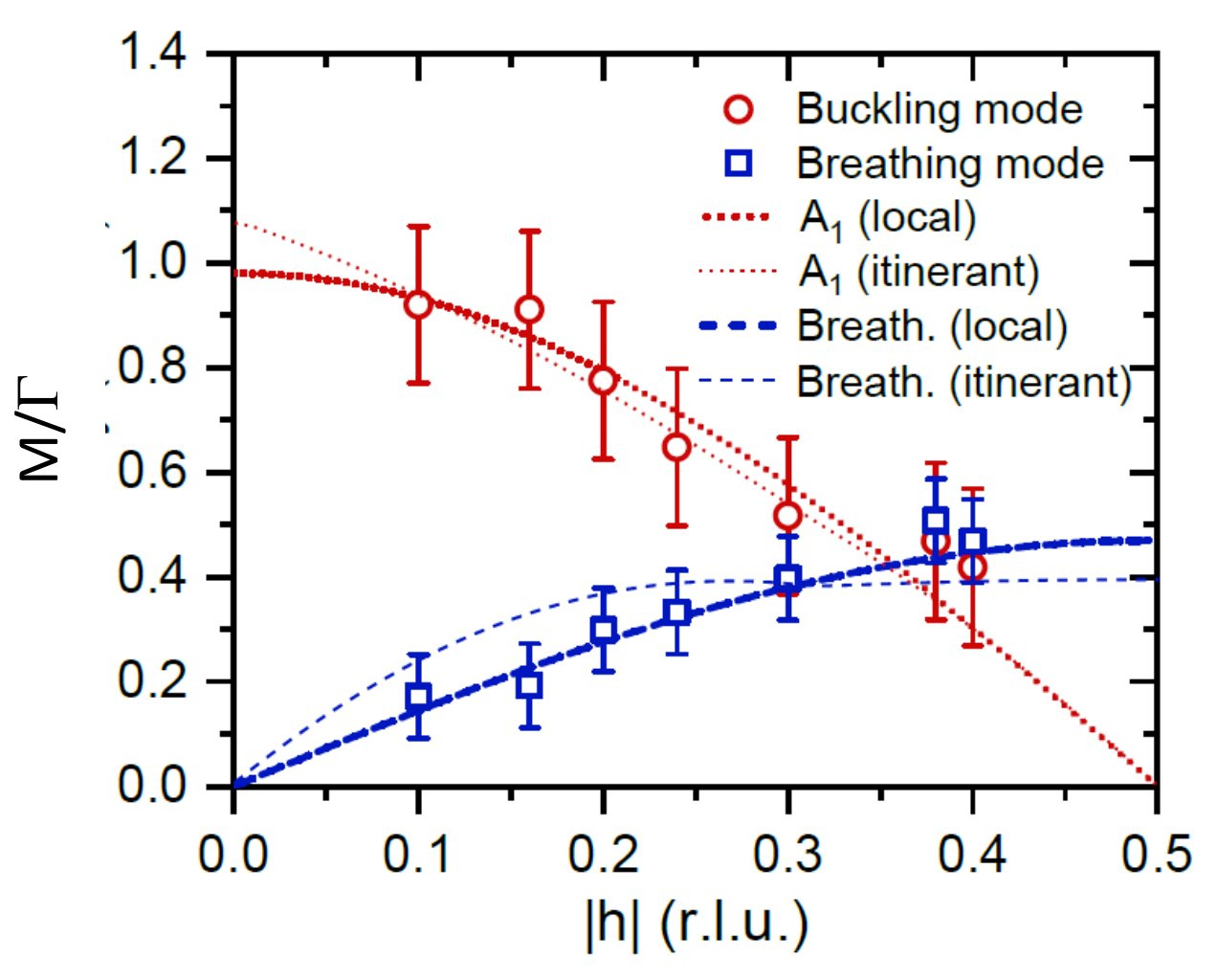}
	\caption{\label{fig:fig16} EPC strength $M$ (in units of $\Gamma$) as a function of momentum transfer for the buckling (red circles) and breathing (blue squares) modes. The momentum dependence of the EPC is reported from Fig.~\ref{fig:fig12}. Lines are best fits to the data obtained from the local (thick) and itinerant (thin) electron models.}
\end{figure}

Before concluding, let us consider briefly the effect of CDWs. This is not the object of our research and we have concentrated the work on $\mathbf{Q}$ values far from the critical CDW wave vector. However, in doped cuprates it is known that the CDWs are ubiquitous, at least in the UD samples and with lower intensity in the OP compounds~\cite{BlancoCanosa2014,Comin2016,Arpaia2019}. We have already shown that in NBCO-OP the role of the CDWs is crucial. More specifically, we stress the following points:
\begin{itemize}
	\item The modification of the spectral weight in the phonon region of NBCO follows the general pattern of  Bi$_2$Sr$_2$CaCu$_2$O$_{8+\delta}$ reported in Ref.~\onlinecite{Chaix2017} in that the increase of the intensity at $h = -0.4$~r.l.u. is analogous to the so-called phonon anomaly~\cite{LeTacon2013,Chaix2017}.
	
	\item At $h = -0.1$~r.l.u., after background subtraction, the phonon intensity is basically the same in the AF and OP samples. This means that our model (especially in the detuning formulation) can be used also in OP systems, provided the momentum transfer is sufficiently small. A more precise assessment is left for future investigations, but it is already clear that the model is meaningful around $h = -0.1$~r.l.u.. This region of the parameter space is of great interest since the coupling to the buckling modes are close to the maximum here, while the coupling of the breathing mode is close to zero. This situation as a whole is favorable to $d$-wave pairing and makes buckling phonons good candidates for a synergistic action together with magnetic fluctuations. Basically, our approach works well in the region of the parameter space that is most relevant to superconductivity in the cuprates. Note that according to Ref.~\onlinecite{Johnston2010} the buckling modes are beneficial to pairing, in particular the out-of-phase B$_1$ mode, whose RIXS intensity vanishes for symmetry reasons (see Appendix~\ref{appendix1}). Since the screening is comparable for the A$_1$ and B$_1$ modes, we argue that the out-of-phase mode couples to the electrons with a strength comparable to the in-phase mode. The effect of the breathing modes, instead, is now considered detrimental, although this has been controversial~\cite{Bulut1996,Shen2002,Ishihara2004,Sandvik2004,Johnston2010}.
	
	\item At large momentum transfer, the situation is completely different: the effect of the CDWs cannot be neglected in the OP sample and, \emph{a fortiori}, in UD compounds. This is remarkable because the CDW signal in OP cuprates is very small and sometimes considered to be zero~\cite{Ghiringhelli2012}; however, the recent discovery that in UD and OP NBCO and YBCO charge density fluctuations persist at all temperatures \cite{Arpaia2019} is consistent with the phenomenology of phonon intensity at high momenta. What remains as a question of interpretation is whether the enhancement of phonon peaks in RIXS is a signal of an actual increase of the EPC, which in turn causes the CDW, or whether the CDW has a different origin and the phonon enhancement at those momenta is the manifestation of the modulated charge density, at constant EPC \cite{Chaix2017,lin2020nature}. In all cases the sensitivity of the EPC to the CDWs is extremely high. 
\end{itemize}

The above arguments clarify the limits of validity of the model itself, i.e., far from CDW-induced effects. It is interesting, however, to try to use the model also to evaluate an effective EPC, $M_\mathrm{eff}$, in the region where the CDWs are important. This very crude approach can be applied to the energy-detuned RIXS spectra plotted in Fig.~\ref{fig:fig15} and measured at $h = -0.4$~r.l.u.. For the breathing mode we obtain $M_\mathrm{eff} \sim 3$ -- 4 times the value in the parent compound, which is a coupling strength more typical of polaronic systems. Note also that, on the basis of the ratio $M_\mathrm{eff}/M$, it would be possible to introduce  a scale to characterize how far we are from the situation free of CDWs.

\section{Conclusions}
\label{sec:conclusions}

We have presented three ways to obtain quantitative information on the EPC with momentum resolution from high energy resolution RIXS spectra. We applied the methods to antiferromagnetic NBCO. We also discussed the evolution of the EPC in underdoped and optimally doped NBCO and its relation to charge order. The three approaches were critically compared by pointing out their merits and limitations, and we showed that often the optimal way to go is a suitable combination of the three methods. Importantly, these approaches can be applied without numerical simulations of the spectra because we have introduced simple procedures based on rescaled variables and on energy detuning. Thus, they can be applied to a variety of strongly correlated materials. The present results have an impact not only on the ``123'' family of cuprates but also on material science in a broad sense. With the upcoming progress in the instrumentation the RIXS studies on electron-phonon coupling, at present in their infancy, will reach maturity.

\begin{acknowledgments} 
RIXS data were collected at the beam line ID32 of the ESRF using the ERIXS spectrometer jointly designed by the ESRF and the Politecnico di Milano. This work was supported by the ERC-P-ReXS project (2016-0790) of the Fondazione CARIPLO and Regione Lombardia, in Italy. R.A. is supported by the Swedish Research Council (VR) under the project ``Evolution of nanoscale charge order in superconducting YBCO nanostructures''. S.J. is supported by the National Science Foundation under Grant No.~DMR-1842056. J.v.d.B. acknowledges financial support from the DFG through the W\"urzburg-Dresden Cluster of Excellence on Complexity and Topology in Quantum Matter -- \textit{ct.qmat} (EXC 2147, project-id 39085490) and through SFB 1143 (project-id 247310070).
\end{acknowledgments}

\appendix

\section{Theoretical arguments on RIXS intensity and selection rules}
\label{appendix1}

We outline here a theoretical approach more general than the one introduced by Ament \emph{et al.} \cite{Ament2011}. We aim at validating the local electron model, which is implicit in our treatment. In fact, we demonstrate that the itinerancy of the electrons does not significantly change the landscape of the EPC in cuprates. We also derive the momentum dependence of the RIXS intensity for the buckling and breathing modes.

The generic form of the EPC is given by the following Hamiltonian:
\begin{equation}
\mathcal{H}_\mathrm{e-ph} = \frac{1}{\sqrt{N}}\sum_{\mathbf{k},\mathbf{q},\sigma,\nu}M^\pdag_\nu(\mathbf{k},\mathbf{q})c_{\mathbf{k}-\mathbf{q},\sigma}^\dagger c^\pdag_{\mathbf{k},\sigma}(b_{\mathbf{q},\nu}^\dagger + b^\pdag_{-\mathbf{q},\nu}).
\end{equation}
Here, $c_{\mathbf{k},\sigma}^\dagger$ ($c^\pdag_{\mathbf{k},\sigma}$) creates (annihilates) an electron with momentum $\mathbf{k}$ and spin $\sigma$ in the $pd$-antibonding band of the CuO$_2$ plane, $b_{\mathbf{q},\nu}^\dagger$ ($b^\pdag_{\mathbf{q},\nu}$) creates (annihilates) a phonon in branch $\nu$ with momentum $\mathbf{q}$, and $M_\nu(\mathbf{k},\mathbf{q})$ is the EPC constant. 

In the general case, the EPC depends on both the electron momentum $\mathbf{k}$ and the phonon momentum $\mathbf{q}$. The latter is related to the momentum transfer $\mathbf{Q}$. To determine how RIXS probes this momentum dependence, we consider the scattering process within the ultra-short core hole lifetime expansion. To first order, the RIXS intensity at the Cu $L_3$ edge is then given by:
\begin{equation}
I(\mathbf{Q}, \omega) \propto |\mathcal{F}_{fg}|^2 
\delta\left(E_g - E_f + \omega\right). 
\end{equation}
Here, the scattering cross section is defined as $\mathcal{F}_{fg} = \langle f | D_\mathrm{out} \mathcal{H}_\mathrm{e-ph} D_\mathrm{in} |g\rangle$, where $|g\rangle$ and $|f\rangle $ are the initial and final states, respectively, with energies $E_g$ and $E_f$, $\omega$ is the energy loss, and $D_\mathrm{in}$ and $D_\mathrm{out}$ are the dipole transition operators. The latter are defined as
\begin{eqnarray*}
D_\mathrm{in} & = & \sum_{i} e^{i\mathbf{q}_\mathrm{in}\cdot \mathbf{R}_i} d_{i,\sigma}^\dagger p^\pdag_{i,\sigma} = \sum_{\mathbf{k}} d_{\mathbf{k}+\mathbf{q}_\mathrm{in},\sigma}^\dagger p^\pdag_{\mathbf{k},\sigma}, \\
D_\mathrm{out} & = & \sum_{i} e^{-i\mathbf{q}_\mathrm{out}\cdot \mathbf{R}_i} p_{i,\sigma}^\dagger d^\pdag_{i,\sigma} = \sum_{\mathbf{k}} p_{\mathbf{k}-\mathbf{q}_\mathrm{out},\sigma}^\dagger d^\pdag_{\mathbf{k},\sigma},
\end{eqnarray*}
where $d_{i,\sigma}^\dagger$ creates an electron in the Cu $3d_{x^2-y^2}$ orbital located at position $\mathbf{R}_i$ and $p_{i,\sigma}^\dagger$ creates an electron in the relevant Cu $2p$ core level, and $\mathbf{q}_\mathrm{in}$ and $\mathbf{q}_\mathrm{out}$ are the incident and scattered photon wave vectors, respectively. Note that we have neglected the polarization-dependent prefactors in the dipole matrix elements for simplicity.

We assume that the ground state can be written in the form $|g\rangle = |\psi_\mathrm{el}, n_{\mathbf{q},\nu} = 0\rangle$, which describes the electronic subsystem with no phonon quanta excited. Here, we are interested in quasi-elastic processes, where the energy transferred into the system excites a phonon. Therefore, we can restrict the final states to only those where one phonon has been excited, i.e., $|f\rangle = |\psi_\mathrm{el}, n_{\mathbf{q},\nu} = 1\rangle$. This assumption is equivalent to the view that the phonon and electron subsystems are not deeply entangled such that there are zero phonons present in the ground state. Under these simplifying assumptions we have:
\begin{widetext}
	\begin{eqnarray}\nonumber
		\mathcal{F}_{fg} & = &  \sum_{\substack{\mathbf{p},\mathbf{p}^\prime \\ \sigma,\sigma^\prime}}\left\langle \psi_\mathrm{el}, n_{\mathbf{q},\nu} = 1 \left| p_{\mathbf{p-q}_\mathrm{out},\sigma}^\dagger d_{\mathbf{p},\sigma}^\pdag \mathcal{H}^\pdag_\mathrm{e-ph} d_{\mathbf{p^\prime+q}_\mathrm{in},\sigma^\prime}^\dagger p^\pdag_{\mathbf{p}^\prime,\sigma^\prime} \right| \psi_\mathrm{el}, n_{\mathbf{q},\nu = 0} \right\rangle \\
		& = & \sum_{\mathbf{p},\sigma} \left\langle \psi_\mathrm{el},n_{\mathbf{q},\nu} = 1 \left| d^\pdag_{\mathbf{p},\sigma} \mathcal{H}^\pdag_\mathrm{e-ph} d_{\mathbf{p+Q},\sigma}^\dagger \right| \psi_\mathrm{el}, n_{\mathbf{q},\nu} = 0 \right\rangle.
	\end{eqnarray}
In the last step we have introduced the momentum transfer ${\bf Q}={\bf q}_\mathrm{in} - {\bf q}_\mathrm{out}$. 

The Cu orbital operator $d_{\mathbf{k},\sigma}^\dagger$ is related to the $c_{\mathbf{k},\sigma}^\dagger$ band operator by $c_{\mathbf{k},\sigma}^\dagger = \phi_\mathrm{Cu}^\star(\mathbf{k}) d_{\mathbf{k},\sigma}^\dagger$, where $\phi_\mathrm{Cu}^\star(\mathbf{k})$ measures the Cu character of the antibonding band. Therefore, the scattering matrix element written in band space is:
	\begin{equation}\label{Eq:A4}
		\mathcal{F}_{fg} = \frac{1}{\sqrt{N}} \sum_{\substack{\mathbf{p,k,q}\\\sigma,\nu}} M^{\phantom\star}_\nu (\mathbf{k,q}) \phi_\mathrm{Cu}^\star(\mathbf{p}+
		\mathbf{Q}) \phi^{\phantom\star}_\mathrm{Cu}(\mathbf{p}) \left\langle \psi_\mathrm{el},n_{\mathbf{q},\nu} = 1 \left| c^\pdag_{\mathbf{p},\sigma} c_{\mathbf{k-q},\sigma}^\dagger c^\pdag_{\mathbf{k},\sigma} c_{\mathbf{p+Q},\sigma}^\dagger \right| \psi_\mathrm{el},n_{\mathbf{q},\nu} = 1 \right\rangle.
	\end{equation}
The expectation value appearing in Eq. (\ref{Eq:A4}) must be evaluated using the correlated many-body wave function. We can, however, simplify the problem by considering either the fully localized or fully itinerant cases.

First, we consider the itinerant case and approximate the many-body wave function 
as the non-interacting Fermi sea. Then, for non-zero values of $\mathbf{Q}$, we require $\mathbf{p} = \mathbf{k-q}$ and $\mathbf{Q} = \mathbf{q}$, and the scattering amplitude simplifies to:
\begin{equation}
\mathcal{F}_{fg} = \frac{1}{\sqrt{N}} \sum_{\mathbf{k},\sigma,\nu} M^{\phantom\star}_\nu (\mathbf{k,Q}) \phi_\mathrm{Cu}^\star(\mathbf{k}) \phi^{\phantom\star}_\mathrm{Cu}(\mathbf{k-Q})[1-n_\mathrm{F}(\epsilon_{\bf{k}-{\bf Q}})]
[1-n_\mathrm{F}(\epsilon_{\bf k})].
\end{equation}
Here, $n_\mathrm{F}(x)$ is the Fermi factor and $\epsilon({\bf k})$ is the band dispersion. 
Similarly, we obtain the localized limit by inverse Fourier transforming Eq. (\ref{Eq:A4}) and then retaining only the local operators inside the expectation value:
\begin{eqnarray}\nonumber
    \mathcal{F}_{fg}
		&=&\frac{1}{N^\frac{5}{2}} \sum_{\substack{\mathbf{p,k,q}\\\sigma,\nu,l}} M^{\phantom\star}_\nu (\mathbf{k,q}) \phi_\mathrm{Cu}^\star(\mathbf{p}+
		\mathbf{Q}) \phi^{\phantom\star}_\mathrm{Cu}(\mathbf{p}) 
		e^{-i (\mathbf{Q}-\mathbf{q})\cdot \mathbf{R}_l}
		\left\langle \psi_\mathrm{el},n_{\mathbf{q},\nu} = 1 \left| (1-n_l) \right| \psi_\mathrm{el},n_{\mathbf{q},\nu} = 1 \right\rangle
		\\
		&=&\frac{1}{N^\frac{3}{2}} \sum_{\substack{\mathbf{p,k}\\\sigma,\nu}} M^{\phantom\star}_\nu (\mathbf{k,Q}) \phi_\mathrm{Cu}^\star(\mathbf{p}+
		\mathbf{Q}) \phi^{\phantom\star}_\mathrm{Cu}(\mathbf{p}).
\end{eqnarray}

Thus, in both limits, RIXS at the Cu $L_3$ edge measures a $\mathbf{k}$-integrated coupling constant weighted by the Cu orbital character of the band and additional phase 
space factors. The involvement of the Cu orbital character is in agreement with conclusions drawn in a previous cluster ED study of a quasi-1D cuprate~\cite{Lee2013}. In the main text we have adopted the localized limit, since we expect strong electron correlations and the sizable core hole potential to localize the excited core electron to the Cu site where it is created. This approach also has the advantage that simple analytic expressions 
for the intensity can be obtained.

Reference~\onlinecite{Johnston2010} argued that  $\phi_\mathrm{Cu}(\mathbf{k})$ can be approximated by a constant $\phi_\mathrm{Cu}(\mathbf{k}) \sim A_\mathrm{Cu}$, which leads to simplified momentum dependencies for the various coupling constants. In this case, $A_\mathrm{Cu}$ can be absorbed into the prefactor of the coupling constant and the scattering amplitude for the localized limit reduces to $F_{fq}=\frac{1}{\sqrt{N}} \sum_{\mathbf{k},\sigma,\nu}M^{\phantom\star}_\nu (\mathbf{k,Q})$. For the in-phase ($+$) and out-of-phase ($-$) Cu-O bond-buckling branches one has ($\mathbf{p=k-Q}$):
	\begin{equation}
		M_{\pm}(\mathbf{k,Q}) \propto \left[\sin\left(\frac{k_x a}{2}\right) \sin\left(\frac{p_x a}{2}\right) \cos\left(\frac{q_y a}{2}\right) \pm \sin\left(\frac{k_y a}{2}\right) \sin\left(\frac{p_y a}{2}\right) \cos\left(\frac{q_x a}{2}\right)\right].
	\end{equation}
Similarly, for the breathing mode:
\begin{equation}
M_\mathrm{breath}(\mathbf{Q}) \propto \left[\sin^2\left(\frac{q_x a}{2}\right) + \sin^2\left(\frac{q_y a}{2}\right)\right]^{1/2}.
\end{equation}
Because the coupling constant for the breathing mode does not depend on the Fermion momentum $\mathbf{k}$, the $\mathbf{k}$-averaging is trivial and $I_\mathrm{breath}(\mathbf{Q}) \propto |M_\mathrm{breath}(\mathbf{Q})|^2$. The situation is different for the buckling modes. Due to the sign change in the $\mathbf{k}$-dependence of $M_{-}(\mathbf{k,Q})$, the intensity for the out-of-phase buckling mode $I_{-}(\mathbf{Q})$ vanishes. For the in-phase buckling phonon branch one arrives at:
\begin{equation}
I_{+}(\mathbf{Q}) \propto \cos^2\left(\frac{q_x a}{2}\right) \cos^2\left(\frac{q_y a}{2}\right).
\end{equation}

Figure~\ref{fig:fig16} of the main text displays the EPC strength $M(\mathbf{Q})$ of the in-phase buckling (red circles) and breathing (blue squares) modes, normalized to the natural width $\Gamma$ of the Cu $L_3$ resonance. The EPC strength is plotted as a function of momentum transfer along the $(1, 0)$ direction of the reciprocal space ($h = q_x a/2\pi$). Note that the EPC strength is proportional to the square root of the RIXS intensity. The best fits to the data obtained in the two limiting cases of local (thick lines) and itinerant (thin lines) electron models are shown for comparison. In the latter case, the band dispersion is taken from Ref.~\onlinecite{Markiewicz2005}. The two approximations yield similar results and describe the the data fairly well, with a small advantage to the local electron picture close to the $X$ point of the BZ.

\end{widetext}

\section{Recovering the phonon signal from RIXS spectra}
\label{appendix2}

\begin{figure*}
	\centering
	\includegraphics[width=0.9\textwidth]{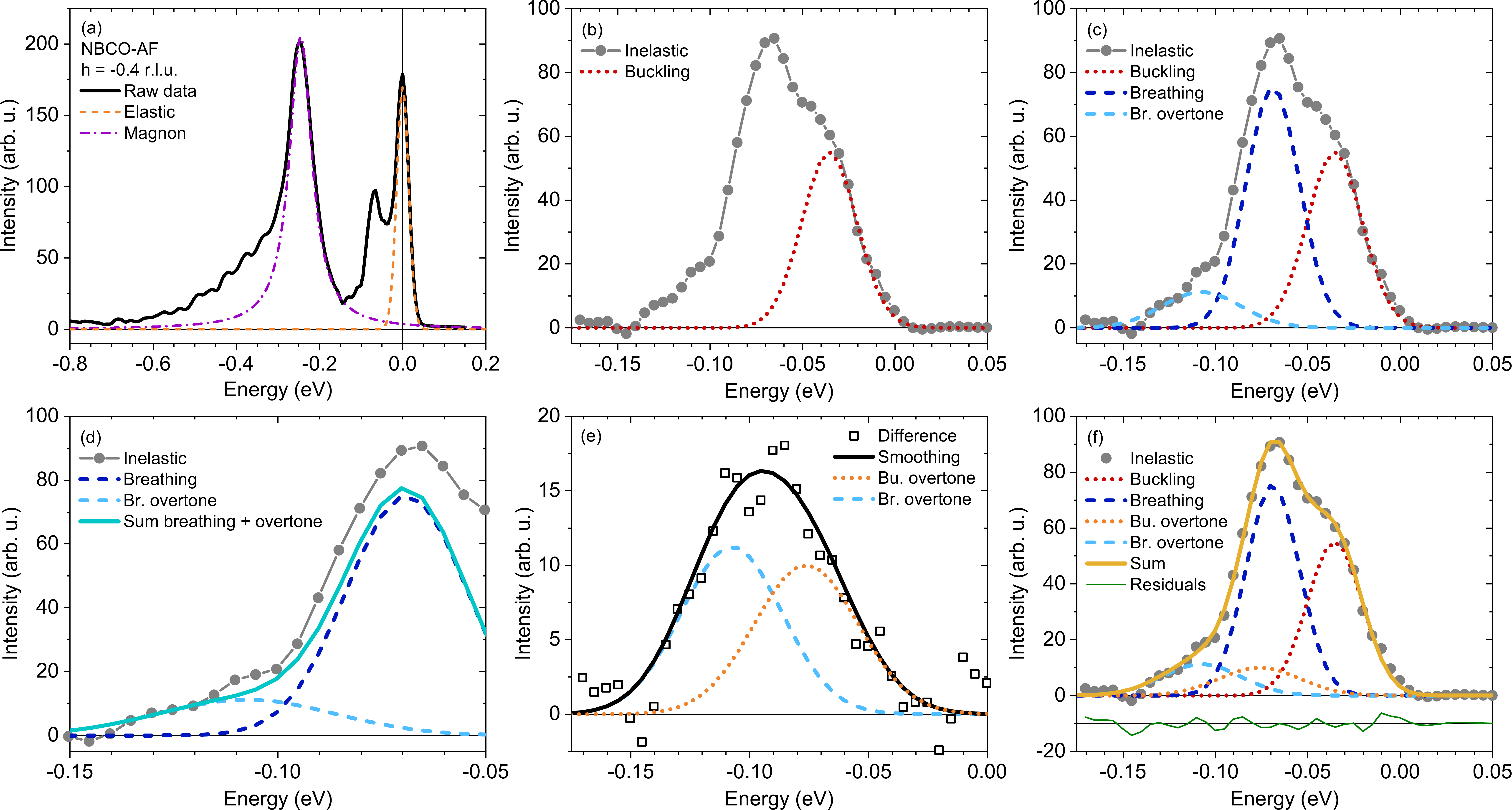}
	\caption{\label{fig:fig17} Detailed presentation of the decomposition of the RIXS spectrum of NBCO-AF, already presented more briefly in Fig. \ref{fig:fig11}. See text of appendix for details.}
\end{figure*}

We summarize here the procedures used to recover the phonon components from the RIXS spectra. These rather technical issues are collected here to improve the flow and cohesion of the main text. Nevertheless, the content of this Appendix must be considered as an integral part of the work. 

We consider the benchmark RIXS spectrum shown in Fig.~\ref{fig:fig17}(a) (black solid line) and we describe its decomposition. For completeness, we show in panel (a) the subtraction of the long and weak tail coming from the magnon peak. This tail is dominated by the Lorentzian shape of the magnon (purple dash dotted line), which is broader than the instrumental line width. Strictly speaking, a Voigt function should be used but the Lorentzian shape is a good approximation. The resolution-limited Gaussian lineshape defining the elastic line is shown as orange dashed line. The resulting inelastic spectrum is plotted as gray circles in panel (b).

We must use the appropriate line shape when decomposing the inelastic phonon region. Since the instrumental line width is much larger than the phonon width, a resolution-limited Gaussian seems to be acceptable. However, as discussed in the main text, the phonon features that we observe may be an average of several phonon modes. Thus, a broadening is generated with respect to the instrumental response function. Our numerical analysis shows that this is a small effect (typically 2 -- 3~meV). To take this effect into account, we leave the line width as a free parameter between 34~meV and 37~meV. Another issue is the energy of the overtone, which in the Einstein model is exactly twice the phonon energy. However, such an extreme situation is very rare in a real material and we use the Einstein model as an approximate description of a system having a tiny dispersion. The factor of two between the phonon energy and its overtone is an upper limit so that also this parameter is left free.

With the above precautions in mind we address the decomposition of the inelastic spectrum by exploiting first the experimental information that is better observed. This is done iteratively with the following logical sequence. Initially, we fit the low-energy shoulder of the buckling mode as shown in panel (b) (dotted red line). The next step is the addition of the breathing component at an energy that is very well defined already in the raw data [dashed blue line in panel (c)]. Moreover, the tail at high energy suggests the approximate position of the breathing overtone (dashed light blue line). The intensity of the main peak and of the overtone of the breathing mode are chosen so that their sum reproduces roughly the turning point of the inelastic signal at around 100~meV [solid line in panel (d), where the energy axis is zoomed into the range of interest]. This is a delicate issue we will discuss in more detail in the next paragraph. By subtracting the two main phonon peaks from the inelastic spectrum, a contribution coming only from the overtones emerges [black squares in panel (e), where both axes are zoomed into the appropriate region]. The black solid line of panel (e) is the smoothed total signal from the overtones. The difference between the black line and the light blue dashed line gives the orange dotted curve representing the buckling overtone. At this time a decomposition cycle is completed, but it is certainly necessary to repeat the procedure to refine the parameters.

The final result is given in panel (f) as a dark yellow line. The residuals are plotted as a thin green line and are very small. The weak point of this procedure comes from the difficulty in finding the breathing weights in panels (c) and (d). As a matter of fact, if this value is changed by 10\%, it is still possible to obtain a good decomposition so that there is some ambiguity in the decomposition. One can improve the procedure, however, if information on $I_1$ or $I_2/I_1$ is exploited, which can be obtained using another method. This is what we have done on the basis of the results from detuning. Once again a suitable mix of the different approaches improves the quality of the decomposition. Note that the energies of the overtones differ from the value of the Einstein model, as expected: for the breathing mode the ratio of the overtone energy and the main peak energy is 1.7 instead of 2, while for the buckling mode it amounts to 1.55. 

\bibliography{biblio}

\begin{thebibliography}{77}%
\makeatletter
\providecommand \@ifxundefined [1]{%
 \@ifx{#1\undefined}
}%
\providecommand \@ifnum [1]{%
 \ifnum #1\expandafter \@firstoftwo
 \else \expandafter \@secondoftwo
 \fi
}%
\providecommand \@ifx [1]{%
 \ifx #1\expandafter \@firstoftwo
 \else \expandafter \@secondoftwo
 \fi
}%
\providecommand \natexlab [1]{#1}%
\providecommand \enquote  [1]{``#1''}%
\providecommand \bibnamefont  [1]{#1}%
\providecommand \bibfnamefont [1]{#1}%
\providecommand \citenamefont [1]{#1}%
\providecommand \href@noop [0]{\@secondoftwo}%
\providecommand \href [0]{\begingroup \@sanitize@url \@href}%
\providecommand \@href[1]{\@@startlink{#1}\@@href}%
\providecommand \@@href[1]{\endgroup#1\@@endlink}%
\providecommand \@sanitize@url [0]{\catcode `\\12\catcode `\$12\catcode
  `\&12\catcode `\#12\catcode `\^12\catcode `\_12\catcode `\%12\relax}%
\providecommand \@@startlink[1]{}%
\providecommand \@@endlink[0]{}%
\providecommand \url  [0]{\begingroup\@sanitize@url \@url }%
\providecommand \@url [1]{\endgroup\@href {#1}{\urlprefix }}%
\providecommand \urlprefix  [0]{URL }%
\providecommand \Eprint [0]{\href }%
\providecommand \doibase [0]{http://dx.doi.org/}%
\providecommand \selectlanguage [0]{\@gobble}%
\providecommand \bibinfo  [0]{\@secondoftwo}%
\providecommand \bibfield  [0]{\@secondoftwo}%
\providecommand \translation [1]{[#1]}%
\providecommand \BibitemOpen [0]{}%
\providecommand \bibitemStop [0]{}%
\providecommand \bibitemNoStop [0]{.\EOS\space}%
\providecommand \EOS [0]{\spacefactor3000\relax}%
\providecommand \BibitemShut  [1]{\csname bibitem#1\endcsname}%
\let\auto@bib@innerbib\@empty
\bibitem [{\citenamefont {Scalapino}(2012)}]{Scalapino2012}%
  \BibitemOpen
  \bibfield  {author} {\bibinfo {author} {\bibfnamefont {D.~J.}\ \bibnamefont
  {Scalapino}},\ }\bibfield  {title} {\enquote {\bibinfo {title} {A common
  thread: The pairing interaction for unconventional superconductors},}\ }\href
  {\doibase 10.1103/RevModPhys.84.1383} {\bibfield  {journal} {\bibinfo
  {journal} {Rev. Mod. Phys.}\ }\textbf {\bibinfo {volume} {84}},\ \bibinfo
  {pages} {1383--1417} (\bibinfo {year} {2012})}\BibitemShut {NoStop}%
\bibitem [{\citenamefont {Savrasov}\ and\ \citenamefont
  {Andersen}(1996)}]{Savrasov1996}%
  \BibitemOpen
  \bibfield  {author} {\bibinfo {author} {\bibfnamefont {S.~Y.}\ \bibnamefont
  {Savrasov}}\ and\ \bibinfo {author} {\bibfnamefont {O.~K.}\ \bibnamefont
  {Andersen}},\ }\bibfield  {title} {\enquote {\bibinfo {title}
  {{Linear-Response Calculation of the Electron-Phonon Coupling in Doped
  CaCu${\mathrm{O}}_{2}$}},}\ }\href {\doibase 10.1103/PhysRevLett.77.4430}
  {\bibfield  {journal} {\bibinfo  {journal} {Phys. Rev. Lett.}\ }\textbf
  {\bibinfo {volume} {77}},\ \bibinfo {pages} {4430--4433} (\bibinfo {year}
  {1996})}\BibitemShut {NoStop}%
\bibitem [{\citenamefont {Andersen}\ \emph {et~al.}(1996)\citenamefont
  {Andersen}, \citenamefont {Savrasov}, \citenamefont {Jepsen},\ and\
  \citenamefont {Liechtenstein}}]{Andersen1996}%
  \BibitemOpen
  \bibfield  {author} {\bibinfo {author} {\bibfnamefont {O.~K.}\ \bibnamefont
  {Andersen}}, \bibinfo {author} {\bibfnamefont {S.~Y.}\ \bibnamefont
  {Savrasov}}, \bibinfo {author} {\bibfnamefont {O.}~\bibnamefont {Jepsen}}, \
  and\ \bibinfo {author} {\bibfnamefont {A.~I.}\ \bibnamefont
  {Liechtenstein}},\ }\bibfield  {title} {\enquote {\bibinfo {title}
  {{Out-of-plane instability and electron-phonon contribution to $s$- and
  $d$-wave pairing in high-temperature superconductors; LDA linear-response
  calculation for doped CaCuO$_2$ and a generic tight-binding model}},}\ }\href
  {\doibase 10.1007/BF00768402} {\bibfield  {journal} {\bibinfo  {journal} {J.
  Low Temp. Phys.}\ }\textbf {\bibinfo {volume} {105}},\ \bibinfo {pages}
  {285--304} (\bibinfo {year} {1996})}\BibitemShut {NoStop}%
\bibitem [{\citenamefont {Sakai}\ \emph {et~al.}(1997)\citenamefont {Sakai},
  \citenamefont {Poilblanc},\ and\ \citenamefont {Scalapino}}]{Sakai1997}%
  \BibitemOpen
  \bibfield  {author} {\bibinfo {author} {\bibfnamefont {T.}~\bibnamefont
  {Sakai}}, \bibinfo {author} {\bibfnamefont {D.}~\bibnamefont {Poilblanc}}, \
  and\ \bibinfo {author} {\bibfnamefont {D.~J.}\ \bibnamefont {Scalapino}},\
  }\bibfield  {title} {\enquote {\bibinfo {title} {{Hole pairing and phonon
  dynamics in generalized two-dimensional $t$-$J$ Holstein models}},}\ }\href
  {\doibase 10.1103/PhysRevB.55.8445} {\bibfield  {journal} {\bibinfo
  {journal} {Phys. Rev. B}\ }\textbf {\bibinfo {volume} {55}},\ \bibinfo
  {pages} {8445--8451} (\bibinfo {year} {1997})}\BibitemShut {NoStop}%
\bibitem [{\citenamefont {Jepsen}\ \emph {et~al.}(1998)\citenamefont {Jepsen},
  \citenamefont {Andersen}, \citenamefont {Dasgupta},\ and\ \citenamefont
  {Savrasov}}]{Jepsen1998}%
  \BibitemOpen
  \bibfield  {author} {\bibinfo {author} {\bibfnamefont {O.}~\bibnamefont
  {Jepsen}}, \bibinfo {author} {\bibfnamefont {O.K.}\ \bibnamefont {Andersen}},
  \bibinfo {author} {\bibfnamefont {I.}~\bibnamefont {Dasgupta}}, \ and\
  \bibinfo {author} {\bibfnamefont {S.}~\bibnamefont {Savrasov}},\ }\bibfield
  {title} {\enquote {\bibinfo {title} {{Buckling and $d$-wave pairing in
  high-$T_\mathrm{c}$ superconductors}},}\ }\href {\doibase
  https://doi.org/10.1016/S0022-3697(98)00089-4} {\bibfield  {journal}
  {\bibinfo  {journal} {J. Phys. Chem. Solids}\ }\textbf {\bibinfo {volume}
  {59}},\ \bibinfo {pages} {1718 -- 1722} (\bibinfo {year} {1998})}\BibitemShut
  {NoStop}%
\bibitem [{\citenamefont {Ishihara}\ and\ \citenamefont
  {Nagaosa}(2004)}]{Ishihara2004}%
  \BibitemOpen
  \bibfield  {author} {\bibinfo {author} {\bibfnamefont {Sumio}\ \bibnamefont
  {Ishihara}}\ and\ \bibinfo {author} {\bibfnamefont {Naoto}\ \bibnamefont
  {Nagaosa}},\ }\bibfield  {title} {\enquote {\bibinfo {title} {Interplay of
  electron-phonon interaction and electron correlation in high-temperature
  superconductivity},}\ }\href {\doibase 10.1103/PhysRevB.69.144520} {\bibfield
   {journal} {\bibinfo  {journal} {Phys. Rev. B}\ }\textbf {\bibinfo {volume}
  {69}},\ \bibinfo {pages} {144520} (\bibinfo {year} {2004})}\BibitemShut
  {NoStop}%
\bibitem [{\citenamefont {Johnston}\ \emph {et~al.}(2010)\citenamefont
  {Johnston}, \citenamefont {Vernay}, \citenamefont {Moritz}, \citenamefont
  {Shen}, \citenamefont {Nagaosa}, \citenamefont {Zaanen},\ and\ \citenamefont
  {Devereaux}}]{Johnston2010}%
  \BibitemOpen
  \bibfield  {author} {\bibinfo {author} {\bibfnamefont {S.}~\bibnamefont
  {Johnston}}, \bibinfo {author} {\bibfnamefont {F.}~\bibnamefont {Vernay}},
  \bibinfo {author} {\bibfnamefont {B.}~\bibnamefont {Moritz}}, \bibinfo
  {author} {\bibfnamefont {Z.-X.}\ \bibnamefont {Shen}}, \bibinfo {author}
  {\bibfnamefont {N.}~\bibnamefont {Nagaosa}}, \bibinfo {author} {\bibfnamefont
  {J.}~\bibnamefont {Zaanen}}, \ and\ \bibinfo {author} {\bibfnamefont {T.~P.}\
  \bibnamefont {Devereaux}},\ }\bibfield  {title} {\enquote {\bibinfo {title}
  {Systematic study of electron-phonon coupling to oxygen modes across the
  cuprates},}\ }\href {\doibase 10.1103/PhysRevB.82.064513} {\bibfield
  {journal} {\bibinfo  {journal} {Phys. Rev. B}\ }\textbf {\bibinfo {volume}
  {82}},\ \bibinfo {pages} {064513} (\bibinfo {year} {2010})}\BibitemShut
  {NoStop}%
\bibitem [{\citenamefont {Fausti}\ \emph {et~al.}(2011)\citenamefont {Fausti},
  \citenamefont {Tobey}, \citenamefont {Dean}, \citenamefont {Kaiser},
  \citenamefont {Dienst}, \citenamefont {Hoffmann}, \citenamefont {Pyon},
  \citenamefont {Takayama}, \citenamefont {Takagi},\ and\ \citenamefont
  {Cavalleri}}]{Fausti2011}%
  \BibitemOpen
  \bibfield  {author} {\bibinfo {author} {\bibfnamefont {D.}~\bibnamefont
  {Fausti}}, \bibinfo {author} {\bibfnamefont {R.~I.}\ \bibnamefont {Tobey}},
  \bibinfo {author} {\bibfnamefont {N.}~\bibnamefont {Dean}}, \bibinfo {author}
  {\bibfnamefont {S.}~\bibnamefont {Kaiser}}, \bibinfo {author} {\bibfnamefont
  {A.}~\bibnamefont {Dienst}}, \bibinfo {author} {\bibfnamefont {M.~C.}\
  \bibnamefont {Hoffmann}}, \bibinfo {author} {\bibfnamefont {S.}~\bibnamefont
  {Pyon}}, \bibinfo {author} {\bibfnamefont {T.}~\bibnamefont {Takayama}},
  \bibinfo {author} {\bibfnamefont {H.}~\bibnamefont {Takagi}}, \ and\ \bibinfo
  {author} {\bibfnamefont {A.}~\bibnamefont {Cavalleri}},\ }\bibfield  {title}
  {\enquote {\bibinfo {title} {Light-induced superconductivity in a
  stripe-ordered cuprate},}\ }\href {\doibase 10.1126/science.1197294}
  {\bibfield  {journal} {\bibinfo  {journal} {Science}\ }\textbf {\bibinfo
  {volume} {331}},\ \bibinfo {pages} {189--191} (\bibinfo {year}
  {2011})}\BibitemShut {NoStop}%
\bibitem [{\citenamefont {Kaiser}\ \emph {et~al.}(2014)\citenamefont {Kaiser},
  \citenamefont {Hunt}, \citenamefont {Nicoletti}, \citenamefont {Hu},
  \citenamefont {Gierz}, \citenamefont {Liu}, \citenamefont {Le~Tacon},
  \citenamefont {Loew}, \citenamefont {Haug}, \citenamefont {Keimer},\ and\
  \citenamefont {Cavalleri}}]{Kaiser2014}%
  \BibitemOpen
  \bibfield  {author} {\bibinfo {author} {\bibfnamefont {S.}~\bibnamefont
  {Kaiser}}, \bibinfo {author} {\bibfnamefont {C.~R.}\ \bibnamefont {Hunt}},
  \bibinfo {author} {\bibfnamefont {D.}~\bibnamefont {Nicoletti}}, \bibinfo
  {author} {\bibfnamefont {W.}~\bibnamefont {Hu}}, \bibinfo {author}
  {\bibfnamefont {I.}~\bibnamefont {Gierz}}, \bibinfo {author} {\bibfnamefont
  {H.~Y.}\ \bibnamefont {Liu}}, \bibinfo {author} {\bibfnamefont
  {M.}~\bibnamefont {Le~Tacon}}, \bibinfo {author} {\bibfnamefont
  {T.}~\bibnamefont {Loew}}, \bibinfo {author} {\bibfnamefont {D.}~\bibnamefont
  {Haug}}, \bibinfo {author} {\bibfnamefont {B.}~\bibnamefont {Keimer}}, \ and\
  \bibinfo {author} {\bibfnamefont {A.}~\bibnamefont {Cavalleri}},\ }\bibfield
  {title} {\enquote {\bibinfo {title} {{Optically induced coherent transport
  far above ${T}_{c}$ in underdoped
  ${\mathrm{YBa}}_{2}{\mathrm{Cu}}_{3}{\mathrm{O}}_{6+\ensuremath{\delta}}$}},}\
  }\href {\doibase 10.1103/PhysRevB.89.184516} {\bibfield  {journal} {\bibinfo
  {journal} {Phys. Rev. B}\ }\textbf {\bibinfo {volume} {89}},\ \bibinfo
  {pages} {184516} (\bibinfo {year} {2014})}\BibitemShut {NoStop}%
\bibitem [{\citenamefont {Hu}\ \emph {et~al.}(2014)\citenamefont {Hu},
  \citenamefont {Kaiser}, \citenamefont {Nicoletti}, \citenamefont {Hunt},
  \citenamefont {Gierz}, \citenamefont {Hoffmann}, \citenamefont {Le~Tacon},
  \citenamefont {Loew}, \citenamefont {Keimer},\ and\ \citenamefont
  {Cavalleri}}]{Hu2014}%
  \BibitemOpen
  \bibfield  {author} {\bibinfo {author} {\bibfnamefont {W.}~\bibnamefont
  {Hu}}, \bibinfo {author} {\bibfnamefont {S.}~\bibnamefont {Kaiser}}, \bibinfo
  {author} {\bibfnamefont {D.}~\bibnamefont {Nicoletti}}, \bibinfo {author}
  {\bibfnamefont {C.~R.}\ \bibnamefont {Hunt}}, \bibinfo {author}
  {\bibfnamefont {I.}~\bibnamefont {Gierz}}, \bibinfo {author} {\bibfnamefont
  {M.~C.}\ \bibnamefont {Hoffmann}}, \bibinfo {author} {\bibfnamefont
  {M.}~\bibnamefont {Le~Tacon}}, \bibinfo {author} {\bibfnamefont
  {T.}~\bibnamefont {Loew}}, \bibinfo {author} {\bibfnamefont {B.}~\bibnamefont
  {Keimer}}, \ and\ \bibinfo {author} {\bibfnamefont {A.}~\bibnamefont
  {Cavalleri}},\ }\bibfield  {title} {\enquote {\bibinfo {title} {{Optically
  enhanced coherent transport in YBa$_2$Cu$_3$O$_{6.5}$ by ultrafast
  redistribution of interlayer coupling}},}\ }\href {\doibase 10.1038/nmat3963}
  {\bibfield  {journal} {\bibinfo  {journal} {Nat. Mater.}\ }\textbf {\bibinfo
  {volume} {13}},\ \bibinfo {pages} {705} (\bibinfo {year} {2014})}\BibitemShut
  {NoStop}%
\bibitem [{\citenamefont {{Liu}}\ \emph {et~al.}(2019)\citenamefont {{Liu}},
  \citenamefont {{F{\"o}rst}}, \citenamefont {{Fechner}}, \citenamefont
  {{Nicoletti}}, \citenamefont {{Porras}}, \citenamefont {{Keimer}},\ and\
  \citenamefont {{Cavalleri}}}]{Liu2019}%
  \BibitemOpen
  \bibfield  {author} {\bibinfo {author} {\bibfnamefont {B.}~\bibnamefont
  {{Liu}}}, \bibinfo {author} {\bibfnamefont {M.}~\bibnamefont {{F{\"o}rst}}},
  \bibinfo {author} {\bibfnamefont {M.}~\bibnamefont {{Fechner}}}, \bibinfo
  {author} {\bibfnamefont {D.}~\bibnamefont {{Nicoletti}}}, \bibinfo {author}
  {\bibfnamefont {J.}~\bibnamefont {{Porras}}}, \bibinfo {author}
  {\bibfnamefont {B.}~\bibnamefont {{Keimer}}}, \ and\ \bibinfo {author}
  {\bibfnamefont {A.}~\bibnamefont {{Cavalleri}}},\ }\bibfield  {title}
  {\enquote {\bibinfo {title} {{Which phonons induce non-equilibrium
  superconductivity in YBa$_2$Cu$_3$O$_{6.5}$?}}}\ }\href@noop {} {\bibfield
  {journal} {\bibinfo  {journal} {arXiv e-prints}\ ,\ \bibinfo {eid}
  {arXiv:1905.08356}} (\bibinfo {year} {2019})},\ \Eprint
  {http://arxiv.org/abs/1905.08356} {arXiv:1905.08356 [cond-mat.supr-con]}
  \BibitemShut {NoStop}%
\bibitem [{\citenamefont {Le~Tacon}\ \emph {et~al.}(2013)\citenamefont
  {Le~Tacon}, \citenamefont {Bosak}, \citenamefont {Souliou}, \citenamefont
  {Dellea}, \citenamefont {Loew}, \citenamefont {Heid}, \citenamefont {Bohnen},
  \citenamefont {Ghiringhelli}, \citenamefont {Krisch},\ and\ \citenamefont
  {Keimer}}]{LeTacon2013}%
  \BibitemOpen
  \bibfield  {author} {\bibinfo {author} {\bibfnamefont {M.}~\bibnamefont
  {Le~Tacon}}, \bibinfo {author} {\bibfnamefont {A.}~\bibnamefont {Bosak}},
  \bibinfo {author} {\bibfnamefont {S.~M.}\ \bibnamefont {Souliou}}, \bibinfo
  {author} {\bibfnamefont {G.}~\bibnamefont {Dellea}}, \bibinfo {author}
  {\bibfnamefont {T.}~\bibnamefont {Loew}}, \bibinfo {author} {\bibfnamefont
  {R.}~\bibnamefont {Heid}}, \bibinfo {author} {\bibfnamefont {K-P.}\
  \bibnamefont {Bohnen}}, \bibinfo {author} {\bibfnamefont {G.}~\bibnamefont
  {Ghiringhelli}}, \bibinfo {author} {\bibfnamefont {M.}~\bibnamefont
  {Krisch}}, \ and\ \bibinfo {author} {\bibfnamefont {B.}~\bibnamefont
  {Keimer}},\ }\bibfield  {title} {\enquote {\bibinfo {title} {{Inelastic X-ray
  scattering in YBa$_2$Cu$_3$O$_{6.6}$ reveals giant phonon anomalies and
  elastic central peak due to charge-density-wave formation}},}\ }\href
  {\doibase 10.1038/nphys2805} {\bibfield  {journal} {\bibinfo  {journal} {Nat.
  Phys.}\ }\textbf {\bibinfo {volume} {10}},\ \bibinfo {pages} {52} (\bibinfo
  {year} {2013})}\BibitemShut {NoStop}%
\bibitem [{\citenamefont {Miao}\ \emph {et~al.}(2018)\citenamefont {Miao},
  \citenamefont {Ishikawa}, \citenamefont {Heid}, \citenamefont {Le~Tacon},
  \citenamefont {Fabbris}, \citenamefont {Meyers}, \citenamefont {Gu},
  \citenamefont {Baron},\ and\ \citenamefont {Dean}}]{Miao2018}%
  \BibitemOpen
  \bibfield  {author} {\bibinfo {author} {\bibfnamefont {H.}~\bibnamefont
  {Miao}}, \bibinfo {author} {\bibfnamefont {D.}~\bibnamefont {Ishikawa}},
  \bibinfo {author} {\bibfnamefont {R.}~\bibnamefont {Heid}}, \bibinfo {author}
  {\bibfnamefont {M.}~\bibnamefont {Le~Tacon}}, \bibinfo {author}
  {\bibfnamefont {G.}~\bibnamefont {Fabbris}}, \bibinfo {author} {\bibfnamefont
  {D.}~\bibnamefont {Meyers}}, \bibinfo {author} {\bibfnamefont {G.~D.}\
  \bibnamefont {Gu}}, \bibinfo {author} {\bibfnamefont {A.~Q.~R.}\ \bibnamefont
  {Baron}}, \ and\ \bibinfo {author} {\bibfnamefont {M.~P.~M.}\ \bibnamefont
  {Dean}},\ }\bibfield  {title} {\enquote {\bibinfo {title} {{Incommensurate
  Phonon Anomaly and the Nature of Charge Density Waves in Cuprates}},}\ }\href
  {\doibase 10.1103/PhysRevX.8.011008} {\bibfield  {journal} {\bibinfo
  {journal} {Phys. Rev. X}\ }\textbf {\bibinfo {volume} {8}},\ \bibinfo {pages}
  {011008} (\bibinfo {year} {2018})}\BibitemShut {NoStop}%
\bibitem [{\citenamefont {McQueeney}\ \emph {et~al.}(1999)\citenamefont
  {McQueeney}, \citenamefont {Petrov}, \citenamefont {Egami}, \citenamefont
  {Yethiraj}, \citenamefont {Shirane},\ and\ \citenamefont
  {Endoh}}]{McQueeney1999}%
  \BibitemOpen
  \bibfield  {author} {\bibinfo {author} {\bibfnamefont {R.~J.}\ \bibnamefont
  {McQueeney}}, \bibinfo {author} {\bibfnamefont {Y.}~\bibnamefont {Petrov}},
  \bibinfo {author} {\bibfnamefont {T.}~\bibnamefont {Egami}}, \bibinfo
  {author} {\bibfnamefont {M.}~\bibnamefont {Yethiraj}}, \bibinfo {author}
  {\bibfnamefont {G.}~\bibnamefont {Shirane}}, \ and\ \bibinfo {author}
  {\bibfnamefont {Y.}~\bibnamefont {Endoh}},\ }\bibfield  {title} {\enquote
  {\bibinfo {title} {{Anomalous Dispersion of LO Phonons in
  ${\mathrm{La}}_{1.85}{\mathrm{Sr}}_{0.15}{\mathrm{CuO}}_{4}$ at Low
  Temperatures}},}\ }\href {\doibase 10.1103/PhysRevLett.82.628} {\bibfield
  {journal} {\bibinfo  {journal} {Phys. Rev. Lett.}\ }\textbf {\bibinfo
  {volume} {82}},\ \bibinfo {pages} {628--631} (\bibinfo {year}
  {1999})}\BibitemShut {NoStop}%
\bibitem [{\citenamefont {Pintschovius}\ \emph {et~al.}(2004)\citenamefont
  {Pintschovius}, \citenamefont {Reznik}, \citenamefont {Reichardt},
  \citenamefont {Endoh}, \citenamefont {Hiraka}, \citenamefont {Tranquada},
  \citenamefont {Uchiyama}, \citenamefont {Masui},\ and\ \citenamefont
  {Tajima}}]{Pintschovius2004}%
  \BibitemOpen
  \bibfield  {author} {\bibinfo {author} {\bibfnamefont {L.}~\bibnamefont
  {Pintschovius}}, \bibinfo {author} {\bibfnamefont {D.}~\bibnamefont
  {Reznik}}, \bibinfo {author} {\bibfnamefont {W.}~\bibnamefont {Reichardt}},
  \bibinfo {author} {\bibfnamefont {Y.}~\bibnamefont {Endoh}}, \bibinfo
  {author} {\bibfnamefont {H.}~\bibnamefont {Hiraka}}, \bibinfo {author}
  {\bibfnamefont {J.~M.}\ \bibnamefont {Tranquada}}, \bibinfo {author}
  {\bibfnamefont {H.}~\bibnamefont {Uchiyama}}, \bibinfo {author}
  {\bibfnamefont {T.}~\bibnamefont {Masui}}, \ and\ \bibinfo {author}
  {\bibfnamefont {S.}~\bibnamefont {Tajima}},\ }\bibfield  {title} {\enquote
  {\bibinfo {title} {{Oxygen phonon branches in
  ${\mathrm{YBa}}_{2}{\mathrm{Cu}}_{3}{\mathrm{O}}_{7}$}},}\ }\href {\doibase
  10.1103/PhysRevB.69.214506} {\bibfield  {journal} {\bibinfo  {journal} {Phys.
  Rev. B}\ }\textbf {\bibinfo {volume} {69}},\ \bibinfo {pages} {214506}
  (\bibinfo {year} {2004})}\BibitemShut {NoStop}%
\bibitem [{\citenamefont {Uchiyama}\ \emph {et~al.}(2004)\citenamefont
  {Uchiyama}, \citenamefont {Baron}, \citenamefont {Tsutsui}, \citenamefont
  {Tanaka}, \citenamefont {Hu}, \citenamefont {Yamamoto}, \citenamefont
  {Tajima},\ and\ \citenamefont {Endoh}}]{Uchiyama2004}%
  \BibitemOpen
  \bibfield  {author} {\bibinfo {author} {\bibfnamefont {H.}~\bibnamefont
  {Uchiyama}}, \bibinfo {author} {\bibfnamefont {A.~Q.~R.}\ \bibnamefont
  {Baron}}, \bibinfo {author} {\bibfnamefont {S.}~\bibnamefont {Tsutsui}},
  \bibinfo {author} {\bibfnamefont {Y.}~\bibnamefont {Tanaka}}, \bibinfo
  {author} {\bibfnamefont {W.-Z.}\ \bibnamefont {Hu}}, \bibinfo {author}
  {\bibfnamefont {A.}~\bibnamefont {Yamamoto}}, \bibinfo {author}
  {\bibfnamefont {S.}~\bibnamefont {Tajima}}, \ and\ \bibinfo {author}
  {\bibfnamefont {Y.}~\bibnamefont {Endoh}},\ }\bibfield  {title} {\enquote
  {\bibinfo {title} {{Softening of Cu-O Bond Stretching Phonons in Tetragonal
  ${\mathrm{HgBa}}_{2}{\mathrm{CuO}}_{4+\ensuremath{\delta}}$}},}\ }\href
  {\doibase 10.1103/PhysRevLett.92.197005} {\bibfield  {journal} {\bibinfo
  {journal} {Phys. Rev. Lett.}\ }\textbf {\bibinfo {volume} {92}},\ \bibinfo
  {pages} {197005} (\bibinfo {year} {2004})}\BibitemShut {NoStop}%
\bibitem [{\citenamefont {Reznik}\ \emph {et~al.}(2006)\citenamefont {Reznik},
  \citenamefont {Pintschovius}, \citenamefont {Ito}, \citenamefont {Iikubo},
  \citenamefont {Sato}, \citenamefont {Goka}, \citenamefont {Fujita},
  \citenamefont {Yamada}, \citenamefont {Gu},\ and\ \citenamefont
  {Tranquada}}]{Reznik2006}%
  \BibitemOpen
  \bibfield  {author} {\bibinfo {author} {\bibfnamefont {D.}~\bibnamefont
  {Reznik}}, \bibinfo {author} {\bibfnamefont {L.}~\bibnamefont
  {Pintschovius}}, \bibinfo {author} {\bibfnamefont {M.}~\bibnamefont {Ito}},
  \bibinfo {author} {\bibfnamefont {S.}~\bibnamefont {Iikubo}}, \bibinfo
  {author} {\bibfnamefont {M.}~\bibnamefont {Sato}}, \bibinfo {author}
  {\bibfnamefont {H.}~\bibnamefont {Goka}}, \bibinfo {author} {\bibfnamefont
  {M.}~\bibnamefont {Fujita}}, \bibinfo {author} {\bibfnamefont
  {K.}~\bibnamefont {Yamada}}, \bibinfo {author} {\bibfnamefont {G.~D.}\
  \bibnamefont {Gu}}, \ and\ \bibinfo {author} {\bibfnamefont {J.~M.}\
  \bibnamefont {Tranquada}},\ }\bibfield  {title} {\enquote {\bibinfo {title}
  {{Electron-phonon coupling reflecting dynamic charge inhomogeneity in copper
  oxide superconductors}},}\ }\href {\doibase 10.1038/nature04704} {\bibfield
  {journal} {\bibinfo  {journal} {Nature}\ }\textbf {\bibinfo {volume} {440}},\
  \bibinfo {pages} {1770} (\bibinfo {year} {2006})}\BibitemShut {NoStop}%
\bibitem [{\citenamefont {Graf}\ \emph {et~al.}(2008)\citenamefont {Graf},
  \citenamefont {d'Astuto}, \citenamefont {Jozwiak}, \citenamefont {Garcia},
  \citenamefont {Saini}, \citenamefont {Krisch}, \citenamefont {Ikeuchi},
  \citenamefont {Baron}, \citenamefont {Eisaki},\ and\ \citenamefont
  {Lanzara}}]{Graf2008}%
  \BibitemOpen
  \bibfield  {author} {\bibinfo {author} {\bibfnamefont {J.}~\bibnamefont
  {Graf}}, \bibinfo {author} {\bibfnamefont {M.}~\bibnamefont {d'Astuto}},
  \bibinfo {author} {\bibfnamefont {C.}~\bibnamefont {Jozwiak}}, \bibinfo
  {author} {\bibfnamefont {D.~R.}\ \bibnamefont {Garcia}}, \bibinfo {author}
  {\bibfnamefont {N.~L.}\ \bibnamefont {Saini}}, \bibinfo {author}
  {\bibfnamefont {M.}~\bibnamefont {Krisch}}, \bibinfo {author} {\bibfnamefont
  {K.}~\bibnamefont {Ikeuchi}}, \bibinfo {author} {\bibfnamefont {A.~Q.~R.}\
  \bibnamefont {Baron}}, \bibinfo {author} {\bibfnamefont {H.}~\bibnamefont
  {Eisaki}}, \ and\ \bibinfo {author} {\bibfnamefont {A.}~\bibnamefont
  {Lanzara}},\ }\bibfield  {title} {\enquote {\bibinfo {title} {{Bond
  Stretching Phonon Softening and Kinks in the Angle-Resolved Photoemission
  Spectra of Optimally Doped
  ${\mathrm{Bi}}_{2}{\mathrm{Sr}}_{1.6}{\mathrm{La}}_{0.4}{\mathrm{Cu}}_{2}{\mathrm{O}}_{6+\ensuremath{\delta}}$
  Superconductors}},}\ }\href {\doibase 10.1103/PhysRevLett.100.227002}
  {\bibfield  {journal} {\bibinfo  {journal} {Phys. Rev. Lett.}\ }\textbf
  {\bibinfo {volume} {100}},\ \bibinfo {pages} {227002} (\bibinfo {year}
  {2008})}\BibitemShut {NoStop}%
\bibitem [{\citenamefont {Chaix}\ \emph {et~al.}(2017)\citenamefont {Chaix},
  \citenamefont {Ghiringhelli}, \citenamefont {Peng}, \citenamefont
  {Hashimoto}, \citenamefont {Moritz}, \citenamefont {Kummer}, \citenamefont
  {Brookes}, \citenamefont {He}, \citenamefont {Chen}, \citenamefont {Ishida},
  \citenamefont {Yoshida}, \citenamefont {Eisaki}, \citenamefont {Salluzzo},
  \citenamefont {Braicovich}, \citenamefont {Shen}, \citenamefont {Devereaux},\
  and\ \citenamefont {Lee}}]{Chaix2017}%
  \BibitemOpen
  \bibfield  {author} {\bibinfo {author} {\bibfnamefont {L.}~\bibnamefont
  {Chaix}}, \bibinfo {author} {\bibfnamefont {G.}~\bibnamefont {Ghiringhelli}},
  \bibinfo {author} {\bibfnamefont {Y.~Y.}\ \bibnamefont {Peng}}, \bibinfo
  {author} {\bibfnamefont {M.}~\bibnamefont {Hashimoto}}, \bibinfo {author}
  {\bibfnamefont {B.}~\bibnamefont {Moritz}}, \bibinfo {author} {\bibfnamefont
  {K.}~\bibnamefont {Kummer}}, \bibinfo {author} {\bibfnamefont {N.~B.}\
  \bibnamefont {Brookes}}, \bibinfo {author} {\bibfnamefont {Y.}~\bibnamefont
  {He}}, \bibinfo {author} {\bibfnamefont {S.}~\bibnamefont {Chen}}, \bibinfo
  {author} {\bibfnamefont {S.}~\bibnamefont {Ishida}}, \bibinfo {author}
  {\bibfnamefont {Y.}~\bibnamefont {Yoshida}}, \bibinfo {author} {\bibfnamefont
  {H.}~\bibnamefont {Eisaki}}, \bibinfo {author} {\bibfnamefont
  {M.}~\bibnamefont {Salluzzo}}, \bibinfo {author} {\bibfnamefont
  {L.}~\bibnamefont {Braicovich}}, \bibinfo {author} {\bibfnamefont {Z.-X.}\
  \bibnamefont {Shen}}, \bibinfo {author} {\bibfnamefont {T.~P.}\ \bibnamefont
  {Devereaux}}, \ and\ \bibinfo {author} {\bibfnamefont {W.-S.}\ \bibnamefont
  {Lee}},\ }\bibfield  {title} {\enquote {\bibinfo {title} {{Dispersive charge
  density wave excitations in Bi$_2$Sr$_2$CaCu$_2$O$_{8+\delta}$}},}\ }\href
  {\doibase 10.1038/nphys4157} {\bibfield  {journal} {\bibinfo  {journal} {Nat.
  Phys.}\ }\textbf {\bibinfo {volume} {13}},\ \bibinfo {pages} {952} (\bibinfo
  {year} {2017})}\BibitemShut {NoStop}%
\bibitem [{\citenamefont {Kastner}\ \emph {et~al.}(1998)\citenamefont
  {Kastner}, \citenamefont {Birgeneau}, \citenamefont {Shirane},\ and\
  \citenamefont {Endoh}}]{Kastner1998}%
  \BibitemOpen
  \bibfield  {author} {\bibinfo {author} {\bibfnamefont {M.~A.}\ \bibnamefont
  {Kastner}}, \bibinfo {author} {\bibfnamefont {R.~J.}\ \bibnamefont
  {Birgeneau}}, \bibinfo {author} {\bibfnamefont {G.}~\bibnamefont {Shirane}},
  \ and\ \bibinfo {author} {\bibfnamefont {Y.}~\bibnamefont {Endoh}},\
  }\bibfield  {title} {\enquote {\bibinfo {title} {Magnetic, transport, and
  optical properties of monolayer copper oxides},}\ }\href {\doibase
  10.1103/RevModPhys.70.897} {\bibfield  {journal} {\bibinfo  {journal} {Rev.
  Mod. Phys.}\ }\textbf {\bibinfo {volume} {70}},\ \bibinfo {pages} {897--928}
  (\bibinfo {year} {1998})}\BibitemShut {NoStop}%
\bibitem [{\citenamefont {Alexandrov}(2000)}]{Alexandrov2000}%
  \BibitemOpen
  \bibfield  {author} {\bibinfo {author} {\bibfnamefont {A.~S.}\ \bibnamefont
  {Alexandrov}},\ }\bibfield  {title} {\enquote {\bibinfo {title} {Polaron
  dynamics and bipolaron condensation in cuprates},}\ }\href {\doibase
  10.1103/PhysRevB.61.12315} {\bibfield  {journal} {\bibinfo  {journal} {Phys.
  Rev. B}\ }\textbf {\bibinfo {volume} {61}},\ \bibinfo {pages} {12315--12327}
  (\bibinfo {year} {2000})}\BibitemShut {NoStop}%
\bibitem [{\citenamefont {R\"osch}\ \emph {et~al.}(2005)\citenamefont
  {R\"osch}, \citenamefont {Gunnarsson}, \citenamefont {Zhou}, \citenamefont
  {Yoshida}, \citenamefont {Sasagawa}, \citenamefont {Fujimori}, \citenamefont
  {Hussain}, \citenamefont {Shen},\ and\ \citenamefont {Uchida}}]{Rosch2004}%
  \BibitemOpen
  \bibfield  {author} {\bibinfo {author} {\bibfnamefont {O.}~\bibnamefont
  {R\"osch}}, \bibinfo {author} {\bibfnamefont {O.}~\bibnamefont {Gunnarsson}},
  \bibinfo {author} {\bibfnamefont {X.~J.}\ \bibnamefont {Zhou}}, \bibinfo
  {author} {\bibfnamefont {T.}~\bibnamefont {Yoshida}}, \bibinfo {author}
  {\bibfnamefont {T.}~\bibnamefont {Sasagawa}}, \bibinfo {author}
  {\bibfnamefont {A.}~\bibnamefont {Fujimori}}, \bibinfo {author}
  {\bibfnamefont {Z.}~\bibnamefont {Hussain}}, \bibinfo {author} {\bibfnamefont
  {Z.-X.}\ \bibnamefont {Shen}}, \ and\ \bibinfo {author} {\bibfnamefont
  {S.}~\bibnamefont {Uchida}},\ }\bibfield  {title} {\enquote {\bibinfo {title}
  {{Polaronic Behavior of Undoped High-${T}_\mathrm{c}$ Cuprate Superconductors
  from Angle-Resolved Photoemission Spectra}},}\ }\href {\doibase
  10.1103/PhysRevLett.95.227002} {\bibfield  {journal} {\bibinfo  {journal}
  {Phys. Rev. Lett.}\ }\textbf {\bibinfo {volume} {95}},\ \bibinfo {pages}
  {227002} (\bibinfo {year} {2005})}\BibitemShut {NoStop}%
\bibitem [{\citenamefont {Shen}\ \emph {et~al.}(2007)\citenamefont {Shen},
  \citenamefont {Ronning}, \citenamefont {Meevasana}, \citenamefont {Lu},
  \citenamefont {Ingle}, \citenamefont {Baumberger}, \citenamefont {Lee},
  \citenamefont {Miller}, \citenamefont {Kohsaka}, \citenamefont {Azuma},
  \citenamefont {Takano}, \citenamefont {Takagi},\ and\ \citenamefont
  {Shen}}]{Shen2007}%
  \BibitemOpen
  \bibfield  {author} {\bibinfo {author} {\bibfnamefont {K.~M.}\ \bibnamefont
  {Shen}}, \bibinfo {author} {\bibfnamefont {F.}~\bibnamefont {Ronning}},
  \bibinfo {author} {\bibfnamefont {W.}~\bibnamefont {Meevasana}}, \bibinfo
  {author} {\bibfnamefont {D.~H.}\ \bibnamefont {Lu}}, \bibinfo {author}
  {\bibfnamefont {N.~J.~C.}\ \bibnamefont {Ingle}}, \bibinfo {author}
  {\bibfnamefont {F.}~\bibnamefont {Baumberger}}, \bibinfo {author}
  {\bibfnamefont {W.~S.}\ \bibnamefont {Lee}}, \bibinfo {author} {\bibfnamefont
  {L.~L.}\ \bibnamefont {Miller}}, \bibinfo {author} {\bibfnamefont
  {Y.}~\bibnamefont {Kohsaka}}, \bibinfo {author} {\bibfnamefont
  {M.}~\bibnamefont {Azuma}}, \bibinfo {author} {\bibfnamefont
  {M.}~\bibnamefont {Takano}}, \bibinfo {author} {\bibfnamefont
  {H.}~\bibnamefont {Takagi}}, \ and\ \bibinfo {author} {\bibfnamefont {Z.-X.}\
  \bibnamefont {Shen}},\ }\bibfield  {title} {\enquote {\bibinfo {title}
  {{Angle-resolved photoemission studies of lattice polaron formation in the
  cuprate ${\mathrm{Ca}}_{2}\mathrm{Cu}{\mathrm{O}}_{2}{\mathrm{Cl}}_{2}$}},}\
  }\href {\doibase 10.1103/PhysRevB.75.075115} {\bibfield  {journal} {\bibinfo
  {journal} {Phys. Rev. B}\ }\textbf {\bibinfo {volume} {75}},\ \bibinfo
  {pages} {075115} (\bibinfo {year} {2007})}\BibitemShut {NoStop}%
\bibitem [{\citenamefont {Bohnen}\ \emph {et~al.}(2003)\citenamefont {Bohnen},
  \citenamefont {Heid},\ and\ \citenamefont {Krauss}}]{Bohnen2003}%
  \BibitemOpen
  \bibfield  {author} {\bibinfo {author} {\bibfnamefont {K.-P.}\ \bibnamefont
  {Bohnen}}, \bibinfo {author} {\bibfnamefont {R.}~\bibnamefont {Heid}}, \ and\
  \bibinfo {author} {\bibfnamefont {M.}~\bibnamefont {Krauss}},\ }\bibfield
  {title} {\enquote {\bibinfo {title} {{Phonon dispersion and electron-phonon
  interaction for YBa$_2$Cu$_3$O$_7$ from first-principles calculations}},}\
  }\href {http://stacks.iop.org/0295-5075/64/i=1/a=104} {\bibfield  {journal}
  {\bibinfo  {journal} {EPL}\ }\textbf {\bibinfo {volume} {64}},\ \bibinfo
  {pages} {104} (\bibinfo {year} {2003})}\BibitemShut {NoStop}%
\bibitem [{\citenamefont {Giustino}\ \emph {et~al.}(2008)\citenamefont
  {Giustino}, \citenamefont {Cohen},\ and\ \citenamefont
  {Louie}}]{Giustino2008}%
  \BibitemOpen
  \bibfield  {author} {\bibinfo {author} {\bibfnamefont {Feliciano}\
  \bibnamefont {Giustino}}, \bibinfo {author} {\bibfnamefont {Marvin~L.}\
  \bibnamefont {Cohen}}, \ and\ \bibinfo {author} {\bibfnamefont {Steven~G.}\
  \bibnamefont {Louie}},\ }\bibfield  {title} {\enquote {\bibinfo {title}
  {Small phonon contribution to the photoemission kink in the copper oxide
  superconductors},}\ }\href {\doibase 10.1038/nature06874} {\bibfield
  {journal} {\bibinfo  {journal} {Nature}\ }\textbf {\bibinfo {volume} {452}},\
  \bibinfo {pages} {975} (\bibinfo {year} {2008})}\BibitemShut {NoStop}%
\bibitem [{\citenamefont {Reznik}\ \emph {et~al.}(2008)\citenamefont {Reznik},
  \citenamefont {Sangiovanni}, \citenamefont {Gunnarsson},\ and\ \citenamefont
  {Devereaux}}]{Reznik2008}%
  \BibitemOpen
  \bibfield  {author} {\bibinfo {author} {\bibfnamefont {D.}~\bibnamefont
  {Reznik}}, \bibinfo {author} {\bibfnamefont {G.}~\bibnamefont {Sangiovanni}},
  \bibinfo {author} {\bibfnamefont {O.}~\bibnamefont {Gunnarsson}}, \ and\
  \bibinfo {author} {\bibfnamefont {T.~P.}\ \bibnamefont {Devereaux}},\
  }\bibfield  {title} {\enquote {\bibinfo {title} {Photoemission kinks and
  phonons in cuprates},}\ }\href {\doibase 10.1038/nature07364} {\bibfield
  {journal} {\bibinfo  {journal} {Nature}\ }\textbf {\bibinfo {volume} {455}},\
  \bibinfo {pages} {E6} (\bibinfo {year} {2008})}\BibitemShut {NoStop}%
\bibitem [{\citenamefont {Heid}\ \emph {et~al.}(2008)\citenamefont {Heid},
  \citenamefont {Bohnen}, \citenamefont {Zeyher},\ and\ \citenamefont
  {Manske}}]{Heid2008}%
  \BibitemOpen
  \bibfield  {author} {\bibinfo {author} {\bibfnamefont {Rolf}\ \bibnamefont
  {Heid}}, \bibinfo {author} {\bibfnamefont {Klaus-Peter}\ \bibnamefont
  {Bohnen}}, \bibinfo {author} {\bibfnamefont {Roland}\ \bibnamefont {Zeyher}},
  \ and\ \bibinfo {author} {\bibfnamefont {Dirk}\ \bibnamefont {Manske}},\
  }\bibfield  {title} {\enquote {\bibinfo {title} {{Momentum Dependence of the
  Electron-Phonon Coupling and Self-Energy Effects in Superconducting
  ${\mathrm{YBa}}_{2}{\mathrm{Cu}}_{3}{\mathrm{O}}_{7}$ within the Local
  Density Approximation}},}\ }\href {\doibase 10.1103/PhysRevLett.100.137001}
  {\bibfield  {journal} {\bibinfo  {journal} {Phys. Rev. Lett.}\ }\textbf
  {\bibinfo {volume} {100}},\ \bibinfo {pages} {137001} (\bibinfo {year}
  {2008})}\BibitemShut {NoStop}%
\bibitem [{\citenamefont {Veenstra}\ \emph {et~al.}(2010)\citenamefont
  {Veenstra}, \citenamefont {Goodvin}, \citenamefont {Berciu},\ and\
  \citenamefont {Damascelli}}]{Veenstra2010}%
  \BibitemOpen
  \bibfield  {author} {\bibinfo {author} {\bibfnamefont {C.~N.}\ \bibnamefont
  {Veenstra}}, \bibinfo {author} {\bibfnamefont {G.~L.}\ \bibnamefont
  {Goodvin}}, \bibinfo {author} {\bibfnamefont {M.}~\bibnamefont {Berciu}}, \
  and\ \bibinfo {author} {\bibfnamefont {A.}~\bibnamefont {Damascelli}},\
  }\bibfield  {title} {\enquote {\bibinfo {title} {Elusive electron-phonon
  coupling in quantitative analyses of the spectral function},}\ }\href
  {\doibase 10.1103/PhysRevB.82.012504} {\bibfield  {journal} {\bibinfo
  {journal} {Phys. Rev. B}\ }\textbf {\bibinfo {volume} {82}},\ \bibinfo
  {pages} {012504} (\bibinfo {year} {2010})}\BibitemShut {NoStop}%
\bibitem [{\citenamefont {Lanzara}\ \emph {et~al.}(2001)\citenamefont
  {Lanzara}, \citenamefont {Bogdanov}, \citenamefont {Zhou}, \citenamefont
  {Kellar}, \citenamefont {Feng}, \citenamefont {Lu}, \citenamefont {Yoshida},
  \citenamefont {Eisaki}, \citenamefont {Fujimori}, \citenamefont {Kishio},
  \citenamefont {Shimoyama}, \citenamefont {Noda}, \citenamefont {Uchida},
  \citenamefont {Hussain},\ and\ \citenamefont {Shen}}]{Lanzara2001}%
  \BibitemOpen
  \bibfield  {author} {\bibinfo {author} {\bibfnamefont {A.}~\bibnamefont
  {Lanzara}}, \bibinfo {author} {\bibfnamefont {P.~V.}\ \bibnamefont
  {Bogdanov}}, \bibinfo {author} {\bibfnamefont {X.~J.}\ \bibnamefont {Zhou}},
  \bibinfo {author} {\bibfnamefont {S.~A.}\ \bibnamefont {Kellar}}, \bibinfo
  {author} {\bibfnamefont {D.~L.}\ \bibnamefont {Feng}}, \bibinfo {author}
  {\bibfnamefont {E.~D.}\ \bibnamefont {Lu}}, \bibinfo {author} {\bibfnamefont
  {T.}~\bibnamefont {Yoshida}}, \bibinfo {author} {\bibfnamefont
  {H.}~\bibnamefont {Eisaki}}, \bibinfo {author} {\bibfnamefont
  {A.}~\bibnamefont {Fujimori}}, \bibinfo {author} {\bibfnamefont
  {K.}~\bibnamefont {Kishio}}, \bibinfo {author} {\bibfnamefont {J.-I.}\
  \bibnamefont {Shimoyama}}, \bibinfo {author} {\bibfnamefont {T.}~\bibnamefont
  {Noda}}, \bibinfo {author} {\bibfnamefont {S.}~\bibnamefont {Uchida}},
  \bibinfo {author} {\bibfnamefont {Z.}~\bibnamefont {Hussain}}, \ and\
  \bibinfo {author} {\bibfnamefont {Z.-X.}\ \bibnamefont {Shen}},\ }\bibfield
  {title} {\enquote {\bibinfo {title} {Evidence for ubiquitous strong
  electron-phonon coupling in high-temperature superconductors},}\ }\href
  {\doibase 10.1038/35087518} {\bibfield  {journal} {\bibinfo  {journal}
  {Nature}\ }\textbf {\bibinfo {volume} {412}},\ \bibinfo {pages} {510}
  (\bibinfo {year} {2001})}\BibitemShut {NoStop}%
\bibitem [{\citenamefont {Zhou}\ \emph {et~al.}(2005)\citenamefont {Zhou},
  \citenamefont {Shi}, \citenamefont {Yoshida}, \citenamefont {Cuk},
  \citenamefont {Yang}, \citenamefont {Brouet}, \citenamefont {Nakamura},
  \citenamefont {Mannella}, \citenamefont {Komiya}, \citenamefont {Ando},
  \citenamefont {Zhou}, \citenamefont {Ti}, \citenamefont {Xiong},
  \citenamefont {Zhao}, \citenamefont {Sasagawa}, \citenamefont {Kakeshita},
  \citenamefont {Eisaki}, \citenamefont {Uchida}, \citenamefont {Fujimori},
  \citenamefont {Zhang}, \citenamefont {Plummer}, \citenamefont {Laughlin},
  \citenamefont {Hussain},\ and\ \citenamefont {Shen}}]{Zhou2005}%
  \BibitemOpen
  \bibfield  {author} {\bibinfo {author} {\bibfnamefont {X.~J.}\ \bibnamefont
  {Zhou}}, \bibinfo {author} {\bibfnamefont {Junren}\ \bibnamefont {Shi}},
  \bibinfo {author} {\bibfnamefont {T.}~\bibnamefont {Yoshida}}, \bibinfo
  {author} {\bibfnamefont {T.}~\bibnamefont {Cuk}}, \bibinfo {author}
  {\bibfnamefont {W.~L.}\ \bibnamefont {Yang}}, \bibinfo {author}
  {\bibfnamefont {V.}~\bibnamefont {Brouet}}, \bibinfo {author} {\bibfnamefont
  {J.}~\bibnamefont {Nakamura}}, \bibinfo {author} {\bibfnamefont
  {N.}~\bibnamefont {Mannella}}, \bibinfo {author} {\bibfnamefont {Seiki}\
  \bibnamefont {Komiya}}, \bibinfo {author} {\bibfnamefont {Yoichi}\
  \bibnamefont {Ando}}, \bibinfo {author} {\bibfnamefont {F.}~\bibnamefont
  {Zhou}}, \bibinfo {author} {\bibfnamefont {W.~X.}\ \bibnamefont {Ti}},
  \bibinfo {author} {\bibfnamefont {J.~W.}\ \bibnamefont {Xiong}}, \bibinfo
  {author} {\bibfnamefont {Z.~X.}\ \bibnamefont {Zhao}}, \bibinfo {author}
  {\bibfnamefont {T.}~\bibnamefont {Sasagawa}}, \bibinfo {author}
  {\bibfnamefont {T.}~\bibnamefont {Kakeshita}}, \bibinfo {author}
  {\bibfnamefont {H.}~\bibnamefont {Eisaki}}, \bibinfo {author} {\bibfnamefont
  {S.}~\bibnamefont {Uchida}}, \bibinfo {author} {\bibfnamefont
  {A.}~\bibnamefont {Fujimori}}, \bibinfo {author} {\bibfnamefont {Zhenyu}\
  \bibnamefont {Zhang}}, \bibinfo {author} {\bibfnamefont {E.~W.}\ \bibnamefont
  {Plummer}}, \bibinfo {author} {\bibfnamefont {R.~B.}\ \bibnamefont
  {Laughlin}}, \bibinfo {author} {\bibfnamefont {Z.}~\bibnamefont {Hussain}}, \
  and\ \bibinfo {author} {\bibfnamefont {Z.-X.}\ \bibnamefont {Shen}},\
  }\bibfield  {title} {\enquote {\bibinfo {title} {{Multiple Bosonic Mode
  Coupling in the Electron Self-Energy of
  $({\mathrm{La}}_{2\ensuremath{-}x}{\mathrm{Sr}}_{x}){\mathrm{CuO}}_{4}$}},}\
  }\href {\doibase 10.1103/PhysRevLett.95.117001} {\bibfield  {journal}
  {\bibinfo  {journal} {Phys. Rev. Lett.}\ }\textbf {\bibinfo {volume} {95}},\
  \bibinfo {pages} {117001} (\bibinfo {year} {2005})}\BibitemShut {NoStop}%
\bibitem [{\citenamefont {Cuk}\ \emph {et~al.}(2005)\citenamefont {Cuk},
  \citenamefont {Lu}, \citenamefont {Zhou}, \citenamefont {Shen}, \citenamefont
  {Devereaux},\ and\ \citenamefont {Nagaosa}}]{Cuk2005}%
  \BibitemOpen
  \bibfield  {author} {\bibinfo {author} {\bibfnamefont {T.}~\bibnamefont
  {Cuk}}, \bibinfo {author} {\bibfnamefont {D.~H.}\ \bibnamefont {Lu}},
  \bibinfo {author} {\bibfnamefont {X.~J.}\ \bibnamefont {Zhou}}, \bibinfo
  {author} {\bibfnamefont {Z.-X.}\ \bibnamefont {Shen}}, \bibinfo {author}
  {\bibfnamefont {T.~P.}\ \bibnamefont {Devereaux}}, \ and\ \bibinfo {author}
  {\bibfnamefont {N.}~\bibnamefont {Nagaosa}},\ }\bibfield  {title} {\enquote
  {\bibinfo {title} {{A review of electron-phonon coupling seen in the
  high-$T_\mathrm{c}$ superconductors by angle-resolved photoemission studies
  (ARPES)}},}\ }\href {\doibase 10.1002/pssb.200404959} {\bibfield  {journal}
  {\bibinfo  {journal} {Phys. Status Solidi B}\ }\textbf {\bibinfo {volume}
  {242}},\ \bibinfo {pages} {11--29} (\bibinfo {year} {2005})}\BibitemShut
  {NoStop}%
\bibitem [{\citenamefont {Johnston}\ \emph {et~al.}(2012)\citenamefont
  {Johnston}, \citenamefont {Vishik}, \citenamefont {Lee}, \citenamefont
  {Schmitt}, \citenamefont {Uchida}, \citenamefont {Fujita}, \citenamefont
  {Ishida}, \citenamefont {Nagaosa}, \citenamefont {Shen},\ and\ \citenamefont
  {Devereaux}}]{Johnston2012}%
  \BibitemOpen
  \bibfield  {author} {\bibinfo {author} {\bibfnamefont {S.}~\bibnamefont
  {Johnston}}, \bibinfo {author} {\bibfnamefont {I.~M.}\ \bibnamefont
  {Vishik}}, \bibinfo {author} {\bibfnamefont {W.~S.}\ \bibnamefont {Lee}},
  \bibinfo {author} {\bibfnamefont {F.}~\bibnamefont {Schmitt}}, \bibinfo
  {author} {\bibfnamefont {S.}~\bibnamefont {Uchida}}, \bibinfo {author}
  {\bibfnamefont {K.}~\bibnamefont {Fujita}}, \bibinfo {author} {\bibfnamefont
  {S.}~\bibnamefont {Ishida}}, \bibinfo {author} {\bibfnamefont
  {N.}~\bibnamefont {Nagaosa}}, \bibinfo {author} {\bibfnamefont {Z.~X.}\
  \bibnamefont {Shen}}, \ and\ \bibinfo {author} {\bibfnamefont {T.~P.}\
  \bibnamefont {Devereaux}},\ }\bibfield  {title} {\enquote {\bibinfo {title}
  {{Evidence for the Importance of Extended Coulomb Interactions and Forward
  Scattering in Cuprate Superconductors}},}\ }\href {\doibase
  10.1103/PhysRevLett.108.166404} {\bibfield  {journal} {\bibinfo  {journal}
  {Phys. Rev. Lett.}\ }\textbf {\bibinfo {volume} {108}},\ \bibinfo {pages}
  {166404} (\bibinfo {year} {2012})}\BibitemShut {NoStop}%
\bibitem [{\citenamefont {Shen}\ \emph {et~al.}(2002)\citenamefont {Shen},
  \citenamefont {Lanzara}, \citenamefont {Ishihara},\ and\ \citenamefont
  {Nagaosa}}]{Shen2002}%
  \BibitemOpen
  \bibfield  {author} {\bibinfo {author} {\bibfnamefont {Z.-X.}\ \bibnamefont
  {Shen}}, \bibinfo {author} {\bibfnamefont {A.}~\bibnamefont {Lanzara}},
  \bibinfo {author} {\bibfnamefont {S.}~\bibnamefont {Ishihara}}, \ and\
  \bibinfo {author} {\bibfnamefont {N.}~\bibnamefont {Nagaosa}},\ }\bibfield
  {title} {\enquote {\bibinfo {title} {Role of the electron-phonon interaction
  in the strongly correlated cuprate superconductors},}\ }\href {\doibase
  10.1080/13642810208220725} {\bibfield  {journal} {\bibinfo  {journal}
  {Philos. Mag. B}\ }\textbf {\bibinfo {volume} {82}},\ \bibinfo {pages}
  {1349--1368} (\bibinfo {year} {2002})}\BibitemShut {NoStop}%
\bibitem [{\citenamefont {Bulut}\ and\ \citenamefont
  {Scalapino}(1996)}]{Bulut1996}%
  \BibitemOpen
  \bibfield  {author} {\bibinfo {author} {\bibfnamefont {N.}~\bibnamefont
  {Bulut}}\ and\ \bibinfo {author} {\bibfnamefont {D.~J.}\ \bibnamefont
  {Scalapino}},\ }\bibfield  {title} {\enquote {\bibinfo {title}
  {{${d}_{{x}^{2}\ensuremath{-}{y}^{2}}$ symmetry and the pairing
  mechanism}},}\ }\href {\doibase 10.1103/PhysRevB.54.14971} {\bibfield
  {journal} {\bibinfo  {journal} {Phys. Rev. B}\ }\textbf {\bibinfo {volume}
  {54}},\ \bibinfo {pages} {14971--14973} (\bibinfo {year} {1996})}\BibitemShut
  {NoStop}%
\bibitem [{\citenamefont {Sandvik}\ \emph {et~al.}(2004)\citenamefont
  {Sandvik}, \citenamefont {Scalapino},\ and\ \citenamefont
  {Bickers}}]{Sandvik2004}%
  \BibitemOpen
  \bibfield  {author} {\bibinfo {author} {\bibfnamefont {A.~W.}\ \bibnamefont
  {Sandvik}}, \bibinfo {author} {\bibfnamefont {D.~J.}\ \bibnamefont
  {Scalapino}}, \ and\ \bibinfo {author} {\bibfnamefont {N.~E.}\ \bibnamefont
  {Bickers}},\ }\bibfield  {title} {\enquote {\bibinfo {title} {{Effect of an
  electron-phonon interaction on the one-electron spectral weight of a $d$-wave
  superconductor}},}\ }\href {\doibase 10.1103/PhysRevB.69.094523} {\bibfield
  {journal} {\bibinfo  {journal} {Phys. Rev. B}\ }\textbf {\bibinfo {volume}
  {69}},\ \bibinfo {pages} {094523} (\bibinfo {year} {2004})}\BibitemShut
  {NoStop}%
\bibitem [{\citenamefont {Pintschovius}(2005)}]{Pintschovius2005}%
  \BibitemOpen
  \bibfield  {author} {\bibinfo {author} {\bibfnamefont {L.}~\bibnamefont
  {Pintschovius}},\ }\bibfield  {title} {\enquote {\bibinfo {title}
  {Electron-phonon coupling effects explored by inelastic neutron
  scattering},}\ }\href {\doibase 10.1002/pssb.200404951} {\bibfield  {journal}
  {\bibinfo  {journal} {Phys. Status Solidi B}\ }\textbf {\bibinfo {volume}
  {242}},\ \bibinfo {pages} {30--50} (\bibinfo {year} {2005})}\BibitemShut
  {NoStop}%
\bibitem [{\citenamefont {Reznik}(2010)}]{Reznik2010}%
  \BibitemOpen
  \bibfield  {author} {\bibinfo {author} {\bibfnamefont {D.}~\bibnamefont
  {Reznik}},\ }\bibfield  {title} {\enquote {\bibinfo {title} {{Giant
  Electron-Phonon Anomaly in Doped La$_2$CuO$_4$ and Other Cuprates}},}\ }\href
  {\doibase 10.1155/2010/523549} {\bibfield  {journal} {\bibinfo  {journal}
  {Adv. Cond. Matter Phys.}\ }\textbf {\bibinfo {volume} {2010}},\ \bibinfo
  {pages} {523549} (\bibinfo {year} {2010})}\BibitemShut {NoStop}%
\bibitem [{\citenamefont {Damascelli}\ \emph {et~al.}(2003)\citenamefont
  {Damascelli}, \citenamefont {Hussain},\ and\ \citenamefont
  {Shen}}]{Damascelli2003}%
  \BibitemOpen
  \bibfield  {author} {\bibinfo {author} {\bibfnamefont {Andrea}\ \bibnamefont
  {Damascelli}}, \bibinfo {author} {\bibfnamefont {Zahid}\ \bibnamefont
  {Hussain}}, \ and\ \bibinfo {author} {\bibfnamefont {Zhi-Xun}\ \bibnamefont
  {Shen}},\ }\bibfield  {title} {\enquote {\bibinfo {title} {Angle-resolved
  photoemission studies of the cuprate superconductors},}\ }\href {\doibase
  10.1103/RevModPhys.75.473} {\bibfield  {journal} {\bibinfo  {journal} {Rev.
  Mod. Phys.}\ }\textbf {\bibinfo {volume} {75}},\ \bibinfo {pages} {473--541}
  (\bibinfo {year} {2003})}\BibitemShut {NoStop}%
\bibitem [{\citenamefont {Lee}\ \emph {et~al.}(2006)\citenamefont {Lee},
  \citenamefont {Fujita}, \citenamefont {McElroy}, \citenamefont {Slezak},
  \citenamefont {Wang}, \citenamefont {Aiura}, \citenamefont {Bando},
  \citenamefont {Ishikado}, \citenamefont {Masui}, \citenamefont {Zhu},
  \citenamefont {Balatsky}, \citenamefont {Eisaki}, \citenamefont {Uchida},\
  and\ \citenamefont {Davis}}]{Lee2006}%
  \BibitemOpen
  \bibfield  {author} {\bibinfo {author} {\bibfnamefont {Jinho}\ \bibnamefont
  {Lee}}, \bibinfo {author} {\bibfnamefont {K.}~\bibnamefont {Fujita}},
  \bibinfo {author} {\bibfnamefont {K.}~\bibnamefont {McElroy}}, \bibinfo
  {author} {\bibfnamefont {J.~A.}\ \bibnamefont {Slezak}}, \bibinfo {author}
  {\bibfnamefont {M.}~\bibnamefont {Wang}}, \bibinfo {author} {\bibfnamefont
  {Y.}~\bibnamefont {Aiura}}, \bibinfo {author} {\bibfnamefont
  {H.}~\bibnamefont {Bando}}, \bibinfo {author} {\bibfnamefont
  {M.}~\bibnamefont {Ishikado}}, \bibinfo {author} {\bibfnamefont
  {T.}~\bibnamefont {Masui}}, \bibinfo {author} {\bibfnamefont {J.-X.}\
  \bibnamefont {Zhu}}, \bibinfo {author} {\bibfnamefont {A.~V.}\ \bibnamefont
  {Balatsky}}, \bibinfo {author} {\bibfnamefont {H.}~\bibnamefont {Eisaki}},
  \bibinfo {author} {\bibfnamefont {S.}~\bibnamefont {Uchida}}, \ and\ \bibinfo
  {author} {\bibfnamefont {J.~C.}\ \bibnamefont {Davis}},\ }\bibfield  {title}
  {\enquote {\bibinfo {title} {{Interplay of electron-lattice interactions and
  superconductivity in Bi$_2$Sr$_2$CaCu$_2$O$_{8+\delta}$}},}\ }\href {\doibase
  10.1038/nature04973} {\bibfield  {journal} {\bibinfo  {journal} {Nature}\
  }\textbf {\bibinfo {volume} {442}},\ \bibinfo {pages} {546} (\bibinfo {year}
  {2006})}\BibitemShut {NoStop}%
\bibitem [{\citenamefont {Brookes}\ \emph {et~al.}(2018)\citenamefont
  {Brookes}, \citenamefont {Yakhou-Harris}, \citenamefont {Kummer},
  \citenamefont {Fondacaro}, \citenamefont {Cezar}, \citenamefont {Betto},
  \citenamefont {Velez-Fort}, \citenamefont {Amorese}, \citenamefont
  {Ghiringhelli}, \citenamefont {Braicovich}, \citenamefont {Barrett},
  \citenamefont {Berruyer}, \citenamefont {Cianciosi}, \citenamefont {Eybert},
  \citenamefont {Marion}, \citenamefont {van~der Linden},\ and\ \citenamefont
  {Zhang}}]{Brookes2018}%
  \BibitemOpen
  \bibfield  {author} {\bibinfo {author} {\bibfnamefont {N.~B.}\ \bibnamefont
  {Brookes}}, \bibinfo {author} {\bibfnamefont {F.}~\bibnamefont
  {Yakhou-Harris}}, \bibinfo {author} {\bibfnamefont {K.}~\bibnamefont
  {Kummer}}, \bibinfo {author} {\bibfnamefont {A.}~\bibnamefont {Fondacaro}},
  \bibinfo {author} {\bibfnamefont {J.C.}\ \bibnamefont {Cezar}}, \bibinfo
  {author} {\bibfnamefont {D.}~\bibnamefont {Betto}}, \bibinfo {author}
  {\bibfnamefont {E.}~\bibnamefont {Velez-Fort}}, \bibinfo {author}
  {\bibfnamefont {A.}~\bibnamefont {Amorese}}, \bibinfo {author} {\bibfnamefont
  {G.}~\bibnamefont {Ghiringhelli}}, \bibinfo {author} {\bibfnamefont
  {L.}~\bibnamefont {Braicovich}}, \bibinfo {author} {\bibfnamefont
  {R.}~\bibnamefont {Barrett}}, \bibinfo {author} {\bibfnamefont
  {G.}~\bibnamefont {Berruyer}}, \bibinfo {author} {\bibfnamefont
  {F.}~\bibnamefont {Cianciosi}}, \bibinfo {author} {\bibfnamefont
  {L.}~\bibnamefont {Eybert}}, \bibinfo {author} {\bibfnamefont
  {P.}~\bibnamefont {Marion}}, \bibinfo {author} {\bibfnamefont
  {P.}~\bibnamefont {van~der Linden}}, \ and\ \bibinfo {author} {\bibfnamefont
  {L.}~\bibnamefont {Zhang}},\ }\bibfield  {title} {\enquote {\bibinfo {title}
  {{The beamline ID32 at the ESRF for soft X-ray high energy resolution
  resonant inelastic X-ray scattering and polarisation dependent X-ray
  absorption spectroscopy}},}\ }\href {\doibase 10.1016/j.nima.2018.07.001}
  {\bibfield  {journal} {\bibinfo  {journal} {Nucl. Instr. Meth. Phys. Res. A}\
  }\textbf {\bibinfo {volume} {903}},\ \bibinfo {pages} {175 -- 192} (\bibinfo
  {year} {2018})}\BibitemShut {NoStop}%
\bibitem [{\citenamefont {Ament}\ \emph {et~al.}(2011)\citenamefont {Ament},
  \citenamefont {van Veenendaal},\ and\ \citenamefont {van~den
  Brink}}]{Ament2011}%
  \BibitemOpen
  \bibfield  {author} {\bibinfo {author} {\bibfnamefont {L.~J.~P.}\
  \bibnamefont {Ament}}, \bibinfo {author} {\bibfnamefont {M.}~\bibnamefont
  {van Veenendaal}}, \ and\ \bibinfo {author} {\bibfnamefont {J.}~\bibnamefont
  {van~den Brink}},\ }\bibfield  {title} {\enquote {\bibinfo {title}
  {{Determining the electron-phonon coupling strength from Resonant Inelastic
  X-ray Scattering at transition metal L-edges}},}\ }\href
  {http://stacks.iop.org/0295-5075/95/i=2/a=27008} {\bibfield  {journal}
  {\bibinfo  {journal} {EPL}\ }\textbf {\bibinfo {volume} {95}},\ \bibinfo
  {pages} {27008} (\bibinfo {year} {2011})}\BibitemShut {NoStop}%
\bibitem [{\citenamefont {Ament}(2010)}]{Ament2010}%
  \BibitemOpen
  \bibfield  {author} {\bibinfo {author} {\bibfnamefont {L.}~\bibnamefont
  {Ament}},\ }\emph {\bibinfo {title} {{Resonant Inelastic X-ray Scattering
  Studies of Elementary Excitations}}},\ \href@noop {} {\bibinfo {type} {{PhD}
  thesis}},\ \bibinfo  {school} {Universiteit Leiden} (\bibinfo {year}
  {2010})\BibitemShut {NoStop}%
\bibitem [{\citenamefont {Devereaux}\ \emph {et~al.}(2016)\citenamefont
  {Devereaux}, \citenamefont {Shvaika}, \citenamefont {Wu}, \citenamefont
  {Wohlfeld}, \citenamefont {Jia}, \citenamefont {Wang}, \citenamefont
  {Moritz}, \citenamefont {Chaix}, \citenamefont {Lee}, \citenamefont {Shen},
  \citenamefont {Ghiringhelli},\ and\ \citenamefont
  {Braicovich}}]{Devereaux2016}%
  \BibitemOpen
  \bibfield  {author} {\bibinfo {author} {\bibfnamefont {T.~P.}\ \bibnamefont
  {Devereaux}}, \bibinfo {author} {\bibfnamefont {A.~M.}\ \bibnamefont
  {Shvaika}}, \bibinfo {author} {\bibfnamefont {K.}~\bibnamefont {Wu}},
  \bibinfo {author} {\bibfnamefont {K.}~\bibnamefont {Wohlfeld}}, \bibinfo
  {author} {\bibfnamefont {C.~J.}\ \bibnamefont {Jia}}, \bibinfo {author}
  {\bibfnamefont {Y.}~\bibnamefont {Wang}}, \bibinfo {author} {\bibfnamefont
  {B.}~\bibnamefont {Moritz}}, \bibinfo {author} {\bibfnamefont
  {L.}~\bibnamefont {Chaix}}, \bibinfo {author} {\bibfnamefont {W.-S.}\
  \bibnamefont {Lee}}, \bibinfo {author} {\bibfnamefont {Z.-X.}\ \bibnamefont
  {Shen}}, \bibinfo {author} {\bibfnamefont {G.}~\bibnamefont {Ghiringhelli}},
  \ and\ \bibinfo {author} {\bibfnamefont {L.}~\bibnamefont {Braicovich}},\
  }\bibfield  {title} {\enquote {\bibinfo {title} {{Directly Characterizing the
  Relative Strength and Momentum Dependence of Electron-Phonon Coupling Using
  Resonant Inelastic X-Ray Scattering}},}\ }\href {\doibase
  10.1103/PhysRevX.6.041019} {\bibfield  {journal} {\bibinfo  {journal} {Phys.
  Rev. X}\ }\textbf {\bibinfo {volume} {6}},\ \bibinfo {pages} {041019}
  (\bibinfo {year} {2016})}\BibitemShut {NoStop}%
\bibitem [{\citenamefont {Geondzhian}\ and\ \citenamefont
  {Gilmore}(2018)}]{Geondzhian2018}%
  \BibitemOpen
  \bibfield  {author} {\bibinfo {author} {\bibfnamefont {Andrey}\ \bibnamefont
  {Geondzhian}}\ and\ \bibinfo {author} {\bibfnamefont {Keith}\ \bibnamefont
  {Gilmore}},\ }\bibfield  {title} {\enquote {\bibinfo {title} {Demonstration
  of resonant inelastic x-ray scattering as a probe of exciton-phonon
  coupling},}\ }\href {\doibase 10.1103/PhysRevB.98.214305} {\bibfield
  {journal} {\bibinfo  {journal} {Phys. Rev. B}\ }\textbf {\bibinfo {volume}
  {98}},\ \bibinfo {pages} {214305} (\bibinfo {year} {2018})}\BibitemShut
  {NoStop}%
\bibitem [{\citenamefont {Lee}\ \emph {et~al.}(2013)\citenamefont {Lee},
  \citenamefont {Johnston}, \citenamefont {Moritz}, \citenamefont {Lee},
  \citenamefont {Yi}, \citenamefont {Zhou}, \citenamefont {Schmitt},
  \citenamefont {Patthey}, \citenamefont {Strocov}, \citenamefont {Kudo},
  \citenamefont {Koike}, \citenamefont {van~den Brink}, \citenamefont
  {Devereaux},\ and\ \citenamefont {Shen}}]{Lee2013}%
  \BibitemOpen
  \bibfield  {author} {\bibinfo {author} {\bibfnamefont {W.~S.}\ \bibnamefont
  {Lee}}, \bibinfo {author} {\bibfnamefont {S.}~\bibnamefont {Johnston}},
  \bibinfo {author} {\bibfnamefont {B.}~\bibnamefont {Moritz}}, \bibinfo
  {author} {\bibfnamefont {J.}~\bibnamefont {Lee}}, \bibinfo {author}
  {\bibfnamefont {M.}~\bibnamefont {Yi}}, \bibinfo {author} {\bibfnamefont
  {K.~J.}\ \bibnamefont {Zhou}}, \bibinfo {author} {\bibfnamefont
  {T.}~\bibnamefont {Schmitt}}, \bibinfo {author} {\bibfnamefont
  {L.}~\bibnamefont {Patthey}}, \bibinfo {author} {\bibfnamefont
  {V.}~\bibnamefont {Strocov}}, \bibinfo {author} {\bibfnamefont
  {K.}~\bibnamefont {Kudo}}, \bibinfo {author} {\bibfnamefont {Y.}~\bibnamefont
  {Koike}}, \bibinfo {author} {\bibfnamefont {J.}~\bibnamefont {van~den
  Brink}}, \bibinfo {author} {\bibfnamefont {T.~P.}\ \bibnamefont {Devereaux}},
  \ and\ \bibinfo {author} {\bibfnamefont {Z.~X.}\ \bibnamefont {Shen}},\
  }\bibfield  {title} {\enquote {\bibinfo {title} {{Role of Lattice Coupling in
  Establishing Electronic and Magnetic Properties in Quasi-One-Dimensional
  Cuprates}},}\ }\href {\doibase 10.1103/PhysRevLett.110.265502} {\bibfield
  {journal} {\bibinfo  {journal} {Phys. Rev. Lett.}\ }\textbf {\bibinfo
  {volume} {110}},\ \bibinfo {pages} {265502} (\bibinfo {year}
  {2013})}\BibitemShut {NoStop}%
\bibitem [{\citenamefont {Johnston}\ \emph {et~al.}(2016)\citenamefont
  {Johnston}, \citenamefont {Monney}, \citenamefont {Bisogni}, \citenamefont
  {Zhou}, \citenamefont {Kraus}, \citenamefont {Behr}, \citenamefont {Strocov},
  \citenamefont {M\'{a}lek}, \citenamefont {Drechsler}, \citenamefont {Geck},
  \citenamefont {Schmitt},\ and\ \citenamefont {van~den Brink}}]{Johnston2016}%
  \BibitemOpen
  \bibfield  {author} {\bibinfo {author} {\bibfnamefont {Steve}\ \bibnamefont
  {Johnston}}, \bibinfo {author} {\bibfnamefont {Claude}\ \bibnamefont
  {Monney}}, \bibinfo {author} {\bibfnamefont {Valentina}\ \bibnamefont
  {Bisogni}}, \bibinfo {author} {\bibfnamefont {Ke-Jin}\ \bibnamefont {Zhou}},
  \bibinfo {author} {\bibfnamefont {Roberto}\ \bibnamefont {Kraus}}, \bibinfo
  {author} {\bibfnamefont {G\"{u}nter}\ \bibnamefont {Behr}}, \bibinfo {author}
  {\bibfnamefont {Vladimir~N.}\ \bibnamefont {Strocov}}, \bibinfo {author}
  {\bibfnamefont {Ji\v{r}i}\ \bibnamefont {M\'{a}lek}}, \bibinfo {author}
  {\bibfnamefont {Stefan-Ludwig}\ \bibnamefont {Drechsler}}, \bibinfo {author}
  {\bibfnamefont {Jochen}\ \bibnamefont {Geck}}, \bibinfo {author}
  {\bibfnamefont {Thorsten}\ \bibnamefont {Schmitt}}, \ and\ \bibinfo {author}
  {\bibfnamefont {Jeroen}\ \bibnamefont {van~den Brink}},\ }\bibfield  {title}
  {\enquote {\bibinfo {title} {{Electron-lattice interactions strongly
  renormalize the charge-transfer energy in the spin-chain cuprate
  Li$_2$CuO$_2$}},}\ }\href {\doibase 10.1038/ncomms10563} {\bibfield
  {journal} {\bibinfo  {journal} {Nat. Commun.}\ }\textbf {\bibinfo {volume}
  {7}},\ \bibinfo {pages} {10563} (\bibinfo {year} {2016})}\BibitemShut
  {NoStop}%
\bibitem [{\citenamefont {Meyers}\ \emph {et~al.}(2018)\citenamefont {Meyers},
  \citenamefont {Nakatsukasa}, \citenamefont {Mu}, \citenamefont {Hao},
  \citenamefont {Yang}, \citenamefont {Cao}, \citenamefont {Fabbris},
  \citenamefont {Miao}, \citenamefont {Pelliciari}, \citenamefont {McNally},
  \citenamefont {Dantz}, \citenamefont {Paris}, \citenamefont {Karapetrova},
  \citenamefont {Choi}, \citenamefont {Haskel}, \citenamefont {Shafer},
  \citenamefont {Arenholz}, \citenamefont {Schmitt}, \citenamefont {Berlijn},
  \citenamefont {Johnston}, \citenamefont {Liu},\ and\ \citenamefont
  {Dean}}]{Meyers2018}%
  \BibitemOpen
  \bibfield  {author} {\bibinfo {author} {\bibfnamefont {D.}~\bibnamefont
  {Meyers}}, \bibinfo {author} {\bibfnamefont {Ken}\ \bibnamefont
  {Nakatsukasa}}, \bibinfo {author} {\bibfnamefont {Sai}\ \bibnamefont {Mu}},
  \bibinfo {author} {\bibfnamefont {Lin}\ \bibnamefont {Hao}}, \bibinfo
  {author} {\bibfnamefont {Junyi}\ \bibnamefont {Yang}}, \bibinfo {author}
  {\bibfnamefont {Yue}\ \bibnamefont {Cao}}, \bibinfo {author} {\bibfnamefont
  {G.}~\bibnamefont {Fabbris}}, \bibinfo {author} {\bibfnamefont
  {Hu}~\bibnamefont {Miao}}, \bibinfo {author} {\bibfnamefont {J.}~\bibnamefont
  {Pelliciari}}, \bibinfo {author} {\bibfnamefont {D.}~\bibnamefont {McNally}},
  \bibinfo {author} {\bibfnamefont {M.}~\bibnamefont {Dantz}}, \bibinfo
  {author} {\bibfnamefont {E.}~\bibnamefont {Paris}}, \bibinfo {author}
  {\bibfnamefont {E.}~\bibnamefont {Karapetrova}}, \bibinfo {author}
  {\bibfnamefont {Yongseong}\ \bibnamefont {Choi}}, \bibinfo {author}
  {\bibfnamefont {D.}~\bibnamefont {Haskel}}, \bibinfo {author} {\bibfnamefont
  {P.}~\bibnamefont {Shafer}}, \bibinfo {author} {\bibfnamefont
  {E.}~\bibnamefont {Arenholz}}, \bibinfo {author} {\bibfnamefont {Thorsten}\
  \bibnamefont {Schmitt}}, \bibinfo {author} {\bibfnamefont {Tom}\ \bibnamefont
  {Berlijn}}, \bibinfo {author} {\bibfnamefont {S.}~\bibnamefont {Johnston}},
  \bibinfo {author} {\bibfnamefont {Jian}\ \bibnamefont {Liu}}, \ and\ \bibinfo
  {author} {\bibfnamefont {M.~P.~M.}\ \bibnamefont {Dean}},\ }\bibfield
  {title} {\enquote {\bibinfo {title} {{Decoupling Carrier Concentration and
  Electron-Phonon Coupling in Oxide Heterostructures Observed with Resonant
  Inelastic X-Ray Scattering}},}\ }\href {\doibase
  10.1103/PhysRevLett.121.236802} {\bibfield  {journal} {\bibinfo  {journal}
  {Phys. Rev. Lett.}\ }\textbf {\bibinfo {volume} {121}},\ \bibinfo {pages}
  {236802} (\bibinfo {year} {2018})}\BibitemShut {NoStop}%
\bibitem [{\citenamefont {Moser}\ \emph {et~al.}(2015)\citenamefont {Moser},
  \citenamefont {Fatale}, \citenamefont {Kr\"uger}, \citenamefont {Berger},
  \citenamefont {Bugnon}, \citenamefont {Magrez}, \citenamefont {Niwa},
  \citenamefont {Miyawaki}, \citenamefont {Harada},\ and\ \citenamefont
  {Grioni}}]{Moser2015}%
  \BibitemOpen
  \bibfield  {author} {\bibinfo {author} {\bibfnamefont {S.}~\bibnamefont
  {Moser}}, \bibinfo {author} {\bibfnamefont {S.}~\bibnamefont {Fatale}},
  \bibinfo {author} {\bibfnamefont {P.}~\bibnamefont {Kr\"uger}}, \bibinfo
  {author} {\bibfnamefont {H.}~\bibnamefont {Berger}}, \bibinfo {author}
  {\bibfnamefont {P.}~\bibnamefont {Bugnon}}, \bibinfo {author} {\bibfnamefont
  {A.}~\bibnamefont {Magrez}}, \bibinfo {author} {\bibfnamefont
  {H.}~\bibnamefont {Niwa}}, \bibinfo {author} {\bibfnamefont {J.}~\bibnamefont
  {Miyawaki}}, \bibinfo {author} {\bibfnamefont {Y.}~\bibnamefont {Harada}}, \
  and\ \bibinfo {author} {\bibfnamefont {M.}~\bibnamefont {Grioni}},\
  }\bibfield  {title} {\enquote {\bibinfo {title} {{Electron-Phonon Coupling in
  the Bulk of Anatase ${\mathrm{TiO}}_{2}$ Measured by Resonant Inelastic X-Ray
  Spectroscopy}},}\ }\href {\doibase 10.1103/PhysRevLett.115.096404} {\bibfield
   {journal} {\bibinfo  {journal} {Phys. Rev. Lett.}\ }\textbf {\bibinfo
  {volume} {115}},\ \bibinfo {pages} {096404} (\bibinfo {year}
  {2015})}\BibitemShut {NoStop}%
\bibitem [{\citenamefont {Fatale}\ \emph {et~al.}(2016)\citenamefont {Fatale},
  \citenamefont {Moser}, \citenamefont {Miyawaki}, \citenamefont {Harada},\
  and\ \citenamefont {Grioni}}]{Fatale2016}%
  \BibitemOpen
  \bibfield  {author} {\bibinfo {author} {\bibfnamefont {S.}~\bibnamefont
  {Fatale}}, \bibinfo {author} {\bibfnamefont {S.}~\bibnamefont {Moser}},
  \bibinfo {author} {\bibfnamefont {J.}~\bibnamefont {Miyawaki}}, \bibinfo
  {author} {\bibfnamefont {Y.}~\bibnamefont {Harada}}, \ and\ \bibinfo {author}
  {\bibfnamefont {M.}~\bibnamefont {Grioni}},\ }\bibfield  {title} {\enquote
  {\bibinfo {title} {{Hybridization and electron-phonon coupling in
  ferroelectric ${\mathrm{BaTiO}}_{3}$ probed by resonant inelastic x-ray
  scattering}},}\ }\href {\doibase 10.1103/PhysRevB.94.195131} {\bibfield
  {journal} {\bibinfo  {journal} {Phys. Rev. B}\ }\textbf {\bibinfo {volume}
  {94}},\ \bibinfo {pages} {195131} (\bibinfo {year} {2016})}\BibitemShut
  {NoStop}%
\bibitem [{\citenamefont {Ghiringhelli}\ \emph {et~al.}(2012)\citenamefont
  {Ghiringhelli}, \citenamefont {Le~Tacon}, \citenamefont {Minola},
  \citenamefont {Blanco-Canosa}, \citenamefont {Mazzoli}, \citenamefont
  {Brookes}, \citenamefont {De~Luca}, \citenamefont {Frano}, \citenamefont
  {Hawthorn}, \citenamefont {He}, \citenamefont {Loew}, \citenamefont {Sala},
  \citenamefont {Peets}, \citenamefont {Salluzzo}, \citenamefont {Schierle},
  \citenamefont {Sutarto}, \citenamefont {Sawatzky}, \citenamefont {Weschke},
  \citenamefont {Keimer},\ and\ \citenamefont {Braicovich}}]{Ghiringhelli2012}%
  \BibitemOpen
  \bibfield  {author} {\bibinfo {author} {\bibfnamefont {G.}~\bibnamefont
  {Ghiringhelli}}, \bibinfo {author} {\bibfnamefont {M.}~\bibnamefont
  {Le~Tacon}}, \bibinfo {author} {\bibfnamefont {M.}~\bibnamefont {Minola}},
  \bibinfo {author} {\bibfnamefont {S.}~\bibnamefont {Blanco-Canosa}}, \bibinfo
  {author} {\bibfnamefont {C.}~\bibnamefont {Mazzoli}}, \bibinfo {author}
  {\bibfnamefont {N.~B.}\ \bibnamefont {Brookes}}, \bibinfo {author}
  {\bibfnamefont {G.~M.}\ \bibnamefont {De~Luca}}, \bibinfo {author}
  {\bibfnamefont {A.}~\bibnamefont {Frano}}, \bibinfo {author} {\bibfnamefont
  {D.~G.}\ \bibnamefont {Hawthorn}}, \bibinfo {author} {\bibfnamefont
  {F.}~\bibnamefont {He}}, \bibinfo {author} {\bibfnamefont {T.}~\bibnamefont
  {Loew}}, \bibinfo {author} {\bibfnamefont {M.~Moretti}\ \bibnamefont {Sala}},
  \bibinfo {author} {\bibfnamefont {D.~C.}\ \bibnamefont {Peets}}, \bibinfo
  {author} {\bibfnamefont {M.}~\bibnamefont {Salluzzo}}, \bibinfo {author}
  {\bibfnamefont {E.}~\bibnamefont {Schierle}}, \bibinfo {author}
  {\bibfnamefont {R.}~\bibnamefont {Sutarto}}, \bibinfo {author} {\bibfnamefont
  {G.~A.}\ \bibnamefont {Sawatzky}}, \bibinfo {author} {\bibfnamefont
  {E.}~\bibnamefont {Weschke}}, \bibinfo {author} {\bibfnamefont
  {B.}~\bibnamefont {Keimer}}, \ and\ \bibinfo {author} {\bibfnamefont
  {L.}~\bibnamefont {Braicovich}},\ }\bibfield  {title} {\enquote {\bibinfo
  {title} {{Long-Range Incommensurate Charge Fluctuations in
  (Y,Nd)Ba$_2$Cu$_3$O$_{6+x}$}},}\ }\href {\doibase 10.1126/science.1223532}
  {\bibfield  {journal} {\bibinfo  {journal} {Science}\ }\textbf {\bibinfo
  {volume} {337}},\ \bibinfo {pages} {821--825} (\bibinfo {year}
  {2012})}\BibitemShut {NoStop}%
\bibitem [{\citenamefont {da~Silva~Neto}\ \emph {et~al.}(2014)\citenamefont
  {da~Silva~Neto}, \citenamefont {Aynajian}, \citenamefont {Frano},
  \citenamefont {Comin}, \citenamefont {Schierle}, \citenamefont {Weschke},
  \citenamefont {Gyenis}, \citenamefont {Wen}, \citenamefont {Schneeloch},
  \citenamefont {Xu}, \citenamefont {Ono}, \citenamefont {Gu}, \citenamefont
  {Le~Tacon},\ and\ \citenamefont {Yazdani}}]{daSilvaNeto2014}%
  \BibitemOpen
  \bibfield  {author} {\bibinfo {author} {\bibfnamefont {Eduardo~H.}\
  \bibnamefont {da~Silva~Neto}}, \bibinfo {author} {\bibfnamefont {Pegor}\
  \bibnamefont {Aynajian}}, \bibinfo {author} {\bibfnamefont {Alex}\
  \bibnamefont {Frano}}, \bibinfo {author} {\bibfnamefont {Riccardo}\
  \bibnamefont {Comin}}, \bibinfo {author} {\bibfnamefont {Enrico}\
  \bibnamefont {Schierle}}, \bibinfo {author} {\bibfnamefont {Eugen}\
  \bibnamefont {Weschke}}, \bibinfo {author} {\bibfnamefont {Andr{\'a}s}\
  \bibnamefont {Gyenis}}, \bibinfo {author} {\bibfnamefont {Jinsheng}\
  \bibnamefont {Wen}}, \bibinfo {author} {\bibfnamefont {John}\ \bibnamefont
  {Schneeloch}}, \bibinfo {author} {\bibfnamefont {Zhijun}\ \bibnamefont {Xu}},
  \bibinfo {author} {\bibfnamefont {Shimpei}\ \bibnamefont {Ono}}, \bibinfo
  {author} {\bibfnamefont {Genda}\ \bibnamefont {Gu}}, \bibinfo {author}
  {\bibfnamefont {Mathieu}\ \bibnamefont {Le~Tacon}}, \ and\ \bibinfo {author}
  {\bibfnamefont {Ali}\ \bibnamefont {Yazdani}},\ }\bibfield  {title} {\enquote
  {\bibinfo {title} {Ubiquitous interplay between charge ordering and
  high-temperature superconductivity in cuprates},}\ }\href {\doibase
  10.1126/science.1243479} {\bibfield  {journal} {\bibinfo  {journal}
  {Science}\ }\textbf {\bibinfo {volume} {343}},\ \bibinfo {pages} {393--396}
  (\bibinfo {year} {2014})}\BibitemShut {NoStop}%
\bibitem [{\citenamefont {da~Silva~Neto}\ \emph {et~al.}(2015)\citenamefont
  {da~Silva~Neto}, \citenamefont {Comin}, \citenamefont {He}, \citenamefont
  {Sutarto}, \citenamefont {Jiang}, \citenamefont {Greene}, \citenamefont
  {Sawatzky},\ and\ \citenamefont {Damascelli}}]{daSilvaNeto2015}%
  \BibitemOpen
  \bibfield  {author} {\bibinfo {author} {\bibfnamefont {Eduardo~H.}\
  \bibnamefont {da~Silva~Neto}}, \bibinfo {author} {\bibfnamefont {Riccardo}\
  \bibnamefont {Comin}}, \bibinfo {author} {\bibfnamefont {Feizhou}\
  \bibnamefont {He}}, \bibinfo {author} {\bibfnamefont {Ronny}\ \bibnamefont
  {Sutarto}}, \bibinfo {author} {\bibfnamefont {Yeping}\ \bibnamefont {Jiang}},
  \bibinfo {author} {\bibfnamefont {Richard~L.}\ \bibnamefont {Greene}},
  \bibinfo {author} {\bibfnamefont {George~A.}\ \bibnamefont {Sawatzky}}, \
  and\ \bibinfo {author} {\bibfnamefont {Andrea}\ \bibnamefont {Damascelli}},\
  }\bibfield  {title} {\enquote {\bibinfo {title} {{Charge ordering in the
  electron-doped superconductor Nd$_{2-x}$Ce$_x$CuO$_4$}},}\ }\href {\doibase
  10.1126/science.1256441} {\bibfield  {journal} {\bibinfo  {journal}
  {Science}\ }\textbf {\bibinfo {volume} {347}},\ \bibinfo {pages} {282--285}
  (\bibinfo {year} {2015})}\BibitemShut {NoStop}%
\bibitem [{\citenamefont {Comin}\ and\ \citenamefont
  {Damascelli}(2016)}]{Comin2016}%
  \BibitemOpen
  \bibfield  {author} {\bibinfo {author} {\bibfnamefont {Riccardo}\
  \bibnamefont {Comin}}\ and\ \bibinfo {author} {\bibfnamefont {Andrea}\
  \bibnamefont {Damascelli}},\ }\bibfield  {title} {\enquote {\bibinfo {title}
  {Resonant x-ray scattering studies of charge order in cuprates},}\ }\href
  {\doibase 10.1146/annurev-conmatphys-031115-011401} {\bibfield  {journal}
  {\bibinfo  {journal} {Annu. Rev. Condens. Matter Phys.}\ }\textbf {\bibinfo
  {volume} {7}},\ \bibinfo {pages} {369--405} (\bibinfo {year}
  {2016})}\BibitemShut {NoStop}%
\bibitem [{\citenamefont {Arpaia}\ \emph {et~al.}(2019)\citenamefont {Arpaia},
  \citenamefont {Caprara}, \citenamefont {Fumagalli}, \citenamefont
  {De~Vecchi}, \citenamefont {Peng}, \citenamefont {Andersson}, \citenamefont
  {Betto}, \citenamefont {De~Luca}, \citenamefont {Brookes}, \citenamefont
  {Lombardi}, \citenamefont {Salluzzo}, \citenamefont {Braicovich},
  \citenamefont {Di~Castro}, \citenamefont {Grilli},\ and\ \citenamefont
  {Ghiringhelli}}]{Arpaia2019}%
  \BibitemOpen
  \bibfield  {author} {\bibinfo {author} {\bibfnamefont {R.}~\bibnamefont
  {Arpaia}}, \bibinfo {author} {\bibfnamefont {S.}~\bibnamefont {Caprara}},
  \bibinfo {author} {\bibfnamefont {R.}~\bibnamefont {Fumagalli}}, \bibinfo
  {author} {\bibfnamefont {G.}~\bibnamefont {De~Vecchi}}, \bibinfo {author}
  {\bibfnamefont {Y.~Y.}\ \bibnamefont {Peng}}, \bibinfo {author}
  {\bibfnamefont {E.}~\bibnamefont {Andersson}}, \bibinfo {author}
  {\bibfnamefont {D.}~\bibnamefont {Betto}}, \bibinfo {author} {\bibfnamefont
  {G.~M.}\ \bibnamefont {De~Luca}}, \bibinfo {author} {\bibfnamefont {N.~B.}\
  \bibnamefont {Brookes}}, \bibinfo {author} {\bibfnamefont {F.}~\bibnamefont
  {Lombardi}}, \bibinfo {author} {\bibfnamefont {M.}~\bibnamefont {Salluzzo}},
  \bibinfo {author} {\bibfnamefont {L.}~\bibnamefont {Braicovich}}, \bibinfo
  {author} {\bibfnamefont {C.}~\bibnamefont {Di~Castro}}, \bibinfo {author}
  {\bibfnamefont {M.}~\bibnamefont {Grilli}}, \ and\ \bibinfo {author}
  {\bibfnamefont {G.}~\bibnamefont {Ghiringhelli}},\ }\bibfield  {title}
  {\enquote {\bibinfo {title} {{Dynamical charge density fluctuations pervading
  the phase diagram of a Cu-based high-Tc superconductor}},}\ }\href {\doibase
  10.1126/science.aav1315} {\bibfield  {journal} {\bibinfo  {journal}
  {Science}\ }\textbf {\bibinfo {volume} {365}},\ \bibinfo {pages} {906}
  (\bibinfo {year} {2019})}\BibitemShut {NoStop}%
\bibitem [{\citenamefont {Rossi}\ \emph {et~al.}(2019)\citenamefont {Rossi},
  \citenamefont {Arpaia}, \citenamefont {Fumagalli}, \citenamefont
  {Moretti~Sala}, \citenamefont {Betto}, \citenamefont {Kummer}, \citenamefont
  {De~Luca}, \citenamefont {van~den Brink}, \citenamefont {Salluzzo},
  \citenamefont {Brookes}, \citenamefont {Braicovich},\ and\ \citenamefont
  {Ghiringhelli}}]{Rossi2019}%
  \BibitemOpen
  \bibfield  {author} {\bibinfo {author} {\bibfnamefont {Matteo}\ \bibnamefont
  {Rossi}}, \bibinfo {author} {\bibfnamefont {Riccardo}\ \bibnamefont
  {Arpaia}}, \bibinfo {author} {\bibfnamefont {Roberto}\ \bibnamefont
  {Fumagalli}}, \bibinfo {author} {\bibfnamefont {Marco}\ \bibnamefont
  {Moretti~Sala}}, \bibinfo {author} {\bibfnamefont {Davide}\ \bibnamefont
  {Betto}}, \bibinfo {author} {\bibfnamefont {Kurt}\ \bibnamefont {Kummer}},
  \bibinfo {author} {\bibfnamefont {Gabriella~M.}\ \bibnamefont {De~Luca}},
  \bibinfo {author} {\bibfnamefont {Jeroen}\ \bibnamefont {van~den Brink}},
  \bibinfo {author} {\bibfnamefont {Marco}\ \bibnamefont {Salluzzo}}, \bibinfo
  {author} {\bibfnamefont {Nicholas~B.}\ \bibnamefont {Brookes}}, \bibinfo
  {author} {\bibfnamefont {Lucio}\ \bibnamefont {Braicovich}}, \ and\ \bibinfo
  {author} {\bibfnamefont {Giacomo}\ \bibnamefont {Ghiringhelli}},\ }\bibfield
  {title} {\enquote {\bibinfo {title} {{Experimental Determination of
  Momentum-Resolved Electron-Phonon Coupling}},}\ }\href {\doibase
  10.1103/PhysRevLett.123.027001} {\bibfield  {journal} {\bibinfo  {journal}
  {Phys. Rev. Lett.}\ }\textbf {\bibinfo {volume} {123}},\ \bibinfo {pages}
  {027001} (\bibinfo {year} {2019})}\BibitemShut {NoStop}%
\bibitem [{\citenamefont {Arpaia}\ \emph {et~al.}(2018)\citenamefont {Arpaia},
  \citenamefont {Andersson}, \citenamefont {Trabaldo}, \citenamefont {Bauch},\
  and\ \citenamefont {Lombardi}}]{Arpaia2018}%
  \BibitemOpen
  \bibfield  {author} {\bibinfo {author} {\bibfnamefont {Riccardo}\
  \bibnamefont {Arpaia}}, \bibinfo {author} {\bibfnamefont {Eric}\ \bibnamefont
  {Andersson}}, \bibinfo {author} {\bibfnamefont {Edoardo}\ \bibnamefont
  {Trabaldo}}, \bibinfo {author} {\bibfnamefont {Thilo}\ \bibnamefont {Bauch}},
  \ and\ \bibinfo {author} {\bibfnamefont {Floriana}\ \bibnamefont
  {Lombardi}},\ }\bibfield  {title} {\enquote {\bibinfo {title} {{Probing the
  phase diagram of cuprates with
  ${\mathrm{YBa}}_{2}{\mathrm{Cu}}_{3}{\mathrm{O}}_{7\ensuremath{-}\ensuremath{\delta}}$
  thin films and nanowires}},}\ }\href {\doibase
  10.1103/PhysRevMaterials.2.024804} {\bibfield  {journal} {\bibinfo  {journal}
  {Phys. Rev. Materials}\ }\textbf {\bibinfo {volume} {2}},\ \bibinfo {pages}
  {024804} (\bibinfo {year} {2018})}\BibitemShut {NoStop}%
\bibitem [{\citenamefont {Salluzzo}\ \emph {et~al.}(2002)\citenamefont
  {Salluzzo}, \citenamefont {Aruta}, \citenamefont {Ausanio}, \citenamefont
  {D'Agostino},\ and\ \citenamefont {Scotti~di Uccio}}]{Salluzzo2002}%
  \BibitemOpen
  \bibfield  {author} {\bibinfo {author} {\bibfnamefont {M.}~\bibnamefont
  {Salluzzo}}, \bibinfo {author} {\bibfnamefont {C.}~\bibnamefont {Aruta}},
  \bibinfo {author} {\bibfnamefont {G.}~\bibnamefont {Ausanio}}, \bibinfo
  {author} {\bibfnamefont {A.}~\bibnamefont {D'Agostino}}, \ and\ \bibinfo
  {author} {\bibfnamefont {U.}~\bibnamefont {Scotti~di Uccio}},\ }\bibfield
  {title} {\enquote {\bibinfo {title} {{Effect of strain on the structure and
  critical temperature in superconducting Nd-doped
  ${\mathrm{YBa}}_{2}{\mathrm{Cu}}_{3}{\mathrm{O}}_{7\ensuremath{-}\ensuremath{\delta}}$}},}\
  }\href {\doibase 10.1103/PhysRevB.66.184518} {\bibfield  {journal} {\bibinfo
  {journal} {Phys. Rev. B}\ }\textbf {\bibinfo {volume} {66}},\ \bibinfo
  {pages} {184518} (\bibinfo {year} {2002})}\BibitemShut {NoStop}%
\bibitem [{\citenamefont {Salluzzo}\ \emph {et~al.}(2005)\citenamefont
  {Salluzzo}, \citenamefont {de~Luca}, \citenamefont {Marr\`e}, \citenamefont
  {Putti}, \citenamefont {Tropeano}, \citenamefont {Scotti~di Uccio},\ and\
  \citenamefont {Vaglio}}]{Salluzzo2005}%
  \BibitemOpen
  \bibfield  {author} {\bibinfo {author} {\bibfnamefont {M.}~\bibnamefont
  {Salluzzo}}, \bibinfo {author} {\bibfnamefont {G.~M.}\ \bibnamefont
  {de~Luca}}, \bibinfo {author} {\bibfnamefont {D.}~\bibnamefont {Marr\`e}},
  \bibinfo {author} {\bibfnamefont {M.}~\bibnamefont {Putti}}, \bibinfo
  {author} {\bibfnamefont {M.}~\bibnamefont {Tropeano}}, \bibinfo {author}
  {\bibfnamefont {U.}~\bibnamefont {Scotti~di Uccio}}, \ and\ \bibinfo {author}
  {\bibfnamefont {R.}~\bibnamefont {Vaglio}},\ }\bibfield  {title} {\enquote
  {\bibinfo {title} {{Thickness effect on the structure and superconductivity
  of ${\mathrm{Nd}}_{1.2}{\mathrm{Ba}}_{1.8}{\mathrm{Cu}}_{3}{\mathrm{O}}_{z}$
  epitaxial films}},}\ }\href {\doibase 10.1103/PhysRevB.72.134521} {\bibfield
  {journal} {\bibinfo  {journal} {Phys. Rev. B}\ }\textbf {\bibinfo {volume}
  {72}},\ \bibinfo {pages} {134521} (\bibinfo {year} {2005})}\BibitemShut
  {NoStop}%
\bibitem [{\citenamefont {Chmaissem}\ \emph {et~al.}(1999)\citenamefont
  {Chmaissem}, \citenamefont {Jorgensen}, \citenamefont {Short}, \citenamefont
  {Knizhnik}, \citenamefont {Eckstein},\ and\ \citenamefont
  {Shaked}}]{Chmaissem1999}%
  \BibitemOpen
  \bibfield  {author} {\bibinfo {author} {\bibfnamefont {O}~\bibnamefont
  {Chmaissem}}, \bibinfo {author} {\bibfnamefont {JD}~\bibnamefont
  {Jorgensen}}, \bibinfo {author} {\bibfnamefont {S}~\bibnamefont {Short}},
  \bibinfo {author} {\bibfnamefont {A}~\bibnamefont {Knizhnik}}, \bibinfo
  {author} {\bibfnamefont {Y}~\bibnamefont {Eckstein}}, \ and\ \bibinfo
  {author} {\bibfnamefont {H}~\bibnamefont {Shaked}},\ }\bibfield  {title}
  {\enquote {\bibinfo {title} {{Scaling of transition temperature and CuO$_2$
  plane buckling in a high-temperature superconductor}},}\ }\href {\doibase
  10.1038/16209} {\bibfield  {journal} {\bibinfo  {journal} {Nature}\ }\textbf
  {\bibinfo {volume} {397}},\ \bibinfo {pages} {45} (\bibinfo {year}
  {1999})}\BibitemShut {NoStop}%
\bibitem [{\citenamefont {Takita}\ \emph {et~al.}(1988)\citenamefont {Takita},
  \citenamefont {Katoh}, \citenamefont {Akinaga}, \citenamefont {Nishino},
  \citenamefont {Ishigaki},\ and\ \citenamefont {Asano}}]{Takita1988}%
  \BibitemOpen
  \bibfield  {author} {\bibinfo {author} {\bibfnamefont {K{\^{o}}ki}\
  \bibnamefont {Takita}}, \bibinfo {author} {\bibfnamefont {Hideo}\
  \bibnamefont {Katoh}}, \bibinfo {author} {\bibfnamefont {Hiroyuki}\
  \bibnamefont {Akinaga}}, \bibinfo {author} {\bibfnamefont {Makoto}\
  \bibnamefont {Nishino}}, \bibinfo {author} {\bibfnamefont {T\={o}ru}\
  \bibnamefont {Ishigaki}}, \ and\ \bibinfo {author} {\bibfnamefont {Hajime}\
  \bibnamefont {Asano}},\ }\bibfield  {title} {\enquote {\bibinfo {title}
  {{X-Ray Diffraction Study on the Crystal Structure of
  Nd$_{1+x}$Ba$_{2-x}$Cu$_3$O$_{7-\delta}$}},}\ }\href {\doibase
  10.1143/jjap.27.l57} {\bibfield  {journal} {\bibinfo  {journal} {Jpn. J.
  Appl. Phys.}\ }\textbf {\bibinfo {volume} {27}},\ \bibinfo {pages} {L57--L60}
  (\bibinfo {year} {1988})}\BibitemShut {NoStop}%
\bibitem [{\citenamefont {Kramer}\ \emph {et~al.}(1994)\citenamefont {Kramer},
  \citenamefont {Yoo}, \citenamefont {McCallum}, \citenamefont {Yelon},
  \citenamefont {Xie},\ and\ \citenamefont {Allenspach}}]{Kramer1994}%
  \BibitemOpen
  \bibfield  {author} {\bibinfo {author} {\bibfnamefont {M.J.}\ \bibnamefont
  {Kramer}}, \bibinfo {author} {\bibfnamefont {S.I.}\ \bibnamefont {Yoo}},
  \bibinfo {author} {\bibfnamefont {R.W.}\ \bibnamefont {McCallum}}, \bibinfo
  {author} {\bibfnamefont {W.B.}\ \bibnamefont {Yelon}}, \bibinfo {author}
  {\bibfnamefont {H.}~\bibnamefont {Xie}}, \ and\ \bibinfo {author}
  {\bibfnamefont {P.}~\bibnamefont {Allenspach}},\ }\bibfield  {title}
  {\enquote {\bibinfo {title} {{Hole filling, charge transfer and
  superconductivity in Nd$_{1+x}$Ba$_{2-x}$Cu$_3$O$_{7+\delta}$}},}\ }\href
  {\doibase https://doi.org/10.1016/0921-4534(94)90027-2} {\bibfield  {journal}
  {\bibinfo  {journal} {Physica C}\ }\textbf {\bibinfo {volume} {219}},\
  \bibinfo {pages} {145 -- 155} (\bibinfo {year} {1994})}\BibitemShut {NoStop}%
\bibitem [{\citenamefont {Izumi}\ \emph {et~al.}(1987)\citenamefont {Izumi},
  \citenamefont {Takekawa}, \citenamefont {Matsui}, \citenamefont {Iyi},
  \citenamefont {Asano}, \citenamefont {Ishigaki},\ and\ \citenamefont
  {Watanabe}}]{Izumi1987}%
  \BibitemOpen
  \bibfield  {author} {\bibinfo {author} {\bibfnamefont {Fujio}\ \bibnamefont
  {Izumi}}, \bibinfo {author} {\bibfnamefont {Shunji}\ \bibnamefont
  {Takekawa}}, \bibinfo {author} {\bibfnamefont {Yoshio}\ \bibnamefont
  {Matsui}}, \bibinfo {author} {\bibfnamefont {Nobuo}\ \bibnamefont {Iyi}},
  \bibinfo {author} {\bibfnamefont {Hajime}\ \bibnamefont {Asano}}, \bibinfo
  {author} {\bibfnamefont {T\={o}ru}\ \bibnamefont {Ishigaki}}, \ and\ \bibinfo
  {author} {\bibfnamefont {Noboru}\ \bibnamefont {Watanabe}},\ }\bibfield
  {title} {\enquote {\bibinfo {title} {{Crystal Structure of the Superconductor
  Ba$_{1.8}$Nd$_{1.2}$Cu$_{3}$O$_{7-y}$}},}\ }\href {\doibase
  10.1143/jjap.26.l1616} {\bibfield  {journal} {\bibinfo  {journal} {Jpn. J.
  Appl. Phys.}\ }\textbf {\bibinfo {volume} {26}},\ \bibinfo {pages}
  {L1616--L1619} (\bibinfo {year} {1987})}\BibitemShut {NoStop}%
\bibitem [{\citenamefont {Hepting}\ \emph {et~al.}(2018)\citenamefont
  {Hepting}, \citenamefont {Chaix}, \citenamefont {Huang}, \citenamefont
  {Fumagalli}, \citenamefont {Peng}, \citenamefont {Moritz}, \citenamefont
  {Kummer}, \citenamefont {Brookes}, \citenamefont {Lee}, \citenamefont
  {Hashimoto}, \citenamefont {Sarkar}, \citenamefont {He}, \citenamefont
  {Rotundu}, \citenamefont {Lee}, \citenamefont {Greene}, \citenamefont
  {Braicovich}, \citenamefont {Ghiringhelli}, \citenamefont {Shen},
  \citenamefont {Devereaux},\ and\ \citenamefont {Lee}}]{Hepting2018}%
  \BibitemOpen
  \bibfield  {author} {\bibinfo {author} {\bibfnamefont {M.}~\bibnamefont
  {Hepting}}, \bibinfo {author} {\bibfnamefont {L.}~\bibnamefont {Chaix}},
  \bibinfo {author} {\bibfnamefont {E.~W.}\ \bibnamefont {Huang}}, \bibinfo
  {author} {\bibfnamefont {R.}~\bibnamefont {Fumagalli}}, \bibinfo {author}
  {\bibfnamefont {Y.~Y.}\ \bibnamefont {Peng}}, \bibinfo {author}
  {\bibfnamefont {B.}~\bibnamefont {Moritz}}, \bibinfo {author} {\bibfnamefont
  {K.}~\bibnamefont {Kummer}}, \bibinfo {author} {\bibfnamefont {N.~B.}\
  \bibnamefont {Brookes}}, \bibinfo {author} {\bibfnamefont {W.~C.}\
  \bibnamefont {Lee}}, \bibinfo {author} {\bibfnamefont {M.}~\bibnamefont
  {Hashimoto}}, \bibinfo {author} {\bibfnamefont {T.}~\bibnamefont {Sarkar}},
  \bibinfo {author} {\bibfnamefont {J.-F.}\ \bibnamefont {He}}, \bibinfo
  {author} {\bibfnamefont {C.~R.}\ \bibnamefont {Rotundu}}, \bibinfo {author}
  {\bibfnamefont {Y.~S.}\ \bibnamefont {Lee}}, \bibinfo {author} {\bibfnamefont
  {R.~L.}\ \bibnamefont {Greene}}, \bibinfo {author} {\bibfnamefont
  {L.}~\bibnamefont {Braicovich}}, \bibinfo {author} {\bibfnamefont
  {G.}~\bibnamefont {Ghiringhelli}}, \bibinfo {author} {\bibfnamefont {Z.~X.}\
  \bibnamefont {Shen}}, \bibinfo {author} {\bibfnamefont {T.~P.}\ \bibnamefont
  {Devereaux}}, \ and\ \bibinfo {author} {\bibfnamefont {W.~S.}\ \bibnamefont
  {Lee}},\ }\bibfield  {title} {\enquote {\bibinfo {title} {{Three-dimensional
  collective charge excitations in electron-doped copper oxide
  superconductors}},}\ }\href {\doibase 10.1038/s41586-018-0648-3} {\bibfield
  {journal} {\bibinfo  {journal} {Nature}\ }\textbf {\bibinfo {volume} {563}},\
  \bibinfo {pages} {374} (\bibinfo {year} {2018})}\BibitemShut {NoStop}%
\bibitem [{\citenamefont {Keski-Rahkonen}\ and\ \citenamefont
  {Krause}(1974)}]{KeskiRahkonen1974}%
  \BibitemOpen
  \bibfield  {author} {\bibinfo {author} {\bibfnamefont {Olavi}\ \bibnamefont
  {Keski-Rahkonen}}\ and\ \bibinfo {author} {\bibfnamefont {Manfred~O.}\
  \bibnamefont {Krause}},\ }\bibfield  {title} {\enquote {\bibinfo {title}
  {Total and partial atomic-level widths},}\ }\href {\doibase
  https://doi.org/10.1016/S0092-640X(74)80020-3} {\bibfield  {journal}
  {\bibinfo  {journal} {At. Data Nucl. Data Tables}\ }\textbf {\bibinfo
  {volume} {14}},\ \bibinfo {pages} {139 -- 146} (\bibinfo {year}
  {1974})}\BibitemShut {NoStop}%
\bibitem [{\citenamefont {Krause}\ and\ \citenamefont
  {Oliver}(1979)}]{Krause1979}%
  \BibitemOpen
  \bibfield  {author} {\bibinfo {author} {\bibfnamefont {M.~O.}\ \bibnamefont
  {Krause}}\ and\ \bibinfo {author} {\bibfnamefont {J.~H.}\ \bibnamefont
  {Oliver}},\ }\bibfield  {title} {\enquote {\bibinfo {title} {{Natural widths
  of atomic K and L levels, K$\alpha$ X-ray lines and several KLL Auger
  lines}},}\ }\href {\doibase 10.1063/1.555595} {\bibfield  {journal} {\bibinfo
   {journal} {J. Phys. Chem. Ref. Data}\ }\textbf {\bibinfo {volume} {8}},\
  \bibinfo {pages} {329--338} (\bibinfo {year} {1979})}\BibitemShut {NoStop}%
\bibitem [{\citenamefont {Vale}\ \emph {et~al.}(2019)\citenamefont {Vale},
  \citenamefont {Dashwood}, \citenamefont {Paris}, \citenamefont {Veiga},
  \citenamefont {Garcia-Fernandez}, \citenamefont {Nag}, \citenamefont
  {Walters}, \citenamefont {Zhou}, \citenamefont {Pietsch}, \citenamefont
  {Jesche}, \citenamefont {Gegenwart}, \citenamefont {Coldea}, \citenamefont
  {Schmitt},\ and\ \citenamefont {McMorrow}}]{Vale2019}%
  \BibitemOpen
  \bibfield  {author} {\bibinfo {author} {\bibfnamefont {J.~G.}\ \bibnamefont
  {Vale}}, \bibinfo {author} {\bibfnamefont {C.~D.}\ \bibnamefont {Dashwood}},
  \bibinfo {author} {\bibfnamefont {E.}~\bibnamefont {Paris}}, \bibinfo
  {author} {\bibfnamefont {L.~S.~I.}\ \bibnamefont {Veiga}}, \bibinfo {author}
  {\bibfnamefont {M.}~\bibnamefont {Garcia-Fernandez}}, \bibinfo {author}
  {\bibfnamefont {A.}~\bibnamefont {Nag}}, \bibinfo {author} {\bibfnamefont
  {A.}~\bibnamefont {Walters}}, \bibinfo {author} {\bibfnamefont {Ke-Jin}\
  \bibnamefont {Zhou}}, \bibinfo {author} {\bibfnamefont {I.-M.}\ \bibnamefont
  {Pietsch}}, \bibinfo {author} {\bibfnamefont {Anton}\ \bibnamefont {Jesche}},
  \bibinfo {author} {\bibfnamefont {P.}~\bibnamefont {Gegenwart}}, \bibinfo
  {author} {\bibfnamefont {R.}~\bibnamefont {Coldea}}, \bibinfo {author}
  {\bibfnamefont {T.}~\bibnamefont {Schmitt}}, \ and\ \bibinfo {author}
  {\bibfnamefont {D.~F.}\ \bibnamefont {McMorrow}},\ }\bibfield  {title}
  {\enquote {\bibinfo {title} {{High-resolution resonant inelastic x-ray
  scattering study of the electron-phonon coupling in honeycomb
  $\ensuremath{\alpha}\ensuremath{-}{\mathrm{Li}}_{2}{\mathrm{IrO}}_{3}$}},}\
  }\href {\doibase 10.1103/PhysRevB.100.224303} {\bibfield  {journal} {\bibinfo
   {journal} {Phys. Rev. B}\ }\textbf {\bibinfo {volume} {100}},\ \bibinfo
  {pages} {224303} (\bibinfo {year} {2019})}\BibitemShut {NoStop}%
\bibitem [{\citenamefont {Fumagalli}\ \emph {et~al.}(2019)\citenamefont
  {Fumagalli}, \citenamefont {Braicovich}, \citenamefont {Minola},
  \citenamefont {Peng}, \citenamefont {Kummer}, \citenamefont {Betto},
  \citenamefont {Rossi}, \citenamefont {Lefran\c{c}ois}, \citenamefont
  {Morawe}, \citenamefont {Salluzzo}, \citenamefont {Suzuki}, \citenamefont
  {Yakhou}, \citenamefont {Le~Tacon}, \citenamefont {Keimer}, \citenamefont
  {Brookes}, \citenamefont {{Moretti Sala}},\ and\ \citenamefont
  {Ghiringhelli}}]{Fumagalli2019}%
  \BibitemOpen
  \bibfield  {author} {\bibinfo {author} {\bibfnamefont {R.}~\bibnamefont
  {Fumagalli}}, \bibinfo {author} {\bibfnamefont {L.}~\bibnamefont
  {Braicovich}}, \bibinfo {author} {\bibfnamefont {M.}~\bibnamefont {Minola}},
  \bibinfo {author} {\bibfnamefont {Y.~Y.}\ \bibnamefont {Peng}}, \bibinfo
  {author} {\bibfnamefont {K.}~\bibnamefont {Kummer}}, \bibinfo {author}
  {\bibfnamefont {D.}~\bibnamefont {Betto}}, \bibinfo {author} {\bibfnamefont
  {M.}~\bibnamefont {Rossi}}, \bibinfo {author} {\bibfnamefont
  {E.}~\bibnamefont {Lefran\c{c}ois}}, \bibinfo {author} {\bibfnamefont
  {C.}~\bibnamefont {Morawe}}, \bibinfo {author} {\bibfnamefont
  {M.}~\bibnamefont {Salluzzo}}, \bibinfo {author} {\bibfnamefont
  {H.}~\bibnamefont {Suzuki}}, \bibinfo {author} {\bibfnamefont
  {F.}~\bibnamefont {Yakhou}}, \bibinfo {author} {\bibfnamefont
  {M.}~\bibnamefont {Le~Tacon}}, \bibinfo {author} {\bibfnamefont
  {B.}~\bibnamefont {Keimer}}, \bibinfo {author} {\bibfnamefont {N.~B.}\
  \bibnamefont {Brookes}}, \bibinfo {author} {\bibfnamefont {M.}~\bibnamefont
  {{Moretti Sala}}}, \ and\ \bibinfo {author} {\bibfnamefont {G.}~\bibnamefont
  {Ghiringhelli}},\ }\bibfield  {title} {\enquote {\bibinfo {title}
  {{Polarization-resolved Cu ${L}_{3}$-edge resonant inelastic x-ray scattering
  of orbital and spin excitations in
  ${\mathrm{NdBa}}_{2}{\mathrm{Cu}}_{3}{\mathrm{O}}_{7\ensuremath{-}\ensuremath{\delta}}$}},}\
  }\href {\doibase 10.1103/PhysRevB.99.134517} {\bibfield  {journal} {\bibinfo
  {journal} {Phys. Rev. B}\ }\textbf {\bibinfo {volume} {99}},\ \bibinfo
  {pages} {134517} (\bibinfo {year} {2019})}\BibitemShut {NoStop}%
\bibitem [{\citenamefont {Moretti~Sala}\ \emph {et~al.}(2011)\citenamefont
  {Moretti~Sala}, \citenamefont {Bisogni}, \citenamefont {Aruta}, \citenamefont
  {Balestrino}, \citenamefont {Berger}, \citenamefont {Brookes}, \citenamefont
  {de~Luca}, \citenamefont {Di~Castro}, \citenamefont {Grioni}, \citenamefont
  {Guarise}, \citenamefont {Medaglia}, \citenamefont {Miletto~Granozio},
  \citenamefont {Minola}, \citenamefont {Perna}, \citenamefont {Radovic},
  \citenamefont {Salluzzo}, \citenamefont {Schmitt}, \citenamefont {Zhou},
  \citenamefont {Braicovich},\ and\ \citenamefont
  {Ghiringhelli}}]{MorettiSala2011}%
  \BibitemOpen
  \bibfield  {author} {\bibinfo {author} {\bibfnamefont {M.}~\bibnamefont
  {Moretti~Sala}}, \bibinfo {author} {\bibfnamefont {V.}~\bibnamefont
  {Bisogni}}, \bibinfo {author} {\bibfnamefont {C.}~\bibnamefont {Aruta}},
  \bibinfo {author} {\bibfnamefont {G.}~\bibnamefont {Balestrino}}, \bibinfo
  {author} {\bibfnamefont {M.}~\bibnamefont {Berger}}, \bibinfo {author}
  {\bibfnamefont {N.~B.}\ \bibnamefont {Brookes}}, \bibinfo {author}
  {\bibfnamefont {G.~M.}\ \bibnamefont {de~Luca}}, \bibinfo {author}
  {\bibfnamefont {D.}~\bibnamefont {Di~Castro}}, \bibinfo {author}
  {\bibfnamefont {M.}~\bibnamefont {Grioni}}, \bibinfo {author} {\bibfnamefont
  {M.}~\bibnamefont {Guarise}}, \bibinfo {author} {\bibfnamefont {P.~G.}\
  \bibnamefont {Medaglia}}, \bibinfo {author} {\bibfnamefont {F.}~\bibnamefont
  {Miletto~Granozio}}, \bibinfo {author} {\bibfnamefont {M.}~\bibnamefont
  {Minola}}, \bibinfo {author} {\bibfnamefont {P.}~\bibnamefont {Perna}},
  \bibinfo {author} {\bibfnamefont {M.}~\bibnamefont {Radovic}}, \bibinfo
  {author} {\bibfnamefont {M.}~\bibnamefont {Salluzzo}}, \bibinfo {author}
  {\bibfnamefont {T.}~\bibnamefont {Schmitt}}, \bibinfo {author} {\bibfnamefont
  {K.~J.}\ \bibnamefont {Zhou}}, \bibinfo {author} {\bibfnamefont
  {L.}~\bibnamefont {Braicovich}}, \ and\ \bibinfo {author} {\bibfnamefont
  {G.}~\bibnamefont {Ghiringhelli}},\ }\bibfield  {title} {\enquote {\bibinfo
  {title} {{Energy and symmetry of $dd$ excitations in undoped layered cuprates
  measured by Cu $L_{3}$ resonant inelastic x-ray scattering}},}\ }\href
  {http://stacks.iop.org/1367-2630/13/i=4/a=043026} {\bibfield  {journal}
  {\bibinfo  {journal} {New J. Phys.}\ }\textbf {\bibinfo {volume} {13}},\
  \bibinfo {pages} {043026} (\bibinfo {year} {2011})}\BibitemShut {NoStop}%
\bibitem [{\citenamefont {Pintschovius}\ and\ \citenamefont
  {Reichardt}(1998)}]{Pintschovius1998}%
  \BibitemOpen
  \bibfield  {author} {\bibinfo {author} {\bibfnamefont {Lothar}\ \bibnamefont
  {Pintschovius}}\ and\ \bibinfo {author} {\bibfnamefont {Winfried}\
  \bibnamefont {Reichardt}},\ }\enquote {\bibinfo {title} {Phonon dispersions
  and phonon density-of-states in copper-oxide superconductors},}\ in\ \href
  {\doibase 10.1007/978-94-015-1284-8_5} {\emph {\bibinfo {booktitle} {Neutron
  Scattering in Layered Copper-Oxide Superconductors}}},\ \bibinfo {editor}
  {edited by\ \bibinfo {editor} {\bibfnamefont {Albert}\ \bibnamefont
  {Furrer}}}\ (\bibinfo  {publisher} {Springer Netherlands},\ \bibinfo
  {address} {Dordrecht},\ \bibinfo {year} {1998})\ pp.\ \bibinfo {pages}
  {165--223}\BibitemShut {NoStop}%
\bibitem [{\citenamefont {Liu}\ \emph {et~al.}(1988)\citenamefont {Liu},
  \citenamefont {Thomsen}, \citenamefont {Kress}, \citenamefont {Cardona},
  \citenamefont {Gegenheimer}, \citenamefont {de~Wette}, \citenamefont {Prade},
  \citenamefont {Kulkarni},\ and\ \citenamefont {Schr\"oder}}]{Liu1988}%
  \BibitemOpen
  \bibfield  {author} {\bibinfo {author} {\bibfnamefont {R.}~\bibnamefont
  {Liu}}, \bibinfo {author} {\bibfnamefont {C.}~\bibnamefont {Thomsen}},
  \bibinfo {author} {\bibfnamefont {W.}~\bibnamefont {Kress}}, \bibinfo
  {author} {\bibfnamefont {M.}~\bibnamefont {Cardona}}, \bibinfo {author}
  {\bibfnamefont {B.}~\bibnamefont {Gegenheimer}}, \bibinfo {author}
  {\bibfnamefont {F.~W.}\ \bibnamefont {de~Wette}}, \bibinfo {author}
  {\bibfnamefont {J.}~\bibnamefont {Prade}}, \bibinfo {author} {\bibfnamefont
  {A.~D.}\ \bibnamefont {Kulkarni}}, \ and\ \bibinfo {author} {\bibfnamefont
  {U.}~\bibnamefont {Schr\"oder}},\ }\bibfield  {title} {\enquote {\bibinfo
  {title} {{Frequencies, eigenvectors, and single-crystal selection rules of
  k=0 phonons in
  ${\mathrm{YBa}}_{2}$${\mathrm{Cu}}_{3}$${\mathrm{O}}_{7\mathrm{\ensuremath{-}}\mathrm{\ensuremath{\delta}}}$:
  Theory and experiment}},}\ }\href {\doibase 10.1103/PhysRevB.37.7971}
  {\bibfield  {journal} {\bibinfo  {journal} {Phys. Rev. B}\ }\textbf {\bibinfo
  {volume} {37}},\ \bibinfo {pages} {7971--7974} (\bibinfo {year}
  {1988})}\BibitemShut {NoStop}%
\bibitem [{\citenamefont {Cardona}\ \emph {et~al.}(1988)\citenamefont
  {Cardona}, \citenamefont {Liu}, \citenamefont {Thomsen}, \citenamefont
  {Bauer}, \citenamefont {Genzel}, \citenamefont {König}, \citenamefont
  {Wittlin}, \citenamefont {Amador}, \citenamefont {Barahona}, \citenamefont
  {Fernández}, \citenamefont {Otero},\ and\ \citenamefont
  {Sáez}}]{Cardona1988}%
  \BibitemOpen
  \bibfield  {author} {\bibinfo {author} {\bibfnamefont {M.}~\bibnamefont
  {Cardona}}, \bibinfo {author} {\bibfnamefont {R.}~\bibnamefont {Liu}},
  \bibinfo {author} {\bibfnamefont {C.}~\bibnamefont {Thomsen}}, \bibinfo
  {author} {\bibfnamefont {M.}~\bibnamefont {Bauer}}, \bibinfo {author}
  {\bibfnamefont {L.}~\bibnamefont {Genzel}}, \bibinfo {author} {\bibfnamefont
  {W.}~\bibnamefont {König}}, \bibinfo {author} {\bibfnamefont
  {A.}~\bibnamefont {Wittlin}}, \bibinfo {author} {\bibfnamefont
  {U.}~\bibnamefont {Amador}}, \bibinfo {author} {\bibfnamefont
  {M.}~\bibnamefont {Barahona}}, \bibinfo {author} {\bibfnamefont
  {F.}~\bibnamefont {Fernández}}, \bibinfo {author} {\bibfnamefont
  {C.}~\bibnamefont {Otero}}, \ and\ \bibinfo {author} {\bibfnamefont
  {R.}~\bibnamefont {Sáez}},\ }\bibfield  {title} {\enquote {\bibinfo {title}
  {{Infrared and Raman spectra of the new superconducting cuprate perovskites
  MBa$_2$Cu$_3$O$_7$, M = Nd, Dy, Er, Tm}},}\ }\href {\doibase
  10.1016/0038-1098(88)90591-1} {\bibfield  {journal} {\bibinfo  {journal}
  {Solid State Commun.}\ }\textbf {\bibinfo {volume} {65}},\ \bibinfo {pages}
  {71 -- 75} (\bibinfo {year} {1988})}\BibitemShut {NoStop}%
\bibitem [{\citenamefont {Yoshida}\ \emph {et~al.}(1990)\citenamefont
  {Yoshida}, \citenamefont {Gotoh}, \citenamefont {Takata}, \citenamefont
  {Koshizuka},\ and\ \citenamefont {Tanaka}}]{Yoshida1990}%
  \BibitemOpen
  \bibfield  {author} {\bibinfo {author} {\bibfnamefont {M.}~\bibnamefont
  {Yoshida}}, \bibinfo {author} {\bibfnamefont {S.}~\bibnamefont {Gotoh}},
  \bibinfo {author} {\bibfnamefont {T.}~\bibnamefont {Takata}}, \bibinfo
  {author} {\bibfnamefont {N.}~\bibnamefont {Koshizuka}}, \ and\ \bibinfo
  {author} {\bibfnamefont {S.}~\bibnamefont {Tanaka}},\ }\bibfield  {title}
  {\enquote {\bibinfo {title} {{Phonon Raman scattering of
  ${\mathrm{NbBa}}_{2}{\mathrm{Cu}}_{3}{\mathrm{O}}_{\mathit{y}}$ and
  ${\mathrm{Nd}}_{1.6}{\mathrm{Ba}}_{1.4}{\mathrm{Cu}}_{3}{\mathrm{O}}_{\mathit{y}}$}},}\
  }\href {\doibase 10.1103/PhysRevB.41.11689} {\bibfield  {journal} {\bibinfo
  {journal} {Phys. Rev. B}\ }\textbf {\bibinfo {volume} {41}},\ \bibinfo
  {pages} {11689--11692} (\bibinfo {year} {1990})}\BibitemShut {NoStop}%
\bibitem [{\citenamefont {Limonov}\ \emph {et~al.}(1998)\citenamefont
  {Limonov}, \citenamefont {Goodilin}, \citenamefont {Yao}, \citenamefont
  {Tajima}, \citenamefont {Shiohara},\ and\ \citenamefont
  {Kitaev}}]{Limonov1998}%
  \BibitemOpen
  \bibfield  {author} {\bibinfo {author} {\bibfnamefont {M.~F.}\ \bibnamefont
  {Limonov}}, \bibinfo {author} {\bibfnamefont {E.~A.}\ \bibnamefont
  {Goodilin}}, \bibinfo {author} {\bibfnamefont {X.}~\bibnamefont {Yao}},
  \bibinfo {author} {\bibfnamefont {S.}~\bibnamefont {Tajima}}, \bibinfo
  {author} {\bibfnamefont {Y.}~\bibnamefont {Shiohara}}, \ and\ \bibinfo
  {author} {\bibfnamefont {Yu.~E.}\ \bibnamefont {Kitaev}},\ }\bibfield
  {title} {\enquote {\bibinfo {title} {{Phonon Raman study of the
  ${\mathrm{NdBa}}_{2}{\mathrm{Cu}}_{3}{\mathrm{O}}_{y}\ensuremath{-}{\mathrm{Nd}}_{2}{\mathrm{Ba}}_{1}{\mathrm{Cu}}_{3}{\mathrm{O}}_{y}$
  system}},}\ }\href {\doibase 10.1103/PhysRevB.58.12368} {\bibfield  {journal}
  {\bibinfo  {journal} {Phys. Rev. B}\ }\textbf {\bibinfo {volume} {58}},\
  \bibinfo {pages} {12368--12376} (\bibinfo {year} {1998})}\BibitemShut
  {NoStop}%
\bibitem [{\citenamefont {Blanco-Canosa}\ \emph {et~al.}(2014)\citenamefont
  {Blanco-Canosa}, \citenamefont {Frano}, \citenamefont {Schierle},
  \citenamefont {Porras}, \citenamefont {Loew}, \citenamefont {Minola},
  \citenamefont {Bluschke}, \citenamefont {Weschke}, \citenamefont {Keimer},\
  and\ \citenamefont {Le~Tacon}}]{BlancoCanosa2014}%
  \BibitemOpen
  \bibfield  {author} {\bibinfo {author} {\bibfnamefont {S.}~\bibnamefont
  {Blanco-Canosa}}, \bibinfo {author} {\bibfnamefont {A.}~\bibnamefont
  {Frano}}, \bibinfo {author} {\bibfnamefont {E.}~\bibnamefont {Schierle}},
  \bibinfo {author} {\bibfnamefont {J.}~\bibnamefont {Porras}}, \bibinfo
  {author} {\bibfnamefont {T.}~\bibnamefont {Loew}}, \bibinfo {author}
  {\bibfnamefont {M.}~\bibnamefont {Minola}}, \bibinfo {author} {\bibfnamefont
  {M.}~\bibnamefont {Bluschke}}, \bibinfo {author} {\bibfnamefont
  {E.}~\bibnamefont {Weschke}}, \bibinfo {author} {\bibfnamefont
  {B.}~\bibnamefont {Keimer}}, \ and\ \bibinfo {author} {\bibfnamefont
  {M.}~\bibnamefont {Le~Tacon}},\ }\bibfield  {title} {\enquote {\bibinfo
  {title} {{Resonant x-ray scattering study of charge-density wave correlations
  in ${\mathrm{YBa}}_{2}{\mathrm{Cu}}_{3}{\mathrm{O}}_{6+x}$}},}\ }\href
  {\doibase 10.1103/PhysRevB.90.054513} {\bibfield  {journal} {\bibinfo
  {journal} {Phys. Rev. B}\ }\textbf {\bibinfo {volume} {90}},\ \bibinfo
  {pages} {054513} (\bibinfo {year} {2014})}\BibitemShut {NoStop}%
\bibitem [{\citenamefont {Devereaux}\ and\ \citenamefont
  {Einzel}(1995)}]{Devereaux1995}%
  \BibitemOpen
  \bibfield  {author} {\bibinfo {author} {\bibfnamefont {T.~P.}\ \bibnamefont
  {Devereaux}}\ and\ \bibinfo {author} {\bibfnamefont {D.}~\bibnamefont
  {Einzel}},\ }\bibfield  {title} {\enquote {\bibinfo {title} {{Electronic
  Raman scattering in superconductors as a probe of anisotropic electron
  pairing}},}\ }\href {\doibase 10.1103/PhysRevB.51.16336} {\bibfield
  {journal} {\bibinfo  {journal} {Phys. Rev. B}\ }\textbf {\bibinfo {volume}
  {51}},\ \bibinfo {pages} {16336--16357} (\bibinfo {year} {1995})}\BibitemShut
  {NoStop}%
\bibitem [{\citenamefont {Lin}\ \emph {et~al.}(2020)\citenamefont {Lin},
  \citenamefont {Miao}, \citenamefont {Mazzone}, \citenamefont {Gu},
  \citenamefont {Nag}, \citenamefont {Walters}, \citenamefont
  {Garcia-Fernandez}, \citenamefont {Barbour}, \citenamefont {Pelliciari},
  \citenamefont {Jarrige}, \citenamefont {Oda}, \citenamefont {Kurosawa},
  \citenamefont {Momono}, \citenamefont {Zhou}, \citenamefont {Bisogni},
  \citenamefont {Liu},\ and\ \citenamefont {Dean}}]{lin2020nature}%
  \BibitemOpen
  \bibfield  {author} {\bibinfo {author} {\bibfnamefont {J.~Q.}\ \bibnamefont
  {Lin}}, \bibinfo {author} {\bibfnamefont {H.}~\bibnamefont {Miao}}, \bibinfo
  {author} {\bibfnamefont {D.~G.}\ \bibnamefont {Mazzone}}, \bibinfo {author}
  {\bibfnamefont {G.~D.}\ \bibnamefont {Gu}}, \bibinfo {author} {\bibfnamefont
  {A.}~\bibnamefont {Nag}}, \bibinfo {author} {\bibfnamefont {A.~C.}\
  \bibnamefont {Walters}}, \bibinfo {author} {\bibfnamefont {M.}~\bibnamefont
  {Garcia-Fernandez}}, \bibinfo {author} {\bibfnamefont {A.}~\bibnamefont
  {Barbour}}, \bibinfo {author} {\bibfnamefont {J.}~\bibnamefont {Pelliciari}},
  \bibinfo {author} {\bibfnamefont {I.}~\bibnamefont {Jarrige}}, \bibinfo
  {author} {\bibfnamefont {M.}~\bibnamefont {Oda}}, \bibinfo {author}
  {\bibfnamefont {K.}~\bibnamefont {Kurosawa}}, \bibinfo {author}
  {\bibfnamefont {N.}~\bibnamefont {Momono}}, \bibinfo {author} {\bibfnamefont
  {K.}~\bibnamefont {Zhou}}, \bibinfo {author} {\bibfnamefont {V.}~\bibnamefont
  {Bisogni}}, \bibinfo {author} {\bibfnamefont {X.}~\bibnamefont {Liu}}, \ and\
  \bibinfo {author} {\bibfnamefont {M.~P.~M.}\ \bibnamefont {Dean}},\
  }\bibfield  {title} {\enquote {\bibinfo {title} {{Nature of the
  charge-density wave excitations in cuprates}},}\ }\href
  {https://arxiv.org/abs/2001.10312} {\  (\bibinfo {year} {2020})},\ \Eprint
  {http://arxiv.org/abs/2001.10312} {arXiv:2001.10312 [cond-mat.supr-con]}
  \BibitemShut {NoStop}%
\bibitem [{\citenamefont {Markiewicz}\ \emph {et~al.}(2005)\citenamefont
  {Markiewicz}, \citenamefont {Sahrakorpi}, \citenamefont {Lindroos},
  \citenamefont {Lin},\ and\ \citenamefont {Bansil}}]{Markiewicz2005}%
  \BibitemOpen
  \bibfield  {author} {\bibinfo {author} {\bibfnamefont {R.~S.}\ \bibnamefont
  {Markiewicz}}, \bibinfo {author} {\bibfnamefont {S.}~\bibnamefont
  {Sahrakorpi}}, \bibinfo {author} {\bibfnamefont {M.}~\bibnamefont
  {Lindroos}}, \bibinfo {author} {\bibfnamefont {Hsin}\ \bibnamefont {Lin}}, \
  and\ \bibinfo {author} {\bibfnamefont {A.}~\bibnamefont {Bansil}},\
  }\bibfield  {title} {\enquote {\bibinfo {title} {{One-band tight-binding
  model parametrization of the high-${T}_{c}$ cuprates including the effect of
  ${k}_{z}$ dispersion}},}\ }\href {\doibase 10.1103/PhysRevB.72.054519}
  {\bibfield  {journal} {\bibinfo  {journal} {Phys. Rev. B}\ }\textbf {\bibinfo
  {volume} {72}},\ \bibinfo {pages} {054519} (\bibinfo {year}
  {2005})}\BibitemShut {NoStop}%
\end{thebibliography}%

\end{document}